\documentclass[11pt]{article}
\usepackage[latin9]{inputenc}
\usepackage{amsfonts}
\usepackage{amsmath}
\usepackage{amssymb}
\usepackage{indentfirst}
\usepackage{graphicx}
\usepackage[colorlinks]{hyperref}
\usepackage{cite}
\usepackage{graphicx}

\numberwithin{equation}{section}
\begin{document}

\begin{titlepage}
\vspace{3cm}
\baselineskip=24pt

\begin{center}
\textbf{\LARGE{$\mathcal{N}$-extended Chern-Simons Carrollian supergravities in $2+1$ spacetime dimensions}}
\par\end{center}{\LARGE \par}

\begin{center}
	\vspace{1cm}
	\textbf{Farhad Ali}$^{\ast}$,
	\textbf{Lucrezia Ravera}$^{\star}$
	\small
	\\[5mm]
	$^{\ast}$\textit{Department of Mathematics, Kohat University of Science and Technology, }\\
	\textit{ 26000 Kohat KPK, Pakistan.}
	\\[2mm]
    $^{\star}$\textit{INFN, Sezione di Milano, }\\
	\textit{ Via Celoria 16, I-20133 Milano, Italy.}
	\\[5mm]
	\footnotesize
	\texttt{farhadali@kust.edu.pk},
	\texttt{lucrezia.ravera@mi.infn.it}
	\par\end{center}
\vskip 20pt
\begin{abstract}
\noindent

In this work we present the ultra-relativistic $\mathcal{N}$-extended AdS Chern-Simons supergravity theories in three spacetime dimensions invariant under $\mathcal{N}$-extended AdS Carroll superalgebras. We first consider the $(2,0)$ and $(1,1)$ cases; subsequently, we generalize our analysis to $\mathcal{N}=(\mathcal{N},0)$, with $\mathcal{N}$ even, and to $\mathcal{N}=(p,q)$, with $p,q>0$.
The $\mathcal{N}$-extended AdS Carroll superalgebras are obtained through the Carrollian (i.e., ultra-relativistic) contraction applied to an $so(2)$ extension of $\mathfrak{osp}(2|2)\otimes \mathfrak{sp}(2)$, to $\mathfrak{osp}(2|1)\otimes \mathfrak{osp}(2,1)$, to an $\mathfrak{so}(\mathcal{N})$ extension of $\mathfrak{osp}(2|\mathcal{N})\otimes \mathfrak{sp}(2)$, and to the direct sum of an $\mathfrak{so}(p) \oplus \mathfrak{so}(q)$ algebra and $\mathfrak{osp}(2|p)\otimes \mathfrak{osp}(2,q)$, respectively. 
We also analyze the flat limit ($\ell \rightarrow \infty$, being $\ell$ the length parameter) of the aforementioned $\mathcal{N}$-extended Chern-Simons AdS Carroll supergravities, in which we recover the ultra-relativistic $\mathcal{N}$-extended (flat) Chern-Simons supergravity theories invariant under $\mathcal{N}$-extended super-Carroll algebras. The flat limit is applied at the level of the superalgebras, Chern-Simons actions, supersymmetry transformation laws, and field equations.

\end{abstract}
\end{titlepage}\newpage {}


\section{Introduction}

The study of spacetime symmetries has proved to be fundamental for analyzing and understanding various physical models. For instance, think of Newtonian gravity, Maxwell's electromagnetism, special and general relativity, string and supergravity theory. Most of these theories are based on relativistic
symmetries. On the other hand, during the years models with non-relativistic symmetries have also been developed and analyzed, and and are still the subject of in-depth studies. 

In this context, Carroll symmetries \cite{LL, Bacry:1968zf}, arising when the velocity of light is sent to zero (i.e., in the ultra-relativistic limit, $c \rightarrow 0$), have attracted some interest over recent years. In fact, models with Carroll symmetries occurred in the literature in the study of tachyon condensation \cite{Gibbons:2002tv}, warped conformal field theories \cite{Hofman:2014loa}, and tensionless strings \cite{Bagchi:2013bga, Bagchi:2015nca, Bagchi:2016yyf, Bagchi:2017cte, Bagchi:2018wsn}. Moreover, a study exploring the Carroll limit corresponding to M2- as well as M3-branes propagating over $D=11$ supergravity backgrounds in M-theory has been recently presented in \cite{Roychowdhury:2019aoi}.

Concerning gravity theories, models of Carrollian (i.e., ultra-relativistic) gravity have been developed and analyzed in \cite{Hartong:2015xda, Bergshoeff:2016soe, Bergshoeff:2017btm}.
In particular, non- and ultra-relativistic Chern-Simons (CS) type actions in $2+1$ dimensions were constructed in \cite{Bergshoeff:2016soe}, where the authors included a spin-$3$ field coupled to gravity.
The geometry of flat and Anti-de Sitter (AdS) Carroll spaces were investigated, both in the bosonic as well as in the supersymmetric case, in \cite{Bergshoeff:2015wma}, were the authors focused on the symmetries of a particle moving in such spaces. Afterwards, in \cite{Matulich:2019cdo}, the AdS Carroll CS gravity theory was discussed for the first time.\footnote{Here let us also mention that the geometric realization of the (A)dS Carroll algebra corresponds to a null surface of (A)dS space (see Refs. \cite{Figueroa-OFarrill:2018ilb, Figueroa-OFarrill:2019sex}).}

Non-relativistic symmetry groups play a remarkable role also in holography \cite{Bagchi:2009my, Christensen:2013lma, Christensen:2013rfa, Hartong:2014oma, Bergshoeff:2014uea, Hartong:2015wxa, Bagchi:2010eg, Bagchi:2012cy, Bagchi:2016bcd, Lodato:2016alv, Bagchi:2019xfx, Duval:2014uva, Duval:2014lpa, Ciambelli:2018xat, Ciambelli:2018wre, Ciambelli:2018ojf, Campoleoni:2018ltl}. 
More specifically, in the work \cite{Bagchi:2010eg}, connections among Carrollian physics, holography of flat space, and the Bondi-Metzner-Sachs (BMS) algebra were discovered and followed up in \cite{Bagchi:2012cy} (see also Refs. \cite{Bagchi:2016bcd, Lodato:2016alv} and \cite{Bagchi:2019xfx}). Besides, conformal extensions of the Carroll group were explored and related to the BMS group in \cite{Duval:2014uva, Duval:2014lpa}, while in \cite{Ciambelli:2018xat, Ciambelli:2018wre, Ciambelli:2018ojf, Campoleoni:2018ltl} it was shewed the way in which Carrollian structures and geometry emerge in the flat holography and fluid/gravity correspondence framework.

Recently, in \cite{Ravera:2019ize} the construction of the three-dimensional $\mathcal{N}=1$ CS supergravity theory invariant under the so-called AdS Carroll superalgebra (ultra-relativistic contraction of the $\mathcal{N}=1$ AdS superalgebra, see \cite{Bergshoeff:2015wma}), together with the study of its flat limit, has been presented for the first time. In the study done in \cite{Ravera:2019ize}, the method introduced in \cite{Concha:2016zdb} was adopted. In \cite{Concha:2016zdb} a generalization of the standard In\"{o}n\"{u}-Wigner (IW) contraction \cite{IW, WW} was presented, consisting in rescaling not only the generators of a Lie (super)algebra but also the arbitrary constants appearing in the components of the invariant tensor; the latter is the key ingredient for formulating an action \`{a} la CS, that is invariant under the IW contracted (super)algebra by construction.
This procedure was further improved in \cite{Ravera:2019ize} by considering dimensionful generators from the
very beginning, on the same lines of \cite{Concha:2018jxx}. As shown in the same paper, this allows to obtain, for instance, the Poincar\'{e} limit from the $\mathfrak{osp}(2|2) \otimes \mathfrak{sp}(2)$ CS supergravity action directly considering the flat limit $\ell \rightarrow \infty$, being $\ell$ the length parameter related to the cosmological constant. 

Moving to higher $\mathcal{N}$, we have that the $\mathcal{N} = 2$ supersymmetric extensions of the Poincar\'{e} and AdS algebras are not unique, and can be subdivided into two inequivalent classes: The $(2, 0)$ and the $(1,1)$ cases.
Here we mention that, in the extension to the $(p,q)$ case, when either $p$ or $q$ is greater than one some subtleties arise. Indeed, even though the $(p, q)$ Poincar\'{e} superalgebra can be derived as an IW contraction of the $(p, q)$ AdS superalgebra, the Poincaré limit applied at the level of the (CS) action requires to enlarge the AdS superalgebra, considering, in particular, a direct sum of an $\mathfrak{so}(p)\oplus \mathfrak{so}(q)$ algebra and the $(p,q)$ AdS superalgebra $\mathfrak{osp}(2|p)\otimes \mathfrak{osp}(2,q)$ \cite{Howe:1995zm, Giacomini:2006dr, deAzcarraga:2011pa}; this is related to the fact that, as it was proven in \cite{Howe:1995zm}, the semi-direct extension of the $\mathfrak{so}(p)\oplus \mathfrak{so}(q)$ automorphism algebra by the $(p,q)$ Poincar\'{e} superalgebra allows to produce a non-degenerate invariant tensor which is used to construct a well-defined three-dimensional CS $(p,q)$ Poincar\'{e} supergravity theory (in particular, when either $p$ or $q$ is greater than one, it is not possible to obtain a non-degenerate invariant tensor without considering this extension).   

It is well assumed that a (super)gravity theory in three spacetime dimensions can be described by a CS action as a gauge theory, providing a useful toy model to approach higher-dimensional models \cite{DK, Deser, PvN, AT1, RPvN, Witten, AT2, NG, Banados:1996hi}. 
In the last decades, diverse three-dimensional supergravity theories have been studied, and, in this context, there has also been a growing interest to extend AdS and Poincar\'{e} supergravity theories to
other symmetries (see \cite{Concha:2018jxx, Concha:2019icz} and references therein).

In the present work, we apply the method of \cite{Concha:2016zdb} with the improvements of \cite{Ravera:2019ize} to develop in a systematic way the ultra-relativistic $\mathcal{N}$-extended AdS CS supergravity theories in three (that is $2+1$) spacetime dimensions invariant under $\mathcal{N}$-extended AdS Carroll superalgebras. In particular, we will distinguish between the two $\mathcal{N}$-extended cases $\mathcal{N}=(\mathcal{N},0)$ and $\mathcal{N}=(p,q)$, generalizing the results presented at the algebraic level in \cite{Bergshoeff:2015wma} and also developing the associated CS supergravity theories in three dimensions. 
More specifically, we start by considering the $(2,0)$ and $(1,1)$ cases, and then generalize our analysis to $\mathcal{N}=(\mathcal{N},0)$, with $\mathcal{N}$ even, and to $\mathcal{N}=(p,q)$, that is $\mathcal{N}=p+q$, with {$p,q>0$}.\footnote{We concentrate on the $\mathcal{N}=(\mathcal{N},0)$ case with $\mathcal{N}$ {even in} order to reproduce a well-defined ultra-relativistic limit at the supersymmetric level (see also \cite{Lukierski:2006tr} and references therein, which deals with similar situations but in the non-relativistic limit $c \rightarrow \infty$).}
The $\mathcal{N}$-extended AdS Carroll superalgebras are obtained through the Carrollian (i.e., ultra-relativistic) contraction applied to an $so(2)$ extension of $\mathfrak{osp}(2|2)\otimes \mathfrak{sp}(2)$, to $\mathfrak{osp}(2|1)\otimes \mathfrak{osp}(2,1)$, to an $\mathfrak{so}(\mathcal{N})$ extension of $\mathfrak{osp}(2|\mathcal{N})\otimes \mathfrak{sp}(2)$ (with $\mathcal{N}$ even), and to the direct sum of an $\mathfrak{so}(p) \oplus \mathfrak{so}(q)$ algebra and $\mathfrak{osp}(2|p)\otimes \mathfrak{osp}(2,q)${,} respectively. Let us mention that the $\mathcal{N}=(\mathcal{N},0)$ case (and thus also the $\mathcal{N}=(2,0)$ one) will be more subtle, since it will require the definitions of new supersymmetry generators in order to properly study the Carroll limit, on the same lines of what was done in \cite{Bergshoeff:2015wma} (see also \cite{Lukierski:2006tr}, which deals with non-relativistic superalgebras, and references therein).
The ultra-relativistic $\mathcal{N}$-extended AdS Carroll supergravity actions are constructed \`{a} la CS, by exploiting the non-vanishing components of the corresponding invariant tensor. The aforementioned actions are all based on a non-degenerate, invariant bilinear form (i.e., an invariant metric).
The results we have obtained in the present work were also open problems suggested in Ref. \cite{Bergshoeff:2015wma}, and represent $\mathcal{N}$-extended generalizations of \cite{Ravera:2019ize}.

Subsequently, we study the flat limit ($\ell \rightarrow \infty$, being $\ell$ the length parameter) of the aforesaid $\mathcal{N}$-extended CS AdS Carroll supergravities, in which we recover the ultra-relativistic $\mathcal{N}$-extended (flat) CS supergravity theories invariant under $\mathcal{N}$-extended super-Carroll algebras. The flat limit is applied at the level of the superalgebras, CS actions, supersymmetry transformation laws, and field equations. 

The remain of the paper is organized as follows: In Section \ref{20case}, we first introduce a new $\mathcal{N}=(2,0)$ AdS Carroll superalgebra, which is obtained as the ultra-relativistic contraction of an $\mathfrak{so}(2)$ extension of $\mathfrak{osp}(2|2)\otimes \mathfrak{sp}(2)$. Here, the $\mathfrak{so}(2)$ extension is necessary in order to end up with an invariant non-degenerate inner product in the ultra-relativistic limit, providing a well-defined CS action. In fact, this allows us to subsequently develop the three-dimensional CS supergravity action invariant under the $(2,0)$ AdS Carroll superalgebra, which we call $(2,0)$ CS AdS Carroll supergravity in $2+1$ dimensions. In Section \ref{11case}, we repeat the same analysis for the $(1,1)$ case, ending up with the CS supergravity action invariant under the $(1,1)$ AdS Carroll superalgebra. Subsequently, we generalize our study to the cases of $\mathcal{N}=(\mathcal{N},0)$, with $\mathcal{N}$ even, and $\mathcal{N}=(p,q)$, with {$p,q>0$}, respectively in {Sections} \ref{ncase} and \ref{pqcase}. In Section \ref{flatlimit}, we discuss the flat limit $\ell \rightarrow \infty$ of the $\mathcal{N}$-extended CS AdS Carroll supergravities introduced in the previous part of the work. Section \ref{conclusions} contains some final comments and remarks.

\section{$(2,0)$ AdS Carroll supergravity in $2+1$ dimensions}\label{20case}

In this section, we first introduce a new $\mathcal{N}=(2,0)$ AdS Carroll superalgebra, which is obtained as the ultra-relativistic contraction of an $\mathfrak{so}(2)$ extension of $\mathfrak{osp}(2|2)\otimes \mathfrak{sp}(2)$. The $\mathfrak{so}(2)$ extension of $\mathfrak{osp}(2|2)\otimes \mathfrak{sp}(2)$ is needed in order to end up with an invariant non-degenerate inner product in the ultra-relativistic limit, namely with a well-defined invariant tensor (this is reminescent of what was done in \cite{Howe:1995zm} in the case of relativistic theories), in such a way to be able to construct a well-defined CS action. Indeed, this allows us to subsequently develop the three-dimensional CS supergravity action invariant under this $\mathcal{N}=(2,0)$ AdS Carroll superalgebra, which we call $(2,0)$ CS AdS Carroll supergravity.

Let us mention, here, that a $\mathcal{N}=(2,0)$ AdS Carroll superalgebra has been first introduced in \cite{Bergshoeff:2015wma}. Nevertheless, due to the degeneracy of the invariant tensor for that superalgebra, one could not construct a well-defined CS action in that case. On the other hand, as we will see in the following, our $\mathcal{N}=(2,0)$ AdS Carroll superalgebra will be different from the one presented in \cite{Bergshoeff:2015wma}, allowing, in particular, the formulation of a three-dimensional CS action in the supergravity context.

\subsection{$\mathcal{N}=(2,0)$ AdS Carroll superalgebra} 

An $\mathcal{N}=(2,0)$ supersymmetric extension of the AdS Carroll algebra was obtained in \cite{Bergshoeff:2015wma} as the ultra-relativistic contraction of $\mathfrak{osp}(2|2)\otimes \mathfrak{sp}(2)$, the latter being generated by the set $\lbrace \tilde{J}_{AB}, \tilde{P}_A , \tilde{Z}^{ij}, \tilde{Q}^i_\alpha \rbrace$, with $A, B , \ldots=0,1,2$, $\alpha=1,2$, and $i=1,2$, where $\tilde{J}_{AB}$ are the Lorentz generators, $\tilde{P}_A$ represent the spacetime translations, $\tilde{Z}^{ij} = \epsilon^{ij} \tilde{Z}$ are internal symmetry generators, and $\tilde{Q}^i_\alpha$ are the supersymmetry generators ($2$-components Majorana spinor charges). The (anti)commutation relations of $\mathfrak{osp}(2|2)\otimes \mathfrak{sp}(2)$ read as follows:
\begin{equation}\label{osp22nomod}
\begin{split}
& \left[ \tilde{J}_{AB}, \tilde{J}_{CD}\right] =\eta _{BC}\tilde{J}_{AD}-\eta _{AC}\tilde{J}_{BD}-\eta
_{BD}\tilde{J}_{AC}+\eta _{AD}\tilde{J}_{BC} \, ,  \\
& \left[ \tilde{J}_{AB},\tilde{P}_{C}\right] =\eta _{BC}\tilde{P}_{A}-\eta _{AC}\tilde{P}_{B} \,  , \\
& \left[ \tilde{P}_A , \tilde{P}_B \right] = \frac{1}{\ell^2} \tilde{J}_{AB} \,  ,   \\
& \left[ \tilde{J}_{AB},\tilde{Q}^i_{\alpha }\right] =-\frac{1}{2}\left( \Gamma _{AB}\tilde{Q}^i\right)_{\alpha } \,  , \quad \quad \left[ \tilde{P}_{A}, \tilde{Q}^i_{\alpha }\right] =-\frac{1}{2 \ell} \left( \Gamma _{A}\tilde{Q}^i\right) _{\alpha } \,  , \\
& \left[ \tilde{Z} , \tilde{Q}^i_{\alpha } \right] = - \epsilon^{ij} \tilde{Q}^j_\alpha \, , \\
& \left\{ \tilde{Q}^i_{\alpha }, \tilde{Q}^j_{\beta }\right\} = \delta^{ij} \left[ - \frac{1}{2 \ell} \left(\Gamma^{AB} C \right)_{\alpha \beta} \tilde{J}_{AB} + \left( \Gamma ^{A}C\right) _{\alpha \beta }\tilde{P}_{A} \right] + \frac{1}{\ell} \epsilon^{ij} C_{\alpha \beta}  \tilde{Z}  \,  ,
\end{split}
\end{equation}
where $\ell$ is a length parameter, $C$ is the charge conjugation matrix, and $\Gamma_A$ and $\Gamma_{AB}$ represent the Dirac matrices in three dimensions. The generators $\tilde{J}_{AB}$, $\tilde{P}_A $, $\tilde{Z}^{ij}$, and $\tilde{Q}^i_\alpha$ have a dual description in terms of $1$-form fields, $\tilde{\omega}^{AB}$ (spin connection), $\tilde{V}^A$ (vielbein), $\tilde{z}_{ij}$ ($1$-form field dual to the generator $\tilde{Z}^{ij}$), and $\tilde{\psi}_i^\alpha$ (gravitinos), respectively. 

Here, we consider the ultra-relativistic contraction of an $\mathfrak{so}(2)$ extension of $\mathfrak{osp}(2|2)\otimes \mathfrak{sp}(2)$, involving an extra generator $\tilde{S}^{ij} = \epsilon^{ij} \tilde{S}$. This will allow the formulation of a well-defined ultra-relativistic CS action, based on a non-degenerate invariant tensor, which would not be possible by considering the $\mathcal{N}=(2,0)$ superalgebra of \cite{Bergshoeff:2015wma}. In particular, we extend \eqref{osp22nomod} by adding the extra $\tilde{S}$ generator and we perform, on the same lines of \cite{Howe:1995zm}, the redefinition
\begin{equation}
\tilde{T} \equiv \tilde{Z} - \ell \tilde{S} \, ,
\end{equation}
to eliminate $\tilde{Z}$ in favour of $\tilde{T}$ (this redefinition is particularly convenient for discussing the flat limit, see also \cite{Howe:1995zm}). Consequently, we rewrite the (anti)commutation relations \eqref{osp22nomod} as follows (we consider dimensionful generators from the very beginning, on the same lines of \cite{Concha:2018jxx}):\footnote{We use the metric $\eta_{AB}$ with the signature $(-,+,+)$.}
\begin{equation}\label{osp22}
\begin{split}
& \left[ \tilde{J}_{AB}, \tilde{J}_{CD}\right] =\eta _{BC}\tilde{J}_{AD}-\eta _{AC}\tilde{J}_{BD}-\eta
_{BD}\tilde{J}_{AC}+\eta _{AD}\tilde{J}_{BC} \, ,  \\
& \left[ \tilde{J}_{AB},\tilde{P}_{C}\right] =\eta _{BC}\tilde{P}_{A}-\eta _{AC}\tilde{P}_{B} \,  , \\
& \left[ \tilde{P}_A , \tilde{P}_B \right] = \frac{1}{\ell^2} \tilde{J}_{AB} \,  ,   \\
& \left[ \tilde{J}_{AB},\tilde{Q}^i_{\alpha }\right] =-\frac{1}{2}\left( \Gamma _{AB}\tilde{Q}^i\right)_{\alpha } \,  , \quad \quad \left[ \tilde{P}_{A}, \tilde{Q}^i_{\alpha }\right] =-\frac{1}{2 \ell} \left( \Gamma _{A}\tilde{Q}^i\right) _{\alpha } \,  , \\
& \left[ \tilde{T} , \tilde{Q}^i_{\alpha } \right] = - \epsilon^{ij} \tilde{Q}^j_\alpha \, , \\
& \left\{ \tilde{Q}^i_{\alpha }, \tilde{Q}^j_{\beta }\right\} = \delta^{ij} \left[ - \frac{1}{2 \ell} \left(\Gamma^{AB} C \right)_{\alpha \beta} \tilde{J}_{AB} + \left( \Gamma ^{A}C\right) _{\alpha \beta }\tilde{P}_{A} \right] + \epsilon^{ij} C_{\alpha \beta} \left( \frac{1}{\ell} \tilde{T} + \tilde{S} \right) \,  .
\end{split}
\end{equation}
Note that, in the flat limit $\ell \rightarrow \infty$, $\tilde{S}$ becomes the central element of the $\mathcal{N}=(2,0)$ Poincar\'{e} superalgebra extended with the extra $so(2)$ generator $\tilde{T}$ (see \cite{Howe:1995zm}).

The non-vanishing components of an invariant tensor for the superalgebra \eqref{osp22}, which will be useful in the sequel, are given by
\begin{equation}\label{invtosp22}
\begin{split}
& \langle \tilde{J}_{AB} \tilde{J}_{CD} \rangle = \alpha_0 \left( \eta_{AD} \eta_{BC} - \eta_{AC} \eta_{BD} \right) \, , \\
& \langle \tilde{J}_{AB} \tilde{P}_{C} \rangle = \alpha_1 \epsilon_{ABC} \,  , \\
& \langle \tilde{P}_{A} \tilde{P}_{B} \rangle = \frac{\alpha_0}{\ell^2} \eta_{AB} \,  , \\
& \langle \tilde{T} \tilde{T} \rangle = - 2 \alpha_0   \, , \\
& \langle \tilde{T} \tilde{S} \rangle = 2 \alpha_1 \, , \\
& \langle \tilde{S} \tilde{S} \rangle = - \frac{2 \alpha_1}{\ell} \, , \\
& \langle \tilde{Q}^i_\alpha \tilde{Q}^j_\beta \rangle = 2 \left(\alpha_1 -\frac{\alpha_0}{\ell} \right) C_{\alpha \beta} \delta^{ij} \,  ,
\end{split}
\end{equation}
where $\epsilon_{ABC}$ is the Levi-Civita symbol in three dimensions and $\alpha_0$ and $\alpha_1$ are arbitrary constants.

To take the Carrollian (i.e., ultra-relativistic) contraction of the superalgebra \eqref{osp22}, we decompose the indices $A, B, \ldots$ as
\begin{equation}\label{indexdec}
A \rightarrow (0,a) \, , \quad a=1,2 \,  .
\end{equation}
This first step induces the following decomposition of the generators:
\begin{equation}\label{gendec}
\tilde{J}_{AB} \rightarrow \lbrace \tilde{J}_{ab}, \tilde{J}_{a0} \equiv \tilde{K}_a \rbrace \,  , \quad
\tilde{P}_A  \rightarrow \lbrace \tilde{P}_a , \tilde{P}_0 \equiv \tilde{H} \rbrace \,  .
\end{equation}
We also have
\begin{equation}\label{gammadec}
\Gamma_{AB} \rightarrow \lbrace \Gamma_{ab} , \Gamma_{a0} \rbrace \,  , \quad \Gamma_A \rightarrow \lbrace \Gamma_a , \Gamma_0 \rbrace \,  .
\end{equation}
Furthermore, we define new supersymmetry charges by
\begin{equation}\label{newsusy22}
\tilde{Q}^{\pm}_\alpha = \frac{1}{\sqrt{2}} \left( \tilde{Q}^1_\alpha \pm  \left(\Gamma_0\right)_{\alpha \beta} \tilde{Q}^2_\beta \right) \, {,}
\end{equation}
{on the same lines of what was done in \cite{Concha:2019mxx} (see also references therein) in the study of supersymmetric non-relativistic models.} Then, we rescale the generators with a parameter $\sigma$ as follows: 
\begin{equation}\label{resc}
\tilde{H} \rightarrow \sigma H \, , \quad \tilde{K}_a \rightarrow \sigma K_a \,  , \quad  \tilde{S} \rightarrow \sigma S  \, , \quad \tilde{Q}^\pm_\alpha \rightarrow \sqrt{\sigma} Q^\pm_\alpha  \, .
\end{equation}  
Taking the limit $\sigma \rightarrow \infty$\footnote{Let us highlight that $\sigma \rightarrow \infty$ corresponds to the limit $\frac{1}{c} \rightarrow \infty$ (where $c$ denotes the velocity of light), that is $c \rightarrow 0$ (ultra-relativistic limit).} and removing the tilde symbol also on the generators that we have not rescaled, we end up with a new $\mathcal{N}=(2,0)$ AdS Carroll superalgebra (differing from the one of \cite{Bergshoeff:2015wma}, due to the presence of the generator $S$), whose non-trivial (anti)commutation relations read as follows:
\begin{equation}\label{adscarrollsuper22}
\begin{split}
& \left[ K_a , J_{bc} \right] = \delta_{ab} K_c - \delta_{ac} K_b \, , \\
& \left[ J_{ab}, P_{c}\right] =\delta_{bc}P_{a}-\delta _{ac}P_{b} \, , \\
& \left[ K_{a}, P_{b}\right] = - \delta_{ab} H \, , \\
& \left[ P_a , P_b \right] = \frac{1}{\ell^2} J_{ab} \, , \\
& \left[ P_a , H \right] = \frac{1}{\ell^2} K_{a} \, , \\
& \left[ J_{ab},Q^\pm_{\alpha }\right] =-\frac{1}{2}\left( \Gamma _{ab} Q^\pm \right)_{\alpha } \, , \\
& \left[ P_{a}, Q^\pm_{\alpha }\right] =-\frac{1}{2 \ell} \left( \Gamma _{a}Q^\mp\right) _{\alpha } \, , \\
& \left[ T, Q^+_{\alpha }\right] = \left( \Gamma _{0} \right)_{\alpha \beta} Q^+_\beta \, , \\
& \left[ T, Q^-_{\alpha }\right] = - \left( \Gamma _{0} \right)_{\alpha \beta} Q^-_\beta \, , \\
& \left\{ Q^+_{\alpha }, Q^+_{\beta }\right\} = \left( \Gamma ^{0}C\right) _{\alpha \beta } \left( H + S \right) \, , \\
& \left\{ Q^+_{\alpha }, Q^-_{\beta }\right\} = - \frac{1}{\ell} \left(\Gamma^{a0} C \right)_{\alpha \beta} K_{a} \, , \\
& \left\{ Q^-_{\alpha }, Q^-_{\beta }\right\} = \left( \Gamma ^{0}C\right) _{\alpha \beta } \left( H - S \right) \, .
\end{split}
\end{equation}
We will now construct a CS supergravity action in three dimensions invariant under the $\mathcal{N}=(2,0)$ AdS Carroll superalgebra \eqref{adscarrollsuper22}: The $(2,0)$ AdS Carroll CS supergravity action.
  
\subsection{$(2,0)$ AdS Carroll supergravity action}

The general form of a three-dimensional CS action is given by
\begin{equation}\label{genCS}
I_{CS}=\frac{k}{4 \pi} \int_\mathcal{M} \Big \langle A dA + \frac{2}{3} A^3 \Big \rangle = \frac{k}{4 \pi} \int_\mathcal{M} \Big \langle A dA + \frac{1}{3} A \left[A,A \right] \Big \rangle \, ,
\end{equation}
where $k=1/(4G)$ is the CS level of the theory,\footnote{For gravitational theories, the CS level is related to the gravitational constant $G$.} $A$ is the gauge connection $1$-form, and $\langle \ldots \rangle$ denotes the invariant tensor. The integral in \eqref{genCS} is over a three-dimensional manifold $\mathcal{M}$.\footnote{In the sequel, we will omit the wedge product ``$\wedge$'' between differential forms.} The CS action \eqref{genCS} can also be rewritten as 
\begin{equation}
I_{CS}=\frac{k}{4 \pi} \int_\mathcal{M} \Big \langle A F - \frac{1}{3} A^3 \Big \rangle \, ,
\end{equation}
in terms of the curvature $2$-form $F= dA + A^2 = dA + \frac{1}{2}\left[A,A \right]$.

In the case of the $\mathcal{N}=(2,0)$ AdS Carroll superalgebra \eqref{adscarrollsuper22}, the connection $1$-form reads\footnote{Here and in the following, for simplicity, we will omit the spinor index $\alpha$.}
\begin{equation}\label{connadscarrollsuper22}
A = \frac{1}{2} \omega^{ab} J_{ab} + k^a K_a + V^a P_a + h H + t T + s S + \psi^+ Q^+ + \psi^- Q^- \,  , 
\end{equation}
where $\omega^{ab}$, $k^a$, $V^a$, $h$, $t$, $s$, $\psi^+$, and $\psi^-$ are the $1$-form fields dual to the generators $J_{ab}$, $K_a$, $P_a$, $H$, $T$, $S$, $Q^+$, and $Q^-$, respectively. 

The corresponding curvature $2$-form $F$ is
\begin{equation}\label{curv2f22}
F = \frac{1}{2} \mathcal{R}^{ab} J_{ab} + \mathcal{K}^a K_a + R^a P_a + \mathcal{H} H + \mathcal{T} T + \mathcal{S} S + \nabla \psi^+ Q^+ + \nabla \psi^- Q^- \, ,
\end{equation}
with
\begin{equation}\label{curvadscarrollsuper22}
\begin{split}
\mathcal{R}^{ab} & = d \omega^{ab} + \frac{1}{\ell^2} V^a V^b = R^{ab} + \frac{1}{\ell^2} V^a V^b  \, , \\
\mathcal{K}^a  & = d k^a + \omega^a_{\phantom{a} b} k^b + \frac{1}{\ell^2} V^a h + \frac{1}{\ell} \bar{\psi}^+ \Gamma^{a0} \psi^- = \mathfrak{K}^a  + \frac{1}{\ell^2} V^a h + \frac{1}{\ell} \bar{\psi}^+ \Gamma^{a0} \psi^- \, , \\
R^a & = d V^a + \omega^a_{\phantom{a}b} V^b \, , \\
\mathcal{H} & = d h + V^a k_a - \frac{1}{2} \bar{\psi}^+ \Gamma^0 \psi^+ - \frac{1}{2} \bar{\psi}^- \Gamma^0 \psi^-   = \mathfrak{H} - \frac{1}{2} \bar{\psi}^+ \Gamma^0 \psi^+ - \frac{1}{2} \bar{\psi}^- \Gamma^0 \psi^-  \,   , \\
\mathcal{T} & = d t  \,   , \\
\mathcal{S} & = d s - \frac{1}{2} \bar{\psi}^+ \Gamma^0 \psi^+ + \frac{1}{2} \bar{\psi}^- \Gamma^0 \psi^- \, , \\
\nabla \psi^+ & = d \psi^+ + \frac{1}{4} \omega^{ab} \Gamma_{ab} \psi^+ + \frac{1}{2 \ell} V^a \Gamma_a \psi^- - t \Gamma_0 \psi^+ \,  , \\
\nabla \psi^- & = d \psi^- + \frac{1}{4} \omega^{ab} \Gamma_{ab} \psi^- + \frac{1}{2 \ell} V^a \Gamma_a \psi^+ + t \Gamma_0 \psi^- \,  .
\end{split}
\end{equation}

Now, in order to formulate a CS action of the form \eqref{genCS} invariant under the $\mathcal{N}=(2,0)$ super-AdS Carroll group, we make use of the connection $1$-form given in \eqref{connadscarrollsuper22} and of the corresponding non-vanishing components of the invariant tensor.

Concerning, in particular, the invariant tensor, we now apply the method of \cite{Concha:2016zdb}, meaning that we rescale not only the generators but also the coefficients appearing in the invariant tensor before applying the, in the present case, ultra-relativistic contraction, in order to end up with a non-trivial invariant tensor for the contracted (super)algebra on which the CS theory will be based.
Specifically, we consider the non-vanishing components of the invariant tensor for the $\mathfrak{so}(2)$ extension of $\mathfrak{osp}(2|2)\otimes \mathfrak{sp}(2)$ (see \eqref{osp22}) given in \eqref{invtosp22}, we decompose the indices as in \eqref{indexdec} and consider the new supersymmetry charges \eqref{newsusy22}, and then we rescale not only the generators in compliance with \eqref{resc} but also the coefficients appearing in \eqref{invtosp22} as follows:
\begin{equation}\label{alsc}
\alpha_0 \rightarrow \alpha_0 \,  , \quad \alpha_1 \rightarrow \sigma \alpha_1 \, .
\end{equation}
Taking the limit $\sigma \rightarrow \infty$, we end up with the following non-vanishing components of an invariant tensor for the $\mathcal{N}=(2,0)$ AdS Carroll superalgebra:
\begin{equation}\label{invadscarrollsuper22}
\begin{split}
& \langle J_{ab} J_{cd} \rangle = \alpha_0 \left( \delta_{ad} \delta_{bc} - \delta_{ac} \delta_{bd} \right) \, , \\
& \langle J_{ab} H \rangle = \alpha_1 \epsilon_{ab} \, , \\
& \langle K_a P_b \rangle = - \alpha_1 \epsilon_{ab} \, , \\
& \langle P_{a} P_{b} \rangle = \frac{\alpha_0}{\ell^2} \delta_{ab} \, , \\
& \langle T T \rangle = - 2 \alpha_0\, , \\
& \langle T S \rangle = 2 \alpha_1 \, , \\
& \langle Q^+_\alpha Q^+_\beta \rangle = \langle Q^-_\alpha Q^-_\beta \rangle = 2 \alpha_1  C_{\alpha \beta} \, .
\end{split}
\end{equation}
This bilinear form is non-degenerate if $\alpha_1 \neq 0$.

After that, using the connection $1$-form in \eqref{connadscarrollsuper22} and the non-vanishing components of the invariant tensor given in \eqref{invadscarrollsuper22} in the general expression \eqref{genCS} for a three-dimensional CS action, we can finally write the $(2,0)$ AdS Carroll CS supergravity action in three spacetime dimensions invariant under \eqref{adscarrollsuper22}, which reads as follows:
\begin{equation}\label{CSAC22}
\begin{split}
I^{(2,0)}_{CS} & = \frac{k}{4 \pi} \int_\mathcal{M} \Bigg \lbrace \frac{\alpha_0}{2} \left( \omega^a_{\phantom{a} b} R^b_{\phantom{b} a}  + \frac{2}{\ell^2} V^a R_a - 4 t dt  \right) + \alpha_1 \bigg( \epsilon_{ab} R^{ab} h - 2 \epsilon_{ab} \mathfrak{K}^a V^b + \frac{1}{\ell^2} \epsilon_{ab} V^a V^b h  \\
& + 4 t ds + 2 \bar{\psi}^+ \nabla \psi^+ + 2 \bar{\psi}^- \nabla \psi^- \bigg) - d \left( \frac{\alpha_1}{2} \epsilon_{ab} \omega^{ab} h - \alpha_1 \epsilon_{ab} k^a V^b - 2 \alpha_1 t s \right) \Bigg \rbrace \,  ,
\end{split}
\end{equation}
written in terms of the curvatures appearing in \eqref{curvadscarrollsuper22}. We can see that in \eqref{CSAC22} we have two different sectors, one proportional to $\alpha_0$ and the other proportional to $\alpha_1$. Observe that the term proportional to $\alpha_0$ corresponds to the exotic Lagrangian involving the Lorentz contribution, a torsional piece, and a contribution from the $1$-form field $t$, while it does not contain any contribution from the $1$-form fields $\psi^+$ and $\psi^-$. Let us mention that the CS action \eqref{CSAC22} can also be rewritten up to boundary terms as 
\begin{equation}\label{CSAC22uptobdy}
\begin{split}
I^{(2,0)}_{CS} & = \frac{k}{4 \pi} \int_\mathcal{M} \Bigg \lbrace \frac{\alpha_0}{2} \left( \omega^a_{\phantom{a} b} R^b_{\phantom{b} a}  + \frac{2}{\ell^2} V^a R_a - 4 t dt  \right) + \alpha_1 \bigg( \epsilon_{ab} R^{ab} h - 2 \epsilon_{ab} \mathfrak{K}^a V^b + \frac{1}{\ell^2} \epsilon_{ab} V^a V^b h  \\
& + 4 t ds + 2 \bar{\psi}^+ \nabla \psi^+ + 2 \bar{\psi}^- \nabla \psi^- \bigg) \Bigg \rbrace \, .
\end{split}
\end{equation}

The CS action \eqref{CSAC22}, characterized by two coupling constants $\alpha_0$ and $\alpha_1$, is invariant by construction under the $\mathcal{N}=(2,0)$ AdS Carroll superalgebra \eqref{adscarrollsuper22}. In particular, the local gauge transformations $\delta_\lambda A = d \lambda + \left[A, \lambda \right]$ with gauge parameter
\begin{equation}\label{gpar22}
\lambda = \frac{1}{2} \lambda^{ab} J_{ab} + \kappa^a K_a + \lambda^a P_a + \tau H + \varphi T + \varsigma S + \varepsilon^+ Q^+ + \varepsilon^- Q^- 
\end{equation}
are given by
\begin{equation}\label{gaugetr22}
\begin{split}
\delta \omega^{ab} & = d \lambda^{ab} + \frac{2}{\ell^2} V^{[a} \lambda^{b]} \, , \\
\delta k^a  & = d \kappa^a - \lambda^a_{\phantom{a} b} k^b + \omega^a_{\phantom{a} b} \kappa^b - \frac{1}{\ell^2} \lambda^a h + \frac{1}{\ell^2} V^a \tau - \frac{1}{\ell} \bar{\varepsilon}^+ \Gamma^{a0} \psi^- - \frac{1}{\ell} \bar{\varepsilon}^- \Gamma^{a0} \psi^+ \, , \\
\delta V^a & = d \lambda^a - \lambda^a_{\phantom{a}b} V^b + \omega^a_{\phantom{a}b} \lambda^b \, , \\
\delta h & = d \tau - \lambda^a k_a + V^a \kappa_a + \bar{\varepsilon}^+ \Gamma^0 \psi^+ + \bar{\varepsilon}^- \Gamma^0 \psi^- \,   , \\
\delta t & = d \varphi \,   , \\
\delta s & = d \varsigma + \bar{\varepsilon}^+ \Gamma^0 \psi^+ - \bar{\varepsilon}^- \Gamma^0 \psi^- \, , \\
\delta \psi^+ & = d \varepsilon^+ - \frac{1}{4} \lambda^{ab} \Gamma_{ab} \psi^+ + \frac{1}{4} \omega^{ab} \Gamma_{ab} \varepsilon^+ - \frac{1}{2 \ell} \lambda^a \Gamma_a \psi^- + \frac{1}{2 \ell} V^a \Gamma_a \varepsilon^- + \varphi \Gamma_0 \psi^+ - t \Gamma_0 \varepsilon^+ \,  , \\
\delta \psi^- & = d \varepsilon^- - \frac{1}{4} \lambda^{ab} \Gamma_{ab} \psi^- + \frac{1}{4} \omega^{ab} \Gamma_{ab} \varepsilon^- - \frac{1}{2 \ell} \lambda^a \Gamma_a \psi^+ + \frac{1}{2 \ell} V^a \Gamma_a \varepsilon^+ - \varphi \Gamma_0 \psi^- + t \Gamma_0 \varepsilon^- \,  .
\end{split}
\end{equation}
Restricting ourselves to supersymmetry, we have
\begin{equation}\label{susytr22}
\begin{split}
\delta \omega^{ab} & = 0 \, , \\
\delta k^a  & = - \frac{1}{\ell} \bar{\varepsilon}^+ \Gamma^{a0} \psi^- - \frac{1}{\ell} \bar{\varepsilon}^- \Gamma^{a0} \psi^+ \, , \\
\delta V^a & = 0 \, , \\
\delta h & = \bar{\varepsilon}^+ \Gamma^0 \psi^+ + \bar{\varepsilon}^- \Gamma^0 \psi^- \,   , \\
\delta t & = 0 \,   , \\
\delta s & = \bar{\varepsilon}^+ \Gamma^0 \psi^+ - \bar{\varepsilon}^- \Gamma^0 \psi^- \, , \\
\delta \psi^+ & = d \varepsilon^+ + \frac{1}{4} \omega^{ab} \Gamma_{ab} \varepsilon^+ + \frac{1}{2 \ell} V^a \Gamma_a \varepsilon^- - t \Gamma_0 \varepsilon^+ \,  , \\
\delta \psi^- & = d \varepsilon^- + \frac{1}{4} \omega^{ab} \Gamma_{ab} \varepsilon^- + \frac{1}{2 \ell} V^a \Gamma_a \varepsilon^+ + t \Gamma_0 \varepsilon^- \,  .
\end{split}
\end{equation}

The equations of motion obtained from the variation of the action \eqref{CSAC22} with respect to the fields $\omega^{ab}$, $k^a$, $V^a$, $h$, $t$, $s$, $\psi^+$, and $\psi^-$ are, respectively,
\begin{equation}\label{eom22}
\begin{split}
\delta \omega^{ab} & : \quad \alpha_0 \mathcal{R}^{ab} + \alpha_1 \epsilon^{ab} \mathcal{H}= 0 \, , \\
\delta k^a & : \quad  \alpha_1 R^a = 0 \, , \\
\delta V^a & : \quad \frac{\alpha_0}{\ell^2} R^a + 2 \alpha_1 \epsilon_{ab} \mathcal{K}^b = 0 \,  , \\
\delta h & : \quad \alpha_1 \mathcal{R}^{ab} = 0  \, , \\
\delta t & : \quad - \alpha_0 \mathcal{T} + \alpha_1 \mathcal{S} = 0  \, , \\
\delta s & : \quad \alpha_1 \mathcal{T} = 0  \, , \\
\delta \psi^+ & : \quad \alpha_1 \nabla \psi^+ = 0 \,  , \\
\delta \psi^- & : \quad \alpha_1 \nabla \psi^- = 0 \,  ,
\end{split}
\end{equation}
up to boundary terms, and we can see that when $\alpha_1 \neq 0$ they reduce to the vanishing of the $(2,0)$ super-AdS Carroll curvature $2$-forms, namely
\begin{equation}\label{eomvac}
\mathcal{R}^{ab} = 0 \,  , \quad \mathcal{K}^a = 0 \, , \quad R^a = 0 \, , \quad \mathcal{H}=0 \, , \quad \mathcal{T} =0 \, , \quad \mathcal{S} = 0 \, , \quad \nabla \psi^+ =0 \, , \quad \nabla \psi^- =0 \, .
\end{equation}
Here, we can also observe that $\alpha_1 \neq 0$ is a sufficient condition to recover \eqref{eomvac}, meaning that one could consistently set $\alpha_0=0$, which corresponds to the vanishing of the exotic term in the CS action \eqref{CSAC22}.

\section{$(1,1)$ AdS Carroll supergravity in $2+1$ dimensions}\label{11case}

In this section, we repeat the analysis done in Section \ref{20case} in the $(1,1)$ case. To this aim, we first review the derivation of the $\mathcal{N}=(1,1)$ AdS Carroll superalgebra introduced in \cite{Bergshoeff:2015wma}, which is obtained as the ultra-relativistic contraction of $\mathfrak{osp}(2|1)\otimes \mathfrak{osp}(2,1)$. Then, we write the non-vanishing components of the invariant tensor of the $\mathcal{N}=(1,1)$ AdS Carroll superalgebra (obtained as the Carrollian contraction of the non-vanishing components of the invariant tensor for $\mathfrak{osp}(2|1)\otimes \mathfrak{osp}(2,1)$). This allows us to construct a three-dimensional CS supergravity action invariant under the $\mathcal{N}=(1,1)$ AdS Carroll superalgebra, which we call the $(1,1)$ AdS Carroll CS supergravity action.

\subsection{Review of the $\mathcal{N}=(1,1)$ AdS Carroll superalgebra}

Let us briefly review the derivation of the $\mathcal{N}=(1,1)$ AdS Carroll superalgebra of \cite{Bergshoeff:2015wma} as the Carrollian contraction of $\mathfrak{osp}(2|1)\otimes \mathfrak{osp}(2,1)$.

The superalgebra $\mathfrak{osp}(2|1)\otimes \mathfrak{osp}(2,1)$ is generated by the set $\lbrace \tilde{J}_{AB}, \tilde{P}_A , \tilde{Q}^+_\alpha , \tilde{Q}^-_\alpha \rbrace$ obeying the following (anti)commutation relations:
\begin{equation}\label{osp11}
\begin{split}
& \left[ \tilde{J}_{AB}, \tilde{J}_{CD}\right] =\eta _{BC}\tilde{J}_{AD}-\eta _{AC}\tilde{J}_{BD}-\eta
_{BD}\tilde{J}_{AC}+\eta _{AD}\tilde{J}_{BC} \, ,  \\
& \left[ \tilde{J}_{AB},\tilde{P}_{C}\right] =\eta _{BC}\tilde{P}_{A}-\eta _{AC}\tilde{P}_{B} \,  , \\
& \left[ \tilde{P}_A , \tilde{P}_B \right] = \frac{1}{\ell^2} \tilde{J}_{AB} \,  ,   \\
& \left[ \tilde{J}_{AB},\tilde{Q}^\pm_{\alpha }\right] =-\frac{1}{2}\left( \Gamma _{AB}\tilde{Q}^\pm\right)_{\alpha } \,  , \quad \quad \left[ \tilde{P}_{A}, \tilde{Q}^\pm_{\alpha }\right] =\mp \frac{1}{2 \ell} \left( \Gamma _{A}\tilde{Q}^\pm \right) _{\alpha } \,  , \\
& \left\{ \tilde{Q}^+_{\alpha }, \tilde{Q}^+_{\beta }\right\} = - \frac{1}{2 \ell} \left(\Gamma^{AB} C \right)_{\alpha \beta} \tilde{J}_{AB} + \left( \Gamma ^{A}C\right) _{\alpha \beta }\tilde{P}_{A} \,  , \\
& \left\{ \tilde{Q}^-_{\alpha }, \tilde{Q}^-_{\beta }\right\} = \frac{1}{2 \ell} \left(\Gamma^{AB} C \right)_{\alpha \beta} \tilde{J}_{AB} + \left( \Gamma ^{A}C\right) _{\alpha \beta }\tilde{P}_{A} \,  .
\end{split}
\end{equation}
Note that by taking the flat limit $\ell \rightarrow \infty$ of \eqref{osp11} one recovers the $\mathcal{N}=(1,1)$ Poincar\'{e} superalgebra.

The non-vanishing components of an invariant tensor for the superalgebra \eqref{osp11} are
\begin{equation}\label{invtosp11}
\begin{split}
& \langle \tilde{J}_{AB} \tilde{J}_{CD} \rangle = \alpha_0 \left( \eta_{AD} \eta_{BC} - \eta_{AC} \eta_{BD} \right) \, , \\
& \langle \tilde{J}_{AB} \tilde{P}_{C} \rangle = \alpha_1 \epsilon_{ABC} \,  , \\
& \langle \tilde{P}_{A} \tilde{P}_{B} \rangle = \frac{\alpha_0}{\ell^2} \eta_{AB} \,  , \\
& \langle \tilde{Q}^+_\alpha \tilde{Q}^+_\beta \rangle = 2 \left(\alpha_1 - \frac{\alpha_0}{\ell} \right) C_{\alpha \beta}  \,  , \\
& \langle \tilde{Q}^-_\alpha \tilde{Q}^-_\beta \rangle = 2 \left(\alpha_1 + \frac{\alpha_0}{\ell} \right) C_{\alpha \beta}  \,  , 
\end{split}
\end{equation}
being $\alpha_0$ and $\alpha_1$ arbitrary independent constants.

Now, to take the Carrollian contraction of the superalgebra \eqref{osp11}, we decompose the indices $A,B,\ldots = 0,1,2$ as in \eqref{indexdec}, which induces the decomposition \eqref{gendec}, together with \eqref{gammadec}.
Then, we rescale the generators with a parameter $\sigma$ as
\begin{equation}\label{resc11}
\tilde{H} \rightarrow \sigma H \, , \quad \tilde{K}_a \rightarrow \sigma K_a \, , \quad \tilde{Q}^\pm_\alpha \rightarrow \sqrt{\sigma} Q^\pm_\alpha  \, .
\end{equation}  
Subsequently, taking the limit $\sigma \rightarrow \infty$ (and removing the tilde symbol also on the generators that we have not rescaled), we end up with the $\mathcal{N}=(1,1)$ AdS Carroll superalgebra introduced in Ref. \cite{Bergshoeff:2015wma}, whose (anti)commutation relations read
\begin{equation}\label{adscarrollsuper11}
\begin{split}
& \left[ K_a , J_{bc} \right] = \delta_{ab} K_c - \delta_{ac} K_b \, , \\
& \left[ J_{ab}, P_{c}\right] =\delta_{bc}P_{a}-\delta _{ac}P_{b} \, , \\
& \left[ K_{a}, P_{b}\right] = - \delta_{ab} H \, , \\
& \left[ P_a , P_b \right] = \frac{1}{\ell^2} J_{ab} \, , \\
& \left[ P_a , H \right] = \frac{1}{\ell^2} K_{a} \, , \\
& \left[ J_{ab},Q^\pm_{\alpha }\right] =-\frac{1}{2}\left( \Gamma _{ab} Q^\pm \right)_{\alpha } \, , \quad \quad \left[ P_{a}, Q^\pm_{\alpha }\right] =\mp \frac{1}{2 \ell} \left( \Gamma _{a}Q^\pm \right) _{\alpha } \, , \\
& \left\{ Q^+_{\alpha }, Q^+_{\beta }\right\} = - \frac{1}{\ell} \left(\Gamma^{a0} C \right)_{\alpha \beta} K_{a} + \left( \Gamma ^{0}C\right) _{\alpha \beta } H \, , \\
& \left\{ Q^-_{\alpha }, Q^-_{\beta }\right\} = \frac{1}{\ell} \left(\Gamma^{a0} C \right)_{\alpha \beta} K_{a} + \left( \Gamma ^{0}C\right) _{\alpha \beta } H \, .
\end{split}
\end{equation}
In the sequel we will construct a CS action in three-dimension invariant under the $\mathcal{N}=(1,1)$ AdS Carroll superalgebra \eqref{adscarrollsuper11}.

\subsection{$(1,1)$ AdS Carroll supergravity action}

We will now construct a three-dimensional CS supergravity action invariant under the $\mathcal{N}=(1,1)$ AdS Carroll superalgebra \eqref{adscarrollsuper11}, which we call the $(1,1)$ AdS Carroll CS supergravity action.

To this aim, we introduce the connection $1$-form $A$ associated with \eqref{adscarrollsuper11}, that is
\begin{equation}\label{connadscarrollsuper11}
A = \frac{1}{2} \omega^{ab} J_{ab} + k^a K_a + V^a P_a + h H + \psi^+ Q^+ + \psi^- Q^- \,  , 
\end{equation}
being $\omega^{ab}$, $k^a$, $V^a$, $h$, $\psi^+$, and $\psi^-$ the $1$-form fields respectively dual to the generators $J_{ab}$, $K_a$, $P_a$, $H$, $Q^+$, and $Q^-$ obeying the (anti)commutation relations given in \eqref{adscarrollsuper11}, and the related curvature $2$-form $F$, which reads
\begin{equation}\label{curv2f11}
F = \frac{1}{2} \mathcal{R}^{ab} J_{ab} + \mathcal{K}^a K_a + R^a P_a + \mathcal{H} H + \nabla \psi^+ Q^+ + \nabla \psi^- Q^- \, ,
\end{equation}
with
\begin{equation}\label{curvadscarrollsuper11}
\begin{split}
\mathcal{R}^{ab} & = d \omega^{ab} + \frac{1}{\ell^2} V^a V^b = R^{ab} + \frac{1}{\ell^2} V^a V^b  \, , \\
\mathcal{K}^a  & = d k^a + \omega^a_{\phantom{a} b} k^b + \frac{1}{\ell^2} V^a h + \frac{1}{2\ell} \bar{\psi}^+ \Gamma^{a0} \psi^+ - \frac{1}{2\ell} \bar{\psi}^- \Gamma^{a0} \psi^-  = \mathfrak{K}^a  + \frac{1}{2\ell} \bar{\psi}^+ \Gamma^{a0} \psi^+ - \frac{1}{2\ell} \bar{\psi}^- \Gamma^{a0} \psi^- \, , \\
R^a & = d V^a + \omega^a_{\phantom{a}b} V^b \, , \\
\mathcal{H} & = d h + V^a k_a - \frac{1}{2} \bar{\psi}^+ \Gamma^0 \psi^+ - \frac{1}{2} \bar{\psi}^- \Gamma^0 \psi^-   = \mathfrak{H} - \frac{1}{2} \bar{\psi}^+ \Gamma^0 \psi^+ - \frac{1}{2} \bar{\psi}^- \Gamma^0 \psi^-  \,   , \\
\nabla \psi^+ & = d \psi^+ + \frac{1}{4} \omega^{ab} \Gamma_{ab} \psi^+ + \frac{1}{2 \ell} V^a \Gamma_a \psi^+ \,  , \\
\nabla \psi^- & = d \psi^- + \frac{1}{4} \omega^{ab} \Gamma_{ab} \psi^- - \frac{1}{2 \ell} V^a \Gamma_a \psi^- \,  .
\end{split}
\end{equation}

Now, we move to the explicit construction of a CS action invariant under the $\mathcal{N}=(1,1)$ super-AdS Carroll group, on the same lines of what we have previously done in Section \ref{20case} for $\mathcal{N}=(2,0)$.
Thus, we consider the non-vanishing components of the (relativistic) invariant tensor given in \eqref{invtosp11}, we decompose the indices as in \eqref{indexdec}, and we rescale not only the generators in compliance with \eqref{resc11} but also the coefficients appearing in \eqref{invtosp11} as in \eqref{alsc}. Then, taking the ultra-relativistic limit $\sigma \rightarrow \infty$, we get the following non-vanishing components of an invariant tensor for the $\mathcal{N}=(1,1)$ AdS Carroll superalgebra:
\begin{equation}\label{invadscarrollsuper11}
\begin{split}
& \langle J_{ab} J_{cd} \rangle = \alpha_0 \left( \delta_{ad} \delta_{bc} - \delta_{ac} \delta_{bd} \right) \, , \\
& \langle J_{ab} H \rangle = \alpha_1 \epsilon_{ab} \, , \\
& \langle K_a P_b \rangle = - \alpha_1 \epsilon_{ab} \, , \\
& \langle P_{a} P_{b} \rangle = \frac{\alpha_0}{\ell^2} \delta_{ab} \, , \\
& \langle Q^+_\alpha Q^+_\beta \rangle = \langle Q^-_\alpha Q^-_\beta \rangle = 2 \alpha_1  C_{\alpha \beta} \, .
\end{split}
\end{equation}
This invariant tensor is non-degenerate if $\alpha_1 \neq 0$.

Substituting the connection $1$-form in \eqref{connadscarrollsuper11} and the non-vanishing components of the invariant tensor \eqref{invadscarrollsuper11} in the general expression \eqref{genCS} for a three-dimensional CS action, we end up with the $(1,1)$ AdS Carroll CS supergravity action in $2+1$ spacetime dimensions, that is
\begin{equation}\label{CSAC11}
\begin{split}
I^{(1,1)}_{CS} & = \frac{k}{4 \pi} \int_\mathcal{M} \Bigg \lbrace \frac{\alpha_0}{2} \left( \omega^a_{\phantom{a} b} R^b_{\phantom{b} a}  + \frac{2}{\ell^2} V^a R_a  \right) + \alpha_1 \bigg( \epsilon_{ab} R^{ab} h - 2 \epsilon_{ab} \mathfrak{K}^a V^b + \frac{1}{\ell^2} \epsilon_{ab} V^a V^b h  \\
& + 2 \bar{\psi}^+ \nabla \psi^+ + 2 \bar{\psi}^- \nabla \psi^- \bigg) - d \left( \frac{\alpha_1}{2} \epsilon_{ab} \omega^{ab} h - \alpha_1 \epsilon_{ab} k^a V^b \right) \Bigg \rbrace \,  ,
\end{split}
\end{equation}
which can also be rewritten omitting boundary terms as follows:
\begin{equation}
\begin{split}
I^{(1,1)}_{CS} & = \frac{k}{4 \pi} \int_\mathcal{M} \Bigg \lbrace \frac{\alpha_0}{2} \left( \omega^a_{\phantom{a} b} R^b_{\phantom{b} a}  + \frac{2}{\ell^2} V^a R_a  \right) + \alpha_1 \bigg( \epsilon_{ab} R^{ab} h - 2 \epsilon_{ab} \mathfrak{K}^a V^b + \frac{1}{\ell^2} \epsilon_{ab} V^a V^b h \\
& + 2 \bar{\psi}^+ \nabla \psi^+ + 2 \bar{\psi}^- \nabla \psi^- \bigg) \Bigg \rbrace \,  .
\end{split}
\end{equation}
The action \eqref{CSAC11} has been written in terms of the curvatures appearing in \eqref{curvadscarrollsuper11}, it involves two different sectors, respectively proportional to $\alpha_0$ (which corresponds to the exotic Lagrangian) and to $\alpha_1$, and it is invariant by construction under the $\mathcal{N}=(1,1)$ AdS Carroll superalgebra \eqref{adscarrollsuper11}. The local gauge transformations $\delta_\lambda A = d \lambda + \left[A, \lambda \right]$ with gauge parameter
\begin{equation}\label{gpar11}
\lambda = \frac{1}{2} \lambda^{ab} J_{ab} + \kappa^a K_a + \lambda^a P_a + \tau H + \varepsilon^+ Q^+ + \varepsilon^- Q^-
\end{equation}
are
\begin{equation}\label{gaugetr11}
\begin{split}
\delta \omega^{ab} & = d \lambda^{ab} + \frac{2}{\ell^2} V^{[a} \lambda^{b]} \, , \\
\delta k^a  & = d \kappa^a - \lambda^a_{\phantom{a} b} k^b + \omega^a_{\phantom{a} b} \kappa^b - \frac{1}{\ell^2} \lambda^a h + \frac{1}{\ell^2} V^a \tau - \frac{1}{\ell} \bar{\varepsilon}^+ \Gamma^{a0} \psi^+ + \frac{1}{\ell} \bar{\varepsilon}^- \Gamma^{a0} \psi^- \, , \\
\delta V^a & = d \lambda^a - \lambda^a_{\phantom{a}b} V^b + \omega^a_{\phantom{a}b} \lambda^b \, , \\
\delta h & = d \tau - \lambda^a k_a + V^a \kappa_a + \bar{\varepsilon}^+ \Gamma^0 \psi^+ + \bar{\varepsilon}^- \Gamma^0 \psi^- \,   , \\
\delta \psi^+ & = d \varepsilon^+ - \frac{1}{4} \lambda^{ab} \Gamma_{ab} \psi^+ + \frac{1}{4} \omega^{ab} \Gamma_{ab} \varepsilon^+ - \frac{1}{2 \ell} \lambda^a \Gamma_a \psi^+ + \frac{1}{2 \ell} V^a \Gamma_a \varepsilon^+ \,  , \\
\delta \psi^- & = d \varepsilon^- - \frac{1}{4} \lambda^{ab} \Gamma_{ab} \psi^- + \frac{1}{4} \omega^{ab} \Gamma_{ab} \varepsilon^- + \frac{1}{2 \ell} \lambda^a \Gamma_a \psi^- - \frac{1}{2 \ell} V^a \Gamma_a \varepsilon^- \,  ,
\end{split}
\end{equation}
and, restricting ourselves to supersymmetry, we are left with the following transformation rules:
\begin{equation}\label{susytr11}
\begin{split}
\delta \omega^{ab} & = 0 \, , \\
\delta k^a  & = - \frac{1}{\ell} \bar{\varepsilon}^+ \Gamma^{a0} \psi^+ + \frac{1}{\ell} \bar{\varepsilon}^- \Gamma^{a0} \psi^- \, , \\
\delta V^a & = 0 \, , \\
\delta h & = \bar{\varepsilon}^+ \Gamma^0 \psi^+ + \bar{\varepsilon}^- \Gamma^0 \psi^- \,   , \\
\delta \psi^+ & = d \varepsilon^+ + \frac{1}{4} \omega^{ab} \Gamma_{ab} \varepsilon^+ + \frac{1}{2 \ell} V^a \Gamma_a \varepsilon^+ \,  , \\
\delta \psi^- & = d \varepsilon^- + \frac{1}{4} \omega^{ab} \Gamma_{ab} \varepsilon^- - \frac{1}{2 \ell} V^a \Gamma_a \varepsilon^- \, .
\end{split}
\end{equation}

The equations of motion obtained from the variation of the action \eqref{CSAC11} with respect to the $1$-form fields $\omega^{ab}$, $k^a$, $V^a$, $h$, $\psi^+$, and $\psi^-$ are
\begin{equation}\label{eom11}
\begin{split}
\delta \omega^{ab} & : \quad \alpha_0 \mathcal{R}^{ab} + \alpha_1 \epsilon^{ab} \mathcal{H}= 0 \, , \\
\delta k^a & : \quad  \alpha_1 R^a = 0 \, , \\
\delta V^a & : \quad \frac{\alpha_0}{\ell^2} R^a + 2 \alpha_1 \epsilon_{ab} \mathcal{K}^b = 0 \,  , \\
\delta h & : \quad \alpha_1 \mathcal{R}^{ab} = 0  \, , \\
\delta \psi^+ & : \quad \alpha_1 \nabla \psi^+ = 0 \,  , \\
\delta \psi^- & : \quad \alpha_1 \nabla \psi^- = 0 \,  ,
\end{split}
\end{equation}
respectively; for $\alpha_1 \neq 0$, they reduce to the vanishing of the $(1,0)$ super-AdS Carroll curvature $2$-forms, namely
\begin{equation}\label{eomvac11}
\mathcal{R}^{ab} = 0 \,  , \quad \mathcal{K}^a = 0 \, , \quad R^a = 0 \, , \quad \mathcal{H}=0 \, , \quad \nabla \psi^+ =0 \, , \quad \nabla \psi^- =0 \, .
\end{equation}
Analogously to what happened in the $(2,0)$ case discussed in Section \ref{20case}, we can see that $\alpha_1 \neq 0$ is a sufficient condition to recover \eqref{eomvac11}, which means that one could consistently set $\alpha_0=0$, making the exotic term in the CS action \eqref{CSAC11} disappear.

\section{$(\mathcal{N},0)$ AdS Carroll supergravity theories in $2+1$ dimensions}\label{ncase}

Now, we generalize our analysis to the $(\mathcal{N},0)$ case, with $\mathcal{N}$ even. First, we present the derivation of the $\mathcal{N}=(\mathcal{N},0)$ AdS Carroll superalgebra as the Carrollian contraction of an $\mathfrak{so}(\mathcal{N})$ extension of $\mathfrak{osp}(2|\mathcal{N})\otimes \mathfrak{sp}(2)$. This also provides us with a non-degenerate invariant tensor in the ultra-relativistic limit. Then, we can subsequently formulate a well-defined three-dimensional CS supergravity action invariant under the aforesaid $\mathcal{N}=(\mathcal{N},0)$ AdS Carroll superalgebra.

\subsection{$\mathcal{N}=(\mathcal{N},0)$ AdS Carroll superalgebra} 

Let us first take the direct sum of $\mathfrak{osp}(2|\mathcal{N})\otimes \mathfrak{sp}(2)$ and an $\mathfrak{so}(\mathcal{N})$ algebra (we consider $\mathcal{N}$ even), that is reminiscent of what was done in Ref. \cite{Howe:1995zm}. In this case, the non-trivial (anti)commutation relations are
\begin{equation}\label{ospnnonmod}
\begin{split}
& \left[ \tilde{J}_{AB}, \tilde{J}_{CD}\right] =\eta _{BC}\tilde{J}_{AD}-\eta _{AC}\tilde{J}_{BD}-\eta
_{BD}\tilde{J}_{AC}+\eta _{AD}\tilde{J}_{BC} \, ,  \\
& \left[ \tilde{J}_{AB},\tilde{P}_{C}\right] =\eta _{BC}\tilde{P}_{A}-\eta _{AC}\tilde{P}_{B} \,  , \quad \quad  \left[ \tilde{P}_A , \tilde{P}_B \right] = \frac{1}{\ell^2} \tilde{J}_{AB} \,  ,   \\
& \left[ \tilde{Z}^{ij} , \tilde{Z}^{kl} \right] = \delta^{jk} \tilde{Z}^{il} - \delta^{ik} \tilde{Z}^{jl}- \delta^{jl} \tilde{Z}^{ik} + \delta^{il} \tilde{Z}^{jk} \, , \\
& \left[ \tilde{S}^{ij} , \tilde{S}^{kl} \right] = - \frac{1}{\ell} \left( \delta^{jk} \tilde{S}^{il} - \delta^{ik} \tilde{S}^{jl}- \delta^{jl} \tilde{S}^{ik} + \delta^{il} \tilde{S}^{jk} \right) \, , \\
& \left[ \tilde{J}_{AB},\tilde{Q}^i_{\alpha }\right] =-\frac{1}{2}\left( \Gamma _{AB}\tilde{Q}^i\right)_{\alpha } \,  , \quad \quad \left[ \tilde{P}_{A}, \tilde{Q}^i_{\alpha }\right] =- \frac{1}{2 \ell} \left( \Gamma _{A}\tilde{Q}^i \right) _{\alpha } \,  , \\
& \left[ \tilde{Z}^{ij}, \tilde{Q}^k_{\alpha }\right] = \delta^{jk} \tilde{Q}^i_\alpha - \delta^{ik} \tilde{Q}^j_\alpha \,  , \\
& \left\{ \tilde{Q}^i_{\alpha }, \tilde{Q}^j_{\beta }\right\} = \delta^{ij} \left[ - \frac{1}{2 \ell} \left(\Gamma^{AB} C \right)_{\alpha \beta} \tilde{J}_{AB} + \left( \Gamma ^{A}C\right) _{\alpha \beta }\tilde{P}_{A} \right] + \frac{1}{\ell} C_{\alpha \beta} \tilde{Z}^{ij} \,  , 
\end{split}
\end{equation}
with $A, B, \ldots=0,1,2$, $i,j,\ldots=1,\ldots,\mathcal{N}$ (where we have considered $\mathcal{N}=2x$, $x=1,\ldots,\frac{\mathcal{N}}{2}$), and where $\tilde{Z}^{ij}=-\tilde{Z}^{ji}$, $\tilde{S}^{ij}=-\tilde{S}^{ij}$. 
Then, we do the following redefinition (on the same lines of \cite{Howe:1995zm}): 
\begin{equation}
\tilde{T}^{ij} \equiv \tilde{Z}^{ij} - \ell \tilde{S}^{ij} \, ,
\end{equation}
which is a generalization of the one performed in Section \ref{20case}. Thus, we can now rewrite the (anti)commutation relations \eqref{ospnnonmod} as
\begin{equation}\label{ospn}
\begin{split}
& \left[ \tilde{J}_{AB}, \tilde{J}_{CD}\right] =\eta _{BC}\tilde{J}_{AD}-\eta _{AC}\tilde{J}_{BD}-\eta
_{BD}\tilde{J}_{AC}+\eta _{AD}\tilde{J}_{BC} \, ,  \\
& \left[ \tilde{J}_{AB},\tilde{P}_{C}\right] =\eta _{BC}\tilde{P}_{A}-\eta _{AC}\tilde{P}_{B} \,  , \\
& \left[ \tilde{P}_A , \tilde{P}_B \right] = \frac{1}{\ell^2} \tilde{J}_{AB} \,  ,   \\
& \left[ \tilde{T}^{ij} , \tilde{T}^{kl} \right] = \delta^{jk} \tilde{T}^{il} - \delta^{ik} \tilde{T}^{jl}- \delta^{jl} \tilde{T}^{ik} + \delta^{il} \tilde{T}^{jk} \, , \\
& \left[ \tilde{T}^{ij} , \tilde{S}^{kl} \right] = \delta^{jk} \tilde{S}^{il} - \delta^{ik} \tilde{S}^{jl}- \delta^{jl} \tilde{S}^{ik} + \delta^{il} \tilde{S}^{jk} \, , \\
& \left[ \tilde{S}^{ij} , \tilde{S}^{kl} \right] = - \frac{1}{\ell} \left( \delta^{jk} \tilde{S}^{il} - \delta^{ik} \tilde{S}^{jl}- \delta^{jl} \tilde{S}^{ik} + \delta^{il} \tilde{S}^{jk} \right) \, , \\
& \left[ \tilde{J}_{AB},\tilde{Q}^i_{\alpha }\right] =-\frac{1}{2}\left( \Gamma _{AB}\tilde{Q}^i\right)_{\alpha } \,  , \\
& \left[ \tilde{P}_{A}, \tilde{Q}^i_{\alpha }\right] =- \frac{1}{2 \ell} \left( \Gamma _{A}\tilde{Q}^i \right) _{\alpha } \,  , \\
& \left[ \tilde{T}^{ij}, \tilde{Q}^k_{\alpha }\right] = \delta^{jk} \tilde{Q}^i_\alpha - \delta^{ik} \tilde{Q}^j_\alpha \,  , \\
& \left\{ \tilde{Q}^i_{\alpha }, \tilde{Q}^j_{\beta }\right\} = \delta^{ij} \left[ - \frac{1}{2 \ell} \left(\Gamma^{AB} C \right)_{\alpha \beta} \tilde{J}_{AB} + \left( \Gamma ^{A}C\right) _{\alpha \beta }\tilde{P}_{A} \right] +  C_{\alpha \beta} \left(\frac{1}{\ell} \tilde{T}^{ij} + \tilde{S}^{ij} \right) \, .
\end{split}
\end{equation}
Observe that, taking the limit $\ell \rightarrow \infty$ of \eqref{ospn}, we get the $\mathcal{N}=(\mathcal{N},0)$ Poincar\'{e} superalgebra involving a semi-direct $\mathfrak{so}(\mathcal{N})$ extension (with $\mathcal{N}$ even).

The non-vanishing components of an invariant tensor for \eqref{ospn}, which will be useful in the sequel, are given by
\begin{equation}\label{invtospn}
\begin{split}
& \langle \tilde{J}_{AB} \tilde{J}_{CD} \rangle = \alpha_0 \left( \eta_{AD} \eta_{BC} - \eta_{AC} \eta_{BD} \right) \, , \\
& \langle \tilde{J}_{AB} \tilde{P}_{C} \rangle = \alpha_1 \epsilon_{ABC} \,  , \\
& \langle \tilde{P}_{A} \tilde{P}_{B} \rangle = \frac{\alpha_0}{\ell^2} \eta_{AB} \,  , \\
& \langle \tilde{T}^{ij} \tilde{T}^{kl} \rangle =  2 \alpha_0 \left( \delta^{il} \delta^{kj} - \delta^{ik} \delta^{lj} \right) \,  , \\
& \langle \tilde{T}^{ij} \tilde{S}^{kl} \rangle = - 2 \alpha_1 \left( \delta^{il} \delta^{kj} - \delta^{ik} \delta^{lj} \right) \,  , \\
& \langle \tilde{S}^{ij} \tilde{S}^{kl} \rangle =  \frac{2 \alpha_1}{\ell} \left( \delta^{il} \delta^{kj} - \delta^{ik} \delta^{lj} \right) \,  , \\
& \langle \tilde{Q}^i_\alpha \tilde{Q}^j_\beta \rangle = 2 \left(\alpha_1 - \frac{\alpha_0}{\ell} \right) C_{\alpha \beta} \delta^{ij} \,  , 
\end{split}
\end{equation}
being $\alpha_0$ and $\alpha_1$ arbitrary independent constants.

In order to take the ultra-relativistic contraction of \eqref{ospn}, we decompose the indices $A,B,\ldots = 0,1,2$ as in \eqref{indexdec}, which induces the decomposition \eqref{gendec}, together with the gamma matrices decomposition \eqref{gammadec}. 

Moreover, we define, on the same lines of what was done in \cite{Lukierski:2006tr} (see also references therein) in the case of non-relativistic theories, new supersymmetry charges by
\begin{equation}\label{newsusyn}
\tilde{Q}^{\pm \, \lambda} _\alpha = \frac{1}{\sqrt{2}} \left( \tilde{Q}^\lambda_\alpha \pm  \left(\Gamma_0\right)_{\alpha \beta} \tilde{Q}^{x + \lambda}_\beta \right) \, ,
\end{equation} 
where in \eqref{newsusyn} we consider $\lambda, \mu , \ldots = 1, \ldots, x$ (these new indices must not be confused with the spinor ones $\alpha, \beta, \ldots = 1,2$), generalizing to the $\mathcal{N}=(\mathcal{N},0)$ case, with $\mathcal{N}$ even, what we have previously done in Section \ref{20case} (see, in particular, \eqref{newsusy22}). {We remind that $x= 1, \ldots , \frac{\mathcal{N}}{2}$.}

This also reflects on the generators $\tilde{T}^{ij}$ and $\tilde{S}^{ij}$, which are now respectively described by 
\begin{equation}\label{dectijsij}
\begin{split}
& \tilde{T}^{\lambda \mu} \, , \quad \tilde{T}'^{\lambda \mu} \equiv \tilde{T}^{\lambda + x \ \mu + x} \, , \quad \tilde{U}^{\lambda \mu} \equiv \tilde{T}^{x+\lambda \ \mu} \, , \quad \tilde{U}'^{\lambda \mu} \equiv \tilde{T}^{\lambda \ x + \mu}  \, , \\
& \tilde{S}^{\lambda \mu} \, , \quad \tilde{S}'^{\lambda \mu} \equiv \tilde{S}^{\lambda + x \ \mu + x} \, , \quad \tilde{V}^{\lambda \mu} \equiv \tilde{S}^{x+\lambda \ \mu} \, , \quad \tilde{V}'^{\lambda \mu} \equiv \tilde{S}^{\lambda \ x + \mu}  \, ,
\end{split}
\end{equation}
which satisfy the symmetry properties
\begin{equation}\label{symmprop}
\begin{split}
& \tilde{T}^{\lambda \mu} = - \tilde{T}^{\mu \lambda} \, , \quad \tilde{T}'^{\lambda \mu} = - \tilde{T}'^{\mu \lambda} \, , \quad \tilde{U}^{\lambda \mu} = - \tilde{U}'^{\mu \lambda} \, , \\
& \tilde{S}^{\lambda \mu} = - \tilde{S}^{\mu \lambda} \, , \quad \tilde{S}'^{\lambda \mu} = - \tilde{S}'^{\mu \lambda} \, , \quad \tilde{V}^{\lambda \mu} = - \tilde{V}'^{\mu \lambda} \, .
\end{split}
\end{equation}

In particular, using \eqref{newsusyn}, \eqref{dectijsij}, and \eqref{symmprop}, together with the decomposition \eqref{gendec} and \eqref{gammadec}, and defining 
\begin{equation}\label{xydef}
\begin{split}
& \tilde{X}^{[\lambda \mu]} \equiv \frac{1}{2} \left( \tilde{T}^{\lambda \mu} + \tilde{T}'^{\lambda \mu} \right) \, , \quad \tilde{X}'^{[\lambda \mu]} \equiv \frac{1}{2} \left( \tilde{T}^{\lambda \mu} - \tilde{T}'^{\lambda \mu} \right) \, , \\
& \tilde{Y}^{[\lambda \mu]} \equiv \frac{1}{2} \left( \tilde{S}^{\lambda \mu} + \tilde{S}'^{\lambda \mu} \right) \, , \quad \tilde{Y}'^{[\lambda \mu]} \equiv \frac{1}{2} \left( \tilde{S}^{\lambda \mu} - \tilde{S}'^{\lambda \mu} \right) \, ,
\end{split}
\end{equation}
the (anti)commutation relations in \eqref{ospn} become
\begin{equation}\label{antinew}
\begin{split}
\lbrace \tilde{Q}^{\pm \, \lambda} _\alpha , \tilde{Q}^{\pm \, \mu} _\beta \rbrace & = \delta^{\lambda \mu} \left[ - \frac{1}{2 \ell} \left( \Gamma^{ab} C \right)_{\alpha \beta} \tilde{J}_{ab} + \left( \Gamma^0 C \right)_{\alpha \beta} \tilde{H} \right] + C_{\alpha \beta} \left( \frac{1}{\ell} \tilde{X}^{[\lambda \mu]} + \tilde{Y}^{[\lambda \mu]} \right) \\
& \mp \left( \Gamma^0 C \right)_{\alpha \beta} \left( \frac{1}{\ell} \tilde{U}^{(\lambda \mu)} + \tilde{V}^{(\lambda \mu)} \right) \, , \\
\lbrace \tilde{Q}^{+ \, \lambda} _\alpha , \tilde{Q}^{- \, \mu} _\beta \rbrace & = \delta^{\lambda \mu} \left[ - \frac{1}{\ell} \left( \Gamma^{a0} C \right)_{\alpha \beta} \tilde{K}_{a} + \left( \Gamma^a C \right)_{\alpha \beta} \tilde{P}_a \right] + C_{\alpha \beta} \left( \frac{1}{\ell} \tilde{X}'^{[\lambda \mu]} + \tilde{Y}'^{[\lambda \mu]} \right) \\
& - \left( \Gamma^0 C \right)_{\alpha \beta} \left( \frac{1}{\ell} \tilde{U}^{[\lambda \mu]} + \tilde{V}^{[\lambda \mu]} \right) \, ,
\end{split}
\end{equation}
where
\begin{equation}\label{uvsas}
\begin{split}
& \tilde{U}^{(\lambda \mu)} = \frac{1}{2} \left( \tilde{U}^{\lambda \mu} + \tilde{U}^{\mu \lambda} \right) \, , \quad \tilde{U}^{[\lambda \mu]} = \frac{1}{2} \left( \tilde{U}^{\lambda \mu} - \tilde{U}^{\mu \lambda} \right) \, , \\
& \tilde{V}^{(\lambda \mu)} = \frac{1}{2} \left( \tilde{V}^{\lambda \mu} + \tilde{V}^{\mu \lambda} \right) \, , \quad \tilde{V}^{[\lambda \mu]} = \frac{1}{2} \left( \tilde{V}^{\lambda \mu} - \tilde{V}^{\mu \lambda} \right) \, .
\end{split}
\end{equation}
The generators $\tilde{X}^{[\lambda \mu]}$ in \eqref{antinew} amount to $\frac{k(k-1)}{2}$ generators, and the same holds for $\tilde{X}'^{[\lambda \mu]}$, $\tilde{U}^{[\lambda \mu]}$, $\tilde{Y}^{[\lambda \mu]}$, $\tilde{Y}'^{[\lambda\mu]}$, and $\tilde{V}^{[\lambda \mu]}$, while the generators $\tilde{U}^{(\lambda \mu)}$ are $\frac{k(k+1)}{2}$ generators, and the same holds for the generators $\tilde{V}^{(\lambda \mu)}$.

Furthermore, from the commutation relations involving the generators $T^{ij}$ and $S^{ij}$, we get the following non-vanishing ones (recall that we have $\tilde{U}'^{\lambda \mu} =- \tilde{U}^{\mu \lambda}$ and $\tilde{V}'^{\lambda \mu} =- \tilde{V}^{\mu \lambda}$):
\begin{equation}\label{newtttsss}
\begin{split}
& \left[ \tilde{T}^{\lambda \mu} , \tilde{T}^{\nu \rho} \right] = \delta^{\mu \nu} \tilde{T}^{\lambda \rho} - \delta^{\lambda \nu} \tilde{T}^{\mu \rho}- \delta^{\mu \rho} \tilde{T}^{\lambda \nu} + \delta^{\lambda \rho} \tilde{T}^{\mu \nu} \, , \\
& \left[ \tilde{T}^{\lambda \mu} , \tilde{U}^{\nu \rho} \right] = \delta^{\mu \rho} \tilde{U}^{\nu \lambda} - \delta^{\lambda \rho} \tilde{U}^{\nu \mu} \, , \\
& \left[ \tilde{T}'^{\lambda \mu} , \tilde{T}'^{\nu \rho} \right] = \delta^{\mu \nu} \tilde{T}'^{\lambda \rho} - \delta^{\lambda \nu} \tilde{T}'^{\mu \rho}- \delta^{\mu \rho} \tilde{T}'^{\lambda \nu} + \delta^{\lambda \rho} \tilde{T}'^{\mu \nu} \, , \\
& \left[ \tilde{T}'^{\lambda \mu} , \tilde{U}^{\nu \rho} \right] = \delta^{\mu \nu} \tilde{U}^{\lambda \rho}- \delta^{\lambda \nu} \tilde{U}^{\mu \rho} \, , \\
& \left[ \tilde{U}^{\lambda \mu} , \tilde{U}^{\nu \rho} \right] = - \delta^{\lambda \nu} \tilde{T}^{\mu \rho} - \delta^{\mu \rho} \tilde{T}'^{\lambda \nu} \, , \\
& \left[ \tilde{T}^{\lambda \mu} , \tilde{S}^{\nu \rho} \right] = \delta^{\mu \nu} \tilde{S}^{\lambda \rho} - \delta^{\lambda \nu} \tilde{S}^{\mu \rho}- \delta^{\mu \rho} \tilde{S}^{\lambda \nu} + \delta^{\lambda \rho} \tilde{S}^{\mu \nu} \, , \\
& \left[ \tilde{T}^{\lambda \mu} , \tilde{V}^{\nu \rho} \right] = \delta^{\mu \rho} \tilde{V}^{\nu \lambda} - \delta^{\lambda \rho} \tilde{V}^{\nu \mu} \, , \\
& \left[ \tilde{T}'^{\lambda \mu} , \tilde{S}'^{\nu \rho} \right] = \delta^{\mu \nu} \tilde{S}'^{\lambda \rho} - \delta^{\lambda \nu} \tilde{S}'^{\mu \rho}- \delta^{\mu \rho} \tilde{S}'^{\lambda \nu} + \delta^{\lambda \rho} \tilde{S}'^{\mu \nu} \, , \\
& \left[ \tilde{T}'^{\lambda \mu} , \tilde{V}^{\nu \rho} \right] = \delta^{\mu \nu} \tilde{V}^{\lambda \rho}- \delta^{\lambda \nu} \tilde{V}^{\mu \rho} \, , \\
& \left[ \tilde{U}^{\lambda \mu} , \tilde{S}^{\nu \rho} \right] = - \delta^{\mu \rho} \tilde{V}^{\lambda \nu} + \delta^{\mu \nu} \tilde{V}^{\lambda \rho} \, , \\
& \left[ \tilde{U}^{\lambda \mu} , \tilde{S}'^{\nu \rho} \right] = - \delta^{\lambda \rho} \tilde{V}^{\nu \mu} + \delta^{\lambda \nu} \tilde{V}^{\rho \mu} \, , \\
& \left[ \tilde{U}^{\lambda \mu} , \tilde{V}^{\nu \rho} \right] =  - \delta^{\lambda \nu} \tilde{S}^{\mu \rho} - \delta^{\mu \rho} \tilde{S}'^{\lambda \nu}  \, , \\
& \left[ \tilde{S}^{\lambda \mu} , \tilde{S}^{\nu \rho} \right] = - \frac{1}{\ell} \left( \delta^{\mu \nu} \tilde{S}^{\lambda \rho} - \delta^{\lambda \nu} \tilde{S}^{\mu \rho}- \delta^{\mu \rho} \tilde{S}^{\lambda \nu} + \delta^{\lambda \rho} \tilde{S}^{\mu \nu} \right) \, , \\
& \left[ \tilde{S}^{\lambda \mu} , \tilde{V}^{\nu \rho} \right] = - \frac{1}{\ell} \left( \delta^{\mu \rho} \tilde{V}^{\nu \lambda} - \delta^{\lambda \rho} \tilde{V}^{\nu \mu} \right) \, , \\
& \left[ \tilde{S}'^{\lambda \mu} , \tilde{S}'^{\nu \rho} \right] = - \frac{1}{\ell} \left( \delta^{\mu \nu} \tilde{S}'^{\lambda \rho} - \delta^{\lambda \nu} \tilde{S}'^{\mu \rho}- \delta^{\mu \rho} \tilde{S}'^{\lambda \nu} + \delta^{\lambda \rho} \tilde{S}'^{\mu \nu} \right) \, , \\
& \left[ \tilde{S}'^{\lambda \mu} , \tilde{V}^{\nu \rho} \right] = - \frac{1}{\ell} \left( \delta^{\mu \nu} \tilde{V}^{\lambda \rho}- \delta^{\lambda \nu} \tilde{V}^{\mu \rho} \right) \, , \\
& \left[ \tilde{V}^{\lambda \mu} , \tilde{V}^{\nu \rho} \right] = \frac{1}{\ell} \left(\delta^{\lambda \nu} \tilde{S}^{\mu \rho} + \delta^{\mu \rho} \tilde{S}'^{\lambda \nu} \right) \, , 
\end{split}
\end{equation}
and 
\begin{equation}\label{newtq}
\begin{split}
& \left[ \tilde{T}^{\lambda \mu}, \tilde{Q}^{\pm \, \nu } _\alpha \right] = \frac{1}{2} \left[\delta^{\mu \nu} \left( \tilde{Q}^{+ \, \lambda } _\alpha + \tilde{Q}^{- \, \lambda } _\alpha \right) - \delta^{\lambda \nu} \left( \tilde{Q}^{+ \, \mu } _\alpha + \tilde{Q}^{- \, \mu } _\alpha \right)   \right] \,  , \\
& \left[ \tilde{T}'^{\lambda \mu}, \tilde{Q}^{\pm \, \nu } _\alpha \right] = \pm \frac{1}{2} \left[\delta^{\mu \nu} \left( \tilde{Q}^{+ \, \lambda } _\alpha - \tilde{Q}^{- \, \lambda } _\alpha \right) - \delta^{\lambda \nu} \left( \tilde{Q}^{+ \, \mu } _\alpha - \tilde{Q}^{- \, \mu } _\alpha \right)   \right]  \,  , \\
& \left[ \tilde{U}^{\lambda \mu}, \tilde{Q}^{\pm \, \nu } _\alpha \right] = \mp \frac{1}{2} \left( \Gamma_0 \right)_{\alpha \beta} \left[ \delta^{\lambda \nu} \left( \tilde{Q}^{+ \, \mu } _\beta + \tilde{Q}^{- \, \ \mu } _\beta \right) \pm \delta^{\mu \nu} \left( \tilde{Q}^{+ \, \lambda } _\beta - \tilde{Q}^{- \, \lambda } _\beta \right) \right]  \, .
\end{split}
\end{equation}

Now, let us rescale the generators with a parameter $\sigma$ as
\begin{equation}\label{rescn}
\tilde{H} \rightarrow \sigma H \, , \quad \tilde{K}_a \rightarrow \sigma K_a \, , \quad \tilde{S}^{\lambda \mu} \rightarrow \sigma S^{\lambda \mu} \, , \quad \tilde{S}'^{\lambda \mu} \rightarrow \sigma S'^{\lambda \mu} \, , \quad \tilde{V}^{\lambda \mu} \rightarrow \sigma V^{\lambda \mu} \, , \quad \tilde{Q}^{\pm \, \lambda}_\alpha \rightarrow \sqrt{\sigma} Q^{\pm \, \lambda}_\alpha  \, ,
\end{equation}  
where we have also removed the tilde symbol on the generators.
Taking the limit $\sigma \rightarrow \infty$ (and removing the tilde symbol also on the generators that we have not rescaled), we end up with the $\mathcal{N}=(\mathcal{N},0)$ AdS Carroll superalgebra (with $\mathcal{N}$ even), whose non-trivial (anti)commutation relations read as follows (recall the definitions \eqref{xydef}, \eqref{uvsas}, since we have expressed the anticommutation relations in terms of the combinations given in that expressions, together with the fact that $\tilde{U}'^{\lambda \mu} =- \tilde{U}^{\mu \lambda}$ and $\tilde{V}'^{\lambda \mu} =- \tilde{V}^{\mu \lambda}$, and the (anti)commutation relations \eqref{antinew}, \eqref{newtttsss}, and \eqref{newtq}):
\begin{equation}\label{adscarrollsupern}
\begin{split}
& \left[ K_a , J_{bc} \right] = \delta_{ab} K_c - \delta_{ac} K_b \, , \\
& \left[ J_{ab}, P_{c}\right] =\delta_{bc}P_{a}-\delta _{ac}P_{b} \, , \\
& \left[ K_{a}, P_{b}\right] = - \delta_{ab} H \, , \\
& \left[ P_a , P_b \right] = \frac{1}{\ell^2} J_{ab} \, , \quad \quad \left[ P_a , H \right] = \frac{1}{\ell^2} K_{a} \, , \\
& \left[ {T}^{\lambda \mu} , {T}^{\nu \rho} \right] = \delta^{\mu \nu} {T}^{\lambda \rho} - \delta^{\lambda \nu} {T}^{\mu \rho}- \delta^{\mu \rho} {T}^{\lambda \nu} + \delta^{\lambda \rho} {T}^{\mu \nu} \, , \\
& \left[ {T}^{\lambda \mu} , {U}^{\nu \rho} \right] = \delta^{\mu \rho} {U}^{\nu \lambda} - \delta^{\lambda \rho} {U}^{\nu \mu} \, , \\
& \left[ {T}'^{\lambda \mu} , {T}'^{\nu \rho} \right] = \delta^{\mu \nu} {T}'^{\lambda \rho} - \delta^{\lambda \nu} {T}'^{\mu \rho}- \delta^{\mu \rho} {T}'^{\lambda \nu} + \delta^{\lambda \rho} {T}'^{\mu \nu} \, , \\
& \left[ {T}'^{\lambda \mu} , {U}^{\nu \rho} \right] = \delta^{\mu \nu} {U}^{\lambda \rho}- \delta^{\lambda \nu} {U}^{\mu \rho} \, , \\
& \left[ {U}^{\lambda \mu} , {U}^{\nu \rho} \right] = - \delta^{\lambda \nu} {T}^{\mu \rho} - \delta^{\mu \rho} {T}'^{\lambda \nu} \, , \\
& \left[ {T}^{\lambda \mu} , {S}^{\nu \rho} \right] = \delta^{\mu \nu} {S}^{\lambda \rho} - \delta^{\lambda \nu} {S}^{\mu \rho}- \delta^{\mu \rho} {S}^{\lambda \nu} + \delta^{\lambda \rho} {S}^{\mu \nu} \, , \\
& \left[ {T}^{\lambda \mu} , {V}^{\nu \rho} \right] = \delta^{\mu \rho} {V}^{\nu \lambda} - \delta^{\lambda \rho} {V}^{\nu \mu} \, , \\
& \left[ {T}'^{\lambda \mu} , {S}'^{\nu \rho} \right] = \delta^{\mu \nu} {S}'^{\lambda \rho} - \delta^{\lambda \nu} {S}'^{\mu \rho}- \delta^{\mu \rho} {S}'^{\lambda \nu} + \delta^{\lambda \rho} {S}'^{\mu \nu} \, , \\
& \left[ {T}'^{\lambda \mu} , {V}^{\nu \rho} \right] = \delta^{\mu \nu} {V}^{\lambda \rho}- \delta^{\lambda \nu} {V}^{\mu \rho} \, , \\
& \left[ {U}^{\lambda \mu} , {S}^{\nu \rho} \right] = - \delta^{\mu \rho} {V}^{\lambda \nu} + \delta^{\mu \nu} {V}^{\lambda \rho} \, , \\
& \left[ {U}^{\lambda \mu} , {S}'^{\nu \rho} \right] = - \delta^{\lambda \rho} {V}^{\nu \mu} + \delta^{\lambda \nu} {V}^{\rho \mu} \, , \\
& \left[ {U}^{\lambda \mu} , {V}^{\nu \rho} \right] =  - \delta^{\lambda \nu} {S}^{\mu \rho} - \delta^{\mu \rho} {S}'^{\lambda \nu}  \, , \\
& \left[ J_{ab}, Q^{\pm \, \lambda} _\alpha \right] =-\frac{1}{2}\left( \Gamma _{ab} Q^{\pm \, \lambda}  \right)_{\alpha } \, , \quad \quad \left[ P_{a}, Q^{\pm \, \lambda} _\alpha \right] =- \frac{1}{2 \ell} \left( \Gamma _{a} Q^{\mp \, \lambda} \right) _\alpha \, , \\
& \left[ {T}^{\lambda \mu}, {Q}^{\pm \, \nu } _\alpha \right] = \frac{1}{2} \left[\delta^{\mu \nu} \left( {Q}^{+ \, \lambda } _\alpha + {Q}^{- \, \lambda } _\alpha \right) - \delta^{\lambda \nu} \left( {Q}^{+ \, \mu } _\alpha + {Q}^{- \, \mu } _\alpha \right)   \right] \,  , \\
& \left[ {T}'^{\lambda \mu}, {Q}^{\pm \, \nu } _\alpha \right] = \pm \frac{1}{2} \left[\delta^{\mu \nu} \left( {Q}^{+ \, \lambda } _\alpha - {Q}^{- \, \lambda } _\alpha \right) - \delta^{\lambda \nu} \left( {Q}^{+ \, \mu } _\alpha - {Q}^{- \, \mu } _\alpha \right)   \right]  \,  , \\
& \left[ {U}^{\lambda \mu}, {Q}^{\pm \, \nu } _\alpha \right] = \mp \frac{1}{2} \left( \Gamma_0 \right)_{\alpha \beta} \left[ \delta^{\lambda \nu} \left( {Q}^{+ \, \mu } _\beta + {Q}^{- \, \ \mu } _\beta \right) \pm \delta^{\mu \nu} \left( {Q}^{+ \, \lambda } _\beta - {Q}^{- \, \lambda } _\beta \right) \right]  \, , 
\end{split}
\end{equation}
\begin{equation}\nonumber
\begin{split}
& \lbrace {Q}^{+ \, \lambda} _\alpha , {Q}^{+ \, \mu} _\beta \rbrace = \left( \Gamma^0 C \right)_{\alpha \beta} \left( \delta^{\lambda \mu}  {H} - {V}^{(\lambda \mu)} \right) + C_{\alpha \beta} {Y}^{[\lambda \mu]} \, , \\
& \lbrace {Q}^{+ \, \lambda} _\alpha , {Q}^{- \, \mu} _\beta \rbrace = - \frac{1}{\ell} \delta^{\lambda \mu} \left( \Gamma^{a0} C \right)_{\alpha \beta} {K}_{a} +  C_{\alpha \beta} {Y}'^{[\lambda \mu]} - \left( \Gamma^0 C \right)_{\alpha \beta} {V}^{[\lambda \mu]} \, , \\
& \lbrace {Q}^{- \, \lambda} _\alpha , {Q}^{- \, \mu} _\beta \rbrace = \left( \Gamma^0 C \right)_{\alpha \beta} \left( \delta^{\lambda \mu}  {H} + {V}^{(\lambda \mu)} \right) + C_{\alpha \beta} {Y}^{[\lambda \mu]}  \, .
\end{split}
\end{equation}
Notice that if we restrict ourselves to the special case $\mathcal{N}=(2,0)$, that is $x=1$, after some algebraic calculations, exploiting the definitions \eqref{dectijsij}, \eqref{xydef}, \eqref{uvsas}, and the symmetry properties \eqref{symmprop}, we exactly reproduce the $\mathcal{N}=(2,0)$ AdS Carroll superalgebra obtained in Section \ref{20case}, given by \eqref{adscarrollsuper22}.\footnote{In particular, when restricting ourselves to $\mathcal{N}=(2,0)$, namely $x=1$, we have $\lambda, \mu , \ldots = 1$ and $T^{\lambda \mu}=T^{11}=0$, $T'^{\lambda \mu}=T'^{11}=0$, $S^{\lambda \mu}=S^{11}=0$, $S'^{\lambda \mu}=S'^{11}=0$, $U^{\lambda \mu}=U^{11}=- T^{12}=- \epsilon^{12} T=-T$, $V^{\lambda \mu}=V^{11}=- S^{12}=- \epsilon^{12} S=-S$, and \eqref{newsusyn} restricts itself to \eqref{newsusy22} (when performing the Carroll limit, we also remove the tilde symbol on the generators); then, one can show that the superalgebra \eqref{adscarrollsupern} reduces to the one given by \eqref{adscarrollsuper22}.}

In the sequel, we will construct a CS action in $2+1$ dimensions invariant under \eqref{adscarrollsupern}.

\subsection{$(\mathcal{N},0)$ AdS Carroll supergravity}

We can now move to the formulation of a three-dimensional CS supergravity action invariant under the superalgebra \eqref{adscarrollsupern}. We call this action $(\mathcal{N},0)$ AdS Carroll CS supergravity action (where, in our analysis, $\mathcal{N}$ is even).

To this aim, let us start by introducing the connection $1$-form $A$ associated with the superalgebra \eqref{adscarrollsupern}, that is
\begin{equation}\label{connadscarrollsupern}
\begin{split}
A & = \frac{1}{2} \omega^{ab} J_{ab} + k^a K_a + V^a P_a + h H \\
& + \frac{1}{2} t^{\lambda \mu} T_{\lambda \mu} + \frac{1}{2} t'^{\lambda \mu} T'_{\lambda \mu} + u^{\lambda \mu} U_{\lambda \mu} + \frac{1}{2} s^{\lambda \mu} S_{\lambda \mu} + \frac{1}{2} s'^{\lambda \mu} S'_{\lambda \mu} + v^{\lambda \mu} V_{\lambda \mu} \\
& + \psi^+_\lambda Q^{+ \, \lambda} + \psi^-_\lambda Q^{- \, \lambda} \, ,
\end{split}
\end{equation}
where $\omega^{ab}$, $k^a$, $V^a$, $h$, $t^{\lambda \mu}$, $t'^{\lambda \mu}$, $u^{\lambda \mu}$, $s^{\lambda \mu}$, $s'^{\lambda \mu}$, $v^{\lambda \mu}$, $\psi^+_\lambda$, and $\psi^-_\lambda$ are the $1$-form fields dual to the generators $J_{ab}$, $K_a$, $P_a$, $H$, $T_{\lambda \mu}$, $T'_{\lambda \mu}$, $U_{\lambda \mu}$, $S_{\lambda \mu}$, $S'_{\lambda \mu}$, $V_{\lambda \mu}$, $Q^{+ \, \lambda}$, and $Q^{- \, \lambda}$, respectively. 

The corresponding curvature $2$-form $F$ is
\begin{equation}\label{curv2fn}
\begin{split}
F & = \frac{1}{2} \mathcal{R}^{ab} J_{ab} + \mathcal{K}^a K_a + R^a P_a + \mathcal{H} H \\
& + \frac{1}{2} \mathcal{T}^{\lambda \mu} T_{\lambda \mu} + \frac{1}{2} \mathcal{T}'^{\lambda \mu} T'_{\lambda \mu} + \mathcal{U}^{\lambda \mu} U_{\lambda \mu} + \frac{1}{2} \mathcal{S}^{\lambda \mu} S_{\lambda \mu} + \frac{1}{2} \mathcal{S}'^{\lambda \mu} S'_{\lambda \mu} + \mathcal{V}^{\lambda \mu} V_{\lambda \mu} \\
& + \nabla \psi^+_\lambda Q^{+ \, \lambda} + \nabla \psi^-_\lambda Q^{- \, \lambda} \, ,
\end{split}
\end{equation}
with
\begin{equation}\label{curvadscarrollsupern}
\begin{split}
\mathcal{R}^{ab} & = d \omega^{ab} + \frac{1}{\ell^2} V^a V^b = R^{ab} + \frac{1}{\ell^2} V^a V^b  \, , \\
\mathcal{K}^a  & = d k^a + \omega^a_{\phantom{a} b} k^b + \frac{1}{\ell^2} V^a h + \frac{1}{\ell} \bar{\psi}^{+ \, \lambda} \Gamma^{a0} \psi^{- \, \lambda} = \mathfrak{K}^a  + \frac{1}{\ell^2} V^a h + \frac{1}{\ell} \bar{\psi}^{+ \, \lambda} \Gamma^{a0} \psi^{- \, \lambda} \, , \\
R^a & = d V^a + \omega^a_{\phantom{a}b} V^b \, , \\
\mathcal{H} & = d h + V^a k_a - \frac{1}{2} \bar{\psi}^{+ \, \lambda} \Gamma^0 \psi^{+ \, \lambda} - \frac{1}{2} \bar{\psi}^{- \, \lambda} \Gamma^0 \psi^{- \, \lambda}   = \mathfrak{H} - \frac{1}{2} \bar{\psi}^{+ \, \lambda} \Gamma^0 \psi^{+ \, \lambda} - \frac{1}{2} \bar{\psi}^{- \, \lambda} \Gamma^0 \psi^{- \, \lambda}  \,   , \\
\mathcal{T}^{\lambda \mu} & = d t^{\lambda \mu} + {t^\lambda}_\nu t^{\nu \mu} + {u'^{[\lambda}}_\nu u^{\nu \vert \mu]}  \,   , \\
\mathcal{T}'^{\lambda \mu} & = d t'^{\lambda \mu} + {t'^\lambda}_\nu t'^{\nu \mu} + {u^{[\lambda}}_{ \nu} u'^{\nu \vert \mu]} \,   , \\
\mathcal{U}^{\lambda \mu} & = d u^{\lambda \mu} + {u^\lambda}_\nu t^{\nu \mu} + {t'^\lambda}_\nu u^{\nu \mu} \,   , \\
\mathcal{S}^{\lambda \mu} & = d s^{\lambda \mu} + 2 {t^\lambda}_\nu s^{\nu \mu} + 2 {u'^{[\lambda}}_\nu v^{\nu \vert \mu]} - \frac{1}{2} \bar{\psi}^{+ \, \lambda} \psi^{+ \, \mu} - \frac{1}{2} \bar{\psi}^{- \, \lambda} \psi^{- \, \mu} - \bar{\psi}^{+ \, [ \lambda} \psi^{- \, \mu ]} \, , \\
\mathcal{S}'^{\lambda \mu} & = d s'^{\lambda \mu} + 2 {t'^\lambda}_\nu s'^{\nu \mu} + 2 {u^{[\lambda}}_\nu  v'^{\nu \vert \mu]} - \frac{1}{2} \bar{\psi}^{+ \, \lambda} \psi^{+ \, \mu} - \frac{1}{2} \bar{\psi}^{- \, \lambda} \psi^{- \, \mu} + \bar{\psi}^{+ \, [ \lambda} \psi^{- \, \mu ]} \, , \\
\mathcal{V}^{\lambda \mu} & = d v^{\lambda \mu} + {v^\lambda}_\nu t^{\nu \mu} + {t'^\lambda}_\nu v^{\nu \mu} + {u^\lambda}_\nu s^{\nu \mu} + {s'^\lambda}_\nu u^{\nu \mu} + \frac{1}{2} \bar{\psi}^{+ \,  \lambda} \Gamma^0 \psi^{+ \, \mu} - \frac{1}{2} \bar{\psi}^{- \, \lambda} \Gamma^0 \psi^{- \, \mu} \\
& + \bar{\psi}^{+ \, [ \lambda} \Gamma^0 \psi^{- \, \mu ]} \,  , \\
\nabla \psi^{+ \, \lambda} & = d \psi^{+ \, \lambda} + \frac{1}{4} \omega^{ab} \Gamma_{ab} \psi^{+ \, \lambda} + \frac{1}{2 \ell} V^a \Gamma_a \psi^{- \, \lambda} + \frac{1}{2} t^{\lambda \mu} \psi^+_\mu + \frac{1}{2} t^{\lambda \mu} \psi^-_\mu + \frac{1}{2} t'^{\lambda \mu} \psi^+_\mu - \frac{1}{2} t'^{\lambda \mu} \psi^-_\mu \\
&  + u^{(\lambda \mu)} \Gamma_0 \psi^+_\mu + u^{[\lambda \mu]} \Gamma_0 \psi^-_\mu  \,  , \\
\nabla \psi^{- \, \lambda} & = d \psi^{- \, \lambda} + \frac{1}{4} \omega^{ab} \Gamma_{ab} \psi^{- \, \lambda} + \frac{1}{2 \ell} V^a \Gamma_a \psi^{+ \, \lambda} + \frac{1}{2} t^{\lambda \mu} \psi^+_\mu + \frac{1}{2} t^{\lambda \mu} \psi^-_\mu - \frac{1}{2} t'^{\lambda \mu} \psi^+_\mu + \frac{1}{2} t'^{\lambda \mu} \psi^-_\mu \\
& - u^{(\lambda \mu)} \Gamma_0 \psi^-_\mu - u^{[\lambda \mu]} \Gamma_0 \psi^+_\mu \, ,
\end{split}
\end{equation}
where we have used
\begin{equation}\label{vu}
\begin{split}
& u^{\lambda \mu} = t^{\lambda + x \ \mu} = - t^{\mu \ \lambda + x} = - u'^{\mu \lambda} \, , \\
& v^{\lambda \mu} = s^{\lambda + x \ \mu} = - s^{\mu \ \lambda + x} = - v'^{\mu \lambda} \, ,
\end{split}
\end{equation}
and where we have
\begin{equation}\label{uvsymmform}
\begin{split}
& u^{(\lambda \mu)} = \frac{1}{2} \left( u^{\lambda \mu} + u^{\mu \lambda} \right) \, , \\
& u^{[\lambda \mu]} = \frac{1}{2} \left( u^{\lambda \mu} - u^{\mu \lambda} \right) \, .
\end{split}
\end{equation}

In order to develop a CS action invariant under \eqref{adscarrollsupern}, we have to consider the non-vanishing components of the invariant tensor in \eqref{invtospn}, decompose the indices as in \eqref{indexdec}, exploit \eqref{newsusyn}, \eqref{dectijsij}, \eqref{symmprop}, and rescale not only the generators in compliance with \eqref{rescn} but also the coefficients appearing in \eqref{invtospn} as in \eqref{alsc}. Consequently, in the ultra-relativistic limit $\sigma \rightarrow \infty$ we get the following non-vanishing components of an invariant tensor for \eqref{adscarrollsupern}:
\begin{equation}\label{invadscarrollsupern}
\begin{split}
& \langle J_{ab} J_{cd} \rangle = \alpha_0 \left( \delta_{ad} \delta_{bc} - \delta_{ac} \delta_{bd} \right) \, , \\
& \langle J_{ab} H \rangle = \alpha_1 \epsilon_{ab} \, , \\
& \langle K_a P_b \rangle = - \alpha_1 \epsilon_{ab} \, , \\
& \langle P_{a} P_{b} \rangle = \frac{\alpha_0}{\ell^2} \delta_{ab} \, , \\
& \langle {T}^{\lambda \mu} {T}^{\nu \rho} \rangle = \langle {T}'^{\lambda \mu} {T}'^{\nu \rho} \rangle  =  2 \alpha_0 \left( \delta^{\lambda \rho} \delta^{\nu \mu} - \delta^{\lambda \nu} \delta^{\rho \mu} \right) \,  , \\
& \langle {U}^{\lambda \mu} {U}^{\nu \rho} \rangle = - 2 \alpha_0 \delta^{\lambda \nu} \delta^{\rho \mu} \, , \\
& \langle {T}^{\lambda \mu} {S}^{\nu \rho} \rangle = \langle {T}'^{\lambda \mu} {S}'^{\nu \rho} \rangle  = - 2 \alpha_1 \left( \delta^{\lambda \rho} \delta^{\nu \mu} - \delta^{\lambda \nu} \delta^{\rho \mu} \right) \,  , \\
& \langle {U}^{\lambda \mu} {V}^{\nu \rho} \rangle = 2 \alpha_1 \delta^{\lambda \nu} \delta^{\rho \mu} \, , \\
& \langle Q^{+ \, \lambda}_\alpha Q^{+ \, \mu}_\beta \rangle = \langle Q^{- \, \lambda}_\alpha Q^{- \, \mu}_\beta \rangle = 2 \alpha_1  C_{\alpha \beta} \delta^{\lambda \mu} \, .
\end{split}
\end{equation}
The invariant tensor for \eqref{adscarrollsupern} above is non-degenerate if $\alpha_1 \neq 0$.

Then, substituting the connection $1$-form in \eqref{connadscarrollsupern} and the non-zero components of the invariant tensor \eqref{invadscarrollsupern} into \eqref{genCS}, we end up with the three-dimensional $(\mathcal{N},0)$ AdS Carroll CS supergravity action, which reads
\begin{equation}\label{CSACn}
\begin{split}
I^{(\mathcal{N},0)}_{CS} & = \frac{k}{4 \pi} \int_\mathcal{M} \Bigg \lbrace \frac{\alpha_0}{2} \bigg( \omega^a_{\phantom{a} b} R^b_{\phantom{b} a}  + \frac{2}{\ell^2} V^a R_a + 2 {t^\lambda}_\mu {d t^\mu}_\lambda + \frac{4}{3} {t^\lambda}_\mu {t^\mu}_\nu {t^\nu}_\lambda + 2 {t'^\lambda}_\mu {d t'^\mu}_\lambda + \frac{4}{3} {t'^\lambda}_\mu {t'^\mu}_\nu {t'^\nu}_\lambda \\
& + 4 {u^\lambda}_\mu {d u'^\mu}_\lambda - 4 t_{\lambda \mu} {u'^{\lambda}}_{\nu} u^{\nu \mu} - 4 t'_{\lambda \mu} {u^{\lambda}}_{\nu} u'^{\nu \mu}  \bigg) + \alpha_1 \bigg[ \epsilon_{ab} R^{ab} h - 2 \epsilon_{ab} \mathfrak{K}^a V^b + \frac{1}{\ell^2} \epsilon_{ab} V^a V^b \\
& - 2 {t^\lambda}_\mu \left( {ds^\mu}_\lambda +  {t^\mu}_\nu {s^\nu}_\lambda \right) - 2 {t'^\lambda}_\mu \left( {ds'^\mu}_\lambda +  {t'^\mu}_\nu {s'^\nu}_\lambda \right) - 4 {u^\lambda}_\mu {d v'^\mu}_\lambda - 2 {u'^\lambda}_\mu {u^\mu}_\nu {s^\nu}_\lambda - 2 {u^\lambda}_\mu {u'^\mu}_\nu {s'^\nu}_\lambda  \\
& -  4 {u'^\lambda}_\mu {v^\mu}_\nu {t^\nu}_\lambda - 4 {u^\lambda}_\mu {v'^\mu}_\nu {t'^\nu}_\lambda + 2 \bar{\psi}^{+ \, \lambda} \nabla \psi^{+ \, \lambda} + 2 \bar{\psi}^{- \, \lambda} \nabla \psi^{- \, \lambda} \bigg] \\
& - d \left( \frac{\alpha_1}{2} \epsilon_{ab} \omega^{ab} h - \alpha_1 \epsilon_{ab} k^a V^b + \alpha_1 {t^\lambda}_\mu {s^\mu}_\lambda + \alpha_1 {t'^\lambda}_\mu {s'^\mu}_\lambda + 2 \alpha_1 {u^\lambda}_\mu {v'^\mu}_\lambda \right) \Bigg \rbrace \,  ,
\end{split}
\end{equation}
where we have also exploited \eqref{vu}. The action \eqref{CSACn} has been written in terms of the curvatures appearing in \eqref{curvadscarrollsupern} and it involves two coupling constants, that are $\alpha_0$ and $\alpha_1$. Up to boundary terms, \eqref{CSACn} can be rewritten as
\begin{equation}
\begin{split}
I^{(\mathcal{N},0)}_{CS} & = \frac{k}{4 \pi} \int_\mathcal{M} \Bigg \lbrace \frac{\alpha_0}{2} \bigg( \omega^a_{\phantom{a} b} R^b_{\phantom{b} a}  + \frac{2}{\ell^2} V^a R_a + 2 {t^\lambda}_\mu {d t^\mu}_\lambda + \frac{4}{3} {t^\lambda}_\mu {t^\mu}_\nu {t^\nu}_\lambda + 2 {t'^\lambda}_\mu {d t'^\mu}_\lambda + \frac{4}{3} {t'^\lambda}_\mu {t'^\mu}_\nu {t'^\nu}_\lambda \\
& + 4 {u^\lambda}_\mu {d u'^\mu}_\lambda - 4 t_{\lambda \mu} {u'^{\lambda}}_{\nu} u^{\nu \mu} - 4 t'_{\lambda \mu} {u^{\lambda}}_{\nu} u'^{\nu \mu}  \bigg) + \alpha_1 \bigg[ \epsilon_{ab} R^{ab} h - 2 \epsilon_{ab} \mathfrak{K}^a V^b + \frac{1}{\ell^2} \epsilon_{ab} V^a V^b \\
& - 2 {t^\lambda}_\mu \left( {ds^\mu}_\lambda +  {t^\mu}_\nu {s^\nu}_\lambda \right) - 2 {t'^\lambda}_\mu \left( {ds'^\mu}_\lambda +  {t'^\mu}_\nu {s'^\nu}_\lambda \right) - 4 {u^\lambda}_\mu {d v'^\mu}_\lambda - 2 {u'^\lambda}_\mu {u^\mu}_\nu {s^\nu}_\lambda - 2 {u^\lambda}_\mu {u'^\mu}_\nu {s'^\nu}_\lambda  \\
& -  4 {u'^\lambda}_\mu {v^\mu}_\nu {t^\nu}_\lambda - 4 {u^\lambda}_\mu {v'^\mu}_\nu {t'^\nu}_\lambda + 2 \bar{\psi}^{+ \, \lambda} \nabla \psi^{+ \, \lambda} + 2 \bar{\psi}^{- \, \lambda} \nabla \psi^{- \, \lambda} \bigg] \Bigg \rbrace \, .
\end{split}
\end{equation}
The contribution proportional to $\alpha_0$ corresponds to the exotic Lagrangian, and, in the present case, it involves, besides the Lorentz and torsional terms, also pieces including the $1$-form fields $t^{\lambda \mu}$, $t'^{\lambda \mu}$, and $u^{\lambda \mu}$. On the other hand, the contribution proportional to $\alpha_1$ also includes terms involving the $1$-form fields $s^{\lambda \mu}$, $s'^{\lambda \mu}$, and $v^{\lambda \mu}$, plus the spinor $1$-form fields $\psi^{+ \, \lambda}$ and $\psi^{- \, \lambda}$.

The CS action \eqref{CSACn} is invariant by construction under \eqref{adscarrollsupern}, and the local gauge transformations $\delta_\lambda A = d \lambda + \left[A, \lambda \right]$ with gauge parameter
\begin{equation}\label{gparn}
\begin{split}
\lambda & = \frac{1}{2} \lambda^{ab} J_{ab} + \kappa^a K_a + \lambda^a P_a + \tau H + \frac{1}{2} \varrho^{\lambda \mu} t_{\lambda \mu} + \frac{1}{2} \varrho'^{\lambda \mu} t'_{\lambda \mu} + \varphi^{\lambda \mu} u_{\lambda \mu} \\
& + \frac{1}{2} \vartheta^{\lambda \mu} s_{\lambda \mu} + \frac{1}{2} \vartheta'^{\lambda \mu} s_{\lambda \mu} + \varsigma^{\lambda \mu} v_{\lambda \mu} + \varepsilon^{+ \, \lambda} Q^+_\lambda + \varepsilon^{- \, \lambda} Q^-_\lambda 
\end{split}
\end{equation}
are given by
\begin{equation}\label{gaugetrn}
\begin{split}
\delta \omega^{ab} & = d \lambda^{ab} + \frac{2}{\ell^2} V^{[a} \lambda^{b]} \, , \\
\delta k^a  & = d \kappa^a - \lambda^a_{\phantom{a} b} k^b + \omega^a_{\phantom{a} b} \kappa^b - \frac{1}{\ell^2} \lambda^a h + \frac{1}{\ell^2} V^a \tau - \frac{1}{\ell} \bar{\varepsilon}^{+ \, \lambda} \Gamma^{a0} \psi^{- \, \lambda} - \frac{1}{\ell} \bar{\varepsilon}^{- \, \lambda} \Gamma^{a0} \psi^{+ \, \lambda} \, , \\
\delta V^a & = d \lambda^a - \lambda^a_{\phantom{a}b} V^b + \omega^a_{\phantom{a}b} \lambda^b \, , \\
\delta h & = d \tau - \lambda^a k_a + V^a \kappa_a + \bar{\varepsilon}^{+ \, \lambda} \Gamma^0 \psi^{+ \, \lambda} + \bar{\varepsilon}^{- \, \lambda} \Gamma^0 \psi^{- \, \lambda} \,   , \\
\delta t^{\lambda \mu} & = d \varrho^{\lambda \mu} - 2 {\varrho^{[\lambda}}_\nu t^{\nu \vert \mu]} - 2 {\varphi'^{[\lambda}}_\nu u^{\nu \vert \mu]} \, , \\
\delta t'^{\lambda \mu} & = d \varrho'^{\lambda \mu} - 2 {\varrho'^{[\lambda}}_\nu t'^{\nu \vert \mu]} - 2 {\varphi^{[\lambda}}_{ \nu} u'^{\nu \vert \mu]} \, , \\
\delta u^{\lambda \mu} & = d \varphi^{\lambda \mu} - {\varphi^\lambda}_\nu t^{\nu \mu} + {u^\lambda}_\nu \varrho^{\nu \mu} - {\varrho'^\lambda}_\nu u^{\nu \mu} + {t'^\lambda}_\nu \varphi^{\nu \mu}  \, , \\
\delta s^{\lambda \mu} & = d \vartheta^{\lambda \mu} - 2 {\varrho^{[\lambda}}_\nu s^{\nu \vert \mu]} + 2 {t^{[\lambda}}_\nu \vartheta^{\nu \vert \mu]} - 2 {\varphi'^{[\lambda}}_\nu v^{\nu \vert \mu]} + 2 {u'^{[\lambda}}_\nu \varsigma^{\nu \vert \mu]} + \bar{\varepsilon}^{+ \, [\lambda} \psi^{+ \, \mu]} + \bar{\varepsilon}^{- \, [\lambda} \psi^{- \, \mu]} \\ 
& + \bar{\varepsilon}^{+ \, [ \lambda} \psi^{- \, \mu ]} + \bar{\varepsilon}^{- \, [ \lambda} \psi^{+ \, \mu ]} \, , \\
\delta s'^{\lambda \mu} & = d \vartheta'^{\lambda \mu} - 2 {\varrho'^{[\lambda}}_\nu s'^{\nu \vert \mu]} + 2 {t'^{[\lambda}}_\nu \vartheta'^{\nu \vert \mu]} - 2 {\varphi^{[\lambda}}_\nu  v'^{\nu \vert \mu]} + 2 {u^{[\lambda}}_\nu  \varsigma'^{\nu \vert \mu]} + \bar{\varepsilon}^{+ \, [\lambda} \psi^{+ \, \mu]} + \bar{\varepsilon}^{- \, [\lambda} \psi^{- \, \mu]} \\
& - \bar{\varepsilon}^{+ \, [ \lambda} \psi^{- \, \mu ]} - \bar{\varepsilon}^{- \, [ \lambda} \psi^{+ \, \mu ]} \, , \\
\delta v^{\lambda \mu} & = d \varsigma^{\lambda \mu} - {\varsigma^\lambda}_\nu t^{\nu \mu} + {v^\lambda}_\nu \varrho^{\nu \mu} - {\varrho'^\lambda}_\nu v^{\nu \mu} + {t'^\lambda}_\nu \varsigma^{\nu \mu} - {\varphi^\lambda}_\nu s^{\nu \mu} + {u^\lambda}_\nu \vartheta^{\nu \mu} - {\vartheta'^\lambda}_\nu u^{\nu \mu} + {s'^\lambda}_\nu \varphi^{\nu \mu}  \\
& - \bar{\varepsilon}^{+ \,  \lambda} \Gamma^0 \psi^{+ \, \mu} + \bar{\varepsilon}^{- \, \lambda} \Gamma^0 \psi^{- \, \mu} - \bar{\varepsilon}^{+ \, [ \lambda} \Gamma^0 \psi^{- \, \mu ]} + \bar{\varepsilon}^{- \, [ \lambda} \Gamma^0 \psi^{+ \, \mu ]} \,  , \\
\delta \psi^{+ \, \lambda} & = d \varepsilon^{+ \, \lambda} - \frac{1}{4} \lambda^{ab} \Gamma_{ab} \psi^{+ \, \lambda} + \frac{1}{4} \omega^{ab} \Gamma_{ab} \varepsilon^{+ \, \lambda} - \frac{1}{2 \ell} \lambda^a \Gamma_a \psi^{- \, \lambda} + \frac{1}{2 \ell} V^a \Gamma_a \varepsilon^{- \, \lambda} \\
& - \frac{1}{2} \varrho^{\lambda \mu} \psi^+_\mu + \frac{1}{2} t^{\lambda \mu} \varepsilon^+_\mu - \frac{1}{2} \varrho^{\lambda \mu} \psi^-_\mu + \frac{1}{2} t^{\lambda \mu} \varepsilon^-_\mu - \frac{1}{2} \varrho'^{\lambda \mu} \psi^+_\mu + \frac{1}{2} t'^{\lambda \mu} \varepsilon^+_\mu + \frac{1}{2} \varrho'^{\lambda \mu} \psi^-_\mu - \frac{1}{2} t'^{\lambda \mu} \varepsilon^-_\mu \\
& - \varphi^{(\lambda \mu)} \Gamma_0 \psi^+_\mu  + u^{(\lambda \mu)} \Gamma_0 \varepsilon^+_\mu - \varphi^{[\lambda \mu]} \Gamma_0 \psi^-_\mu  + u^{[\lambda \mu]} \Gamma_0 \varepsilon^-_\mu  \,  , \\
\delta \psi^{- \, \lambda} & = d \varepsilon^{- \, \lambda} - \frac{1}{4} \lambda^{ab} \Gamma_{ab} \psi^{- \, \lambda} + \frac{1}{4} \omega^{ab} \Gamma_{ab} \varepsilon^{- \, \lambda} - \frac{1}{2 \ell} \lambda^a \Gamma_a \psi^{+ \, \lambda} + \frac{1}{2 \ell} V^a \Gamma_a \varepsilon^{+ \, \lambda} \\
& - \frac{1}{2} \varrho^{\lambda \mu} \psi^+_\mu + \frac{1}{2} t^{\lambda \mu} \varepsilon^+_\mu - \frac{1}{2} \varrho^{\lambda \mu} \psi^-_\mu + \frac{1}{2} t^{\lambda \mu} \varepsilon^-_\mu + \frac{1}{2} \varrho'^{\lambda \mu} \psi^+_\mu - \frac{1}{2} t'^{\lambda \mu} \varepsilon^+_\mu - \frac{1}{2} \varrho'^{\lambda \mu} \psi^-_\mu + \frac{1}{2} t'^{\lambda \mu} \varepsilon^-_\mu \\
& + \varphi^{(\lambda \mu)} \Gamma_0 \psi^-_\mu - u^{(\lambda \mu)} \Gamma_0 \varepsilon^-_\mu + \varphi^{[\lambda \mu]} \Gamma_0 \psi^+_\mu - u^{[\lambda \mu]} \Gamma_0 \varepsilon^+_\mu \, ,
\end{split}
\end{equation}
where we have also used the properties and definitions
\begin{equation}\label{parsymm}
\begin{split}
& \varphi^{\lambda \mu} = - \varphi'^{\mu \lambda} \, , \\
& \varphi^{(\lambda \mu)} \equiv \frac{1}{2} \left( \varphi^{\lambda \mu} + \varphi^{\mu \lambda} \right) \, , \\
& \varphi^{[\lambda \mu]} = \frac{1}{2} \left( \varphi^{\lambda \mu} - \varphi^{\mu \lambda} \right) \, .
\end{split}
\end{equation}
Restricting ourselves to supersymmetry, we get the following supersymmetry transformation laws:
\begin{equation}\label{susytrn}
\begin{split}
\delta \omega^{ab} & = 0 \, , \\
\delta k^a  & = - \frac{1}{\ell} \bar{\varepsilon}^{+ \, \lambda} \Gamma^{a0} \psi^{- \, \lambda} - \frac{1}{\ell} \bar{\varepsilon}^{- \, \lambda} \Gamma^{a0} \psi^{+ \, \lambda} \, , \\
\delta V^a & = 0 \, , \\
\delta h & = \bar{\varepsilon}^{+ \, \lambda} \Gamma^0 \psi^{+ \, \lambda} + \bar{\varepsilon}^{- \, \lambda} \Gamma^0 \psi^{- \, \lambda} \,   , \\
\delta t^{\lambda \mu} & = 0 \, , \quad \quad \delta t'^{\lambda \mu} = 0 \, , \quad \quad \delta u^{\lambda \mu} = 0  \, , \\
\delta s^{\lambda \mu} & = \bar{\varepsilon}^{+ \, [\lambda} \psi^{+ \, \mu]} + \bar{\varepsilon}^{- \, [\lambda} \psi^{- \, \mu]} + \bar{\varepsilon}^{+ \, [ \lambda} \psi^{- \, \mu ]} + \bar{\varepsilon}^{- \, [ \lambda} \psi^{+ \, \mu ]} \, , \\
\delta s'^{\lambda \mu} & = \bar{\varepsilon}^{+ \, [\lambda} \psi^{+ \, \mu]} + \bar{\varepsilon}^{- \, [\lambda} \psi^{- \, \mu]} - \bar{\varepsilon}^{+ \, [ \lambda} \psi^{- \, \mu ]} - \bar{\varepsilon}^{- \, [ \lambda} \psi^{+ \, \mu ]} \, , \\
\delta v^{\lambda \mu} & = - \bar{\varepsilon}^{+ \,  \lambda} \Gamma^0 \psi^{+ \, \mu} + \bar{\varepsilon}^{- \, \lambda} \Gamma^0 \psi^{- \, \mu} - \bar{\varepsilon}^{+ \, [ \lambda} \Gamma^0 \psi^{- \, \mu ]} + \bar{\varepsilon}^{- \, [ \lambda} \Gamma^0 \psi^{+ \, \mu ]} \,  , \\
\delta \psi^{+ \, \lambda} & = d \varepsilon^{+ \, \lambda} + \frac{1}{4} \omega^{ab} \Gamma_{ab} \varepsilon^{+ \, \lambda} + \frac{1}{2 \ell} V^a \Gamma_a \varepsilon^{- \, \lambda} + \frac{1}{2} t^{\lambda \mu} \varepsilon^+_\mu + \frac{1}{2} t^{\lambda \mu} \varepsilon^-_\mu + \frac{1}{2} t'^{\lambda \mu} \varepsilon^+_\mu - \frac{1}{2} t'^{\lambda \mu} \varepsilon^-_\mu \\
& + u^{(\lambda \mu)} \Gamma_0 \varepsilon^+_\mu + u^{[\lambda \mu]} \Gamma_0 \varepsilon^-_\mu  \,  , \\
\delta \psi^{- \, \lambda} & = d \varepsilon^{- \, \lambda} + \frac{1}{4} \omega^{ab} \Gamma_{ab} \varepsilon^{- \, \lambda} + \frac{1}{2 \ell} V^a \Gamma_a \varepsilon^{+ \, \lambda} + \frac{1}{2} t^{\lambda \mu} \varepsilon^+_\mu + \frac{1}{2} t^{\lambda \mu} \varepsilon^-_\mu - \frac{1}{2} t'^{\lambda \mu} \varepsilon^+_\mu + \frac{1}{2} t'^{\lambda \mu} \varepsilon^-_\mu \\
& - u^{(\lambda \mu)} \Gamma_0 \varepsilon^-_\mu - u^{[\lambda \mu]} \Gamma_0 \varepsilon^+_\mu \, .
\end{split}
\end{equation}

Finally, one can prove that from the variation of the action \eqref{CSACn} with respect to the $1$-form fields $\omega^{ab}$, $k^a$, $V^a$, $h$, $t^{\lambda \mu}$, $t'^{\lambda \mu}$, $u^{\lambda \mu}$, $s^{\lambda \mu}$, $s'^{\lambda \mu}$, $v^{\lambda \mu}$, $\psi^{+ \, \lambda}$, and $\psi^{- \, \lambda}$, we get, respectively, the equations of motion
\begin{equation}\label{eomn}
\begin{split}
\delta \omega^{ab} & : \quad \alpha_0 \mathcal{R}^{ab} + \alpha_1 \epsilon^{ab} \mathcal{H}= 0 \, , \\
\delta k^a & : \quad  \alpha_1 R^a = 0 \, , \\
\delta V^a & : \quad \frac{\alpha_0}{\ell^2} R^a + 2 \alpha_1 \epsilon_{ab} \mathcal{K}^b = 0 \,  , \\
\delta h & : \quad \alpha_1 \mathcal{R}^{ab} = 0  \, , \\
\delta t^{\lambda \mu} & : \quad \alpha_0 \mathcal{T}^{\lambda \mu} + \alpha_1  \mathcal{S}^{\lambda \mu}  = 0 \,  , \\
\delta t'^{\lambda \mu} & : \quad \alpha_0 \mathcal{T}'^{\lambda \mu} + \alpha_1 \mathcal{S}'^{\lambda \mu}  = 0 \,  , \\
\delta u^{\lambda \mu} & : \quad  - \alpha_0 \mathcal{U}^{\lambda \mu} + \alpha_1 \mathcal{V}^{\lambda \mu} = 0    \,  , \\
\delta s^{\lambda \mu} & : \quad \alpha_1 \mathcal{T}^{\lambda \mu} = 0 \, , \\
\delta s'^{\lambda \mu} & : \quad \alpha_1 \mathcal{T}'^{\lambda \mu} = 0 \, , \\
\delta v^{\lambda \mu} & : \quad \alpha_1 \mathcal{U}^{\lambda \mu} = 0 \, , \\
\delta \psi^{+ \, \lambda} & : \quad \alpha_1 \nabla \psi^{+ \, \lambda} = 0 \,  , \\
\delta \psi^{- \, \lambda} & : \quad \alpha_1 \nabla \psi^{- \, \lambda} = 0 \,  ,
\end{split}
\end{equation}
written up to boundary contributions. We observe that, for $\alpha_1 \neq 0$, the equations \eqref{eomn} reduce precisely to the vanishing of the $(\mathcal{N},0)$ super-AdS Carroll curvature $2$-forms given in \eqref{curv2fn}, that is to say
\begin{equation}\label{eomvacn}
\begin{split}
& \mathcal{R}^{ab} = 0 \,  , \quad \mathcal{K}^a = 0 \, , \quad R^a = 0 \, , \quad \mathcal{H}=0 \, , \\
& \mathcal{T}^{\lambda \mu} = 0 \, , \quad \mathcal{T}'^{\lambda \mu} = 0 \, , \quad \mathcal{U}^{\lambda \mu} = 0 \, , \\
& \mathcal{S}^{\lambda \mu} = 0 \, , \quad \mathcal{S}'^{\lambda \mu} = 0 \, , \quad \mathcal{V}^{\lambda \mu} = 0 \, , \\
& \nabla \psi^{+ \, \lambda} =0 \, , \quad \nabla \psi^{- \, \lambda} = 0 \, .
\end{split}
\end{equation}
Also in this case, we notice that $\alpha_1 \neq 0$ is a sufficient condition to recover \eqref{eomvacn}, which means that we can consistently impose $\alpha_0=0$ in the CS action \eqref{CSACn}, making the exotic term disappear.

Let us also mention that, restricting ourselves to the special case $\mathcal{N}=(2,0)$, that is $x=1$, after some algebraic calculations, we exactly reproduce the results of Section \ref{20case}, that is to say, in fact, the $(2,0)$ AdS Carroll supergravity theory.

\section{$(p,q)$ AdS Carroll supergravity theories in $2+1$ dimensions}\label{pqcase}

We finally extend our analysis to the $(p,q)$ case, {where both $p$ and $q$ are integers $>0$}. We first derive the $\mathcal{N}=(p,q)$ AdS Carroll superalgebra as the Carrollian contraction of the direct sum of an $\mathfrak{so}(p) \oplus \mathfrak{so}(q)$ algebra and $\mathfrak{osp}(2|p)\otimes \mathfrak{osp}(2,q)$. This allows us to end up with a non-degenerate invariant tensor in the ultra-relativistic limit, and to consequently construct the three-dimensional CS supergravity theory invariant under the aforesaid $\mathcal{N}=(p,q)$ AdS Carroll superalgebra.

\subsection{$\mathcal{N}=(p,q)$ AdS Carroll superalgebra} 

Let us begin by considering the direct sum of the $\mathfrak{osp}(2|p)\otimes \mathfrak{osp}(2,q)$ superalgebra and an $\mathfrak{so}(p) \oplus \mathfrak{so}(q)$ algebra{.} The non-zero (anti)commutation relations are the following ones (see \cite{Howe:1995zm}):
\begin{equation}\label{osppqnonmod}
\begin{split}
& \left[ \tilde{J}_{AB}, \tilde{J}_{CD}\right] =\eta _{BC}\tilde{J}_{AD}-\eta _{AC}\tilde{J}_{BD}-\eta
_{BD}\tilde{J}_{AC}+\eta _{AD}\tilde{J}_{BC} \, ,  \\
& \left[ \tilde{J}_{AB},\tilde{P}_{C}\right] =\eta _{BC}\tilde{P}_{A}-\eta _{AC}\tilde{P}_{B} \,  , \quad \quad \left[ \tilde{P}_A , \tilde{P}_B \right] = \frac{1}{\ell^2} \tilde{J}_{AB} \,  ,   \\
& \left[ \tilde{Z}^{ij} , \tilde{Z}^{kl} \right] = \delta^{jk} \tilde{Z}^{il} - \delta^{ik} \tilde{Z}^{jl}- \delta^{jl} \tilde{Z}^{ik} + \delta^{il} \tilde{Z}^{jk} \, , \\
& \left[ \tilde{Z}^{IJ} , \tilde{Z}^{KL} \right] = \delta^{JK} \tilde{Z}^{IL} - \delta^{IK} \tilde{Z}^{JL}- \delta^{JL} \tilde{Z}^{IK} + \delta^{IL} \tilde{Z}^{JK} \, , \\
& \left[ \tilde{S}^{ij} , \tilde{S}^{kl} \right] = - \frac{1}{\ell} \left( \delta^{jk} \tilde{S}^{il} - \delta^{ik} \tilde{S}^{jl}- \delta^{jl} \tilde{S}^{ik} + \delta^{il} \tilde{S}^{jk} \right) \, , \\
& \left[ \tilde{S}^{IJ} , \tilde{S}^{KL} \right] = - \frac{1}{\ell} \left( \delta^{JK} \tilde{S}^{IL} - \delta^{IK} \tilde{S}^{JL}- \delta^{JL} \tilde{S}^{IK} + \delta^{IL} \tilde{S}^{JK} \right) \, , \\
& \left[ \tilde{J}_{AB},\tilde{Q}^i_{\alpha }\right] =-\frac{1}{2}\left( \Gamma _{AB}\tilde{Q}^i\right)_{\alpha } \,  , \quad \quad \left[ \tilde{J}_{AB},\tilde{Q}^I_{\alpha }\right] =-\frac{1}{2}\left( \Gamma _{AB}\tilde{Q}^I\right)_{\alpha } \,  , \\
& \left[ \tilde{P}_{A}, \tilde{Q}^i_{\alpha }\right] =- \frac{1}{2 \ell} \left( \Gamma _{A}\tilde{Q}^i \right) _{\alpha } \,  , \quad \quad \left[ \tilde{P}_{A}, \tilde{Q}^I_{\alpha }\right] = \frac{1}{2 \ell} \left( \Gamma _{A}\tilde{Q}^I \right) _{\alpha } \,  , 
\end{split} 
\end{equation}
\begin{equation}\nonumber
\begin{split}
& \left[ \tilde{Z}^{ij}, \tilde{Q}^k_{\alpha }\right] = \delta^{jk} \tilde{Q}^i_\alpha - \delta^{ik} \tilde{Q}^j_\alpha \,  , \quad \quad \left[ \tilde{Z}^{IJ}, \tilde{Q}^K_{\alpha }\right] = \delta^{JK} \tilde{Q}^I_\alpha - \delta^{IK} \tilde{Q}^J_\alpha \,  , \\
& \left\{ \tilde{Q}^i_{\alpha }, \tilde{Q}^j_{\beta }\right\} = \delta^{ij} \left[ - \frac{1}{2 \ell} \left(\Gamma^{AB} C \right)_{\alpha \beta} \tilde{J}_{AB} + \left( \Gamma ^{A}C\right) _{\alpha \beta }\tilde{P}_{A} \right] + \frac{1}{\ell} C_{\alpha \beta} \tilde{Z}^{ij} \,  , \\
& \left\{ \tilde{Q}^I_{\alpha }, \tilde{Q}^J_{\beta }\right\} = \delta^{IJ} \left[ \frac{1}{2 \ell} \left(\Gamma^{AB} C \right)_{\alpha \beta} \tilde{J}_{AB} + \left( \Gamma ^{A}C\right) _{\alpha \beta }\tilde{P}_{A} \right] - \frac{1}{\ell} C_{\alpha \beta} \tilde{Z}^{IJ} \,  ,
\end{split}
\end{equation}
with $A, B , \ldots=0,1,2$, $i,j,\ldots=1,\ldots,p$, $I,J,\ldots=1,\ldots, q$, and where $\tilde{Z}^{ij}=-\tilde{Z}^{ji}$, $\tilde{S}^{ij}=-\tilde{S}^{ij}$, and $\tilde{Z}^{IJ} = - \tilde{Z}^{JI}$, $\tilde{S}^{IJ}= - \tilde{S}^{JI}$.

We now perform, on the same lines of \cite{Howe:1995zm}, the redefinition
\begin{equation}
\tilde{T}^{ij} \equiv \tilde{Z}^{ij} - \ell \tilde{S}^{ij} \, , \quad \tilde{T}^{IJ} \equiv \tilde{Z}^{IJ} - \ell \tilde{S}^{IJ} \, ,
\end{equation}
which is analogous to the ones we have done in Sections \ref{20case} and \ref{ncase}. Consequently, we can rewrite the (anti)commutation relations \eqref{osppqnonmod} as
\begin{equation}\label{osppq}
\begin{split}
& \left[ \tilde{J}_{AB}, \tilde{J}_{CD}\right] =\eta _{BC}\tilde{J}_{AD}-\eta _{AC}\tilde{J}_{BD}-\eta
_{BD}\tilde{J}_{AC}+\eta _{AD}\tilde{J}_{BC} \, ,  \\
& \left[ \tilde{J}_{AB},\tilde{P}_{C}\right] =\eta _{BC}\tilde{P}_{A}-\eta _{AC}\tilde{P}_{B} \,  , \\
& \left[ \tilde{P}_A , \tilde{P}_B \right] = \frac{1}{\ell^2} \tilde{J}_{AB} \,  ,   \\
& \left[ \tilde{T}^{ij} , \tilde{T}^{kl} \right] = \delta^{jk} \tilde{T}^{il} - \delta^{ik} \tilde{T}^{jl}- \delta^{jl} \tilde{T}^{ik} + \delta^{il} \tilde{T}^{jk} \, , \\
& \left[ \tilde{T}^{IJ} , \tilde{T}^{KL} \right] = \delta^{JK} \tilde{T}^{IL} - \delta^{IK} \tilde{T}^{JL}- \delta^{JL} \tilde{T}^{IK} + \delta^{IL} \tilde{T}^{JK} \, , \\
& \left[ \tilde{T}^{ij} , \tilde{S}^{kl} \right] = \delta^{jk} \tilde{S}^{il} - \delta^{ik} \tilde{S}^{jl}- \delta^{jl} \tilde{S}^{ik} + \delta^{il} \tilde{S}^{jk} \, , \\
& \left[ \tilde{T}^{IJ} , \tilde{S}^{KL} \right] = \delta^{JK} \tilde{S}^{IL} - \delta^{IK} \tilde{S}^{JL}- \delta^{JL} \tilde{S}^{IK} + \delta^{IL} \tilde{S}^{JK} \, , \\
& \left[ \tilde{S}^{ij} , \tilde{S}^{kl} \right] = - \frac{1}{\ell} \left( \delta^{jk} \tilde{S}^{il} - \delta^{ik} \tilde{S}^{jl}- \delta^{jl} \tilde{S}^{ik} + \delta^{il} \tilde{S}^{jk} \right) \, , \\
& \left[ \tilde{S}^{IJ} , \tilde{S}^{KL} \right] = - \frac{1}{\ell} \left( \delta^{JK} \tilde{S}^{IL} - \delta^{IK} \tilde{S}^{JL}- \delta^{JL} \tilde{S}^{IK} + \delta^{IL} \tilde{S}^{JK} \right) \, , \\
& \left[ \tilde{J}_{AB},\tilde{Q}^i_{\alpha }\right] =-\frac{1}{2}\left( \Gamma _{AB}\tilde{Q}^i\right)_{\alpha } \,  , \\
& \left[ \tilde{J}_{AB},\tilde{Q}^I_{\alpha }\right] =-\frac{1}{2}\left( \Gamma _{AB}\tilde{Q}^I\right)_{\alpha } \,  , \\
& \left[ \tilde{P}_{A}, \tilde{Q}^i_{\alpha }\right] =- \frac{1}{2 \ell} \left( \Gamma _{A}\tilde{Q}^i \right) _{\alpha } \,  , \\
& \left[ \tilde{P}_{A}, \tilde{Q}^I_{\alpha }\right] = \frac{1}{2 \ell} \left( \Gamma _{A}\tilde{Q}^I \right) _{\alpha } \,  , \\
& \left[ \tilde{T}^{ij}, \tilde{Q}^k_{\alpha }\right] = \delta^{jk} \tilde{Q}^i_\alpha - \delta^{ik} \tilde{Q}^j_\alpha \,  , \\
& \left[ \tilde{T}^{IJ}, \tilde{Q}^K_{\alpha }\right] = \delta^{JK} \tilde{Q}^I_\alpha - \delta^{IK} \tilde{Q}^J_\alpha \,  , \\
& \left\{ \tilde{Q}^i_{\alpha }, \tilde{Q}^j_{\beta }\right\} = \delta^{ij} \left[ - \frac{1}{2 \ell} \left(\Gamma^{AB} C \right)_{\alpha \beta} \tilde{J}_{AB} + \left( \Gamma ^{A}C\right) _{\alpha \beta }\tilde{P}_{A} \right] +  C_{\alpha \beta} \left(\frac{1}{\ell} \tilde{T}^{ij} + \tilde{S}^{ij} \right) \,  , \\
& \left\{ \tilde{Q}^I_{\alpha }, \tilde{Q}^J_{\beta }\right\} = \delta^{IJ} \left[ \frac{1}{2 \ell} \left(\Gamma^{AB} C \right)_{\alpha \beta} \tilde{J}_{AB} + \left( \Gamma ^{A}C\right) _{\alpha \beta }\tilde{P}_{A} \right] - C_{\alpha \beta} \left(\frac{1}{\ell} \tilde{T}^{IJ} + \tilde{S}^{IJ} \right) \, .
\end{split}
\end{equation}
Notice that taking the flat limit $\ell \rightarrow \infty$ of \eqref{osppq}, one recovers the $\mathcal{N}=(p,q)$ Poincar\'{e} superalgebra of \cite{Howe:1995zm}{.}

The non-vanishing components of an invariant tensor for the superalgebra \eqref{osppq}, that will be useful in the following study, are given by
\begin{equation}\label{invtosppq}
\begin{split}
& \langle \tilde{J}_{AB} \tilde{J}_{CD} \rangle = \alpha_0 \left( \eta_{AD} \eta_{BC} - \eta_{AC} \eta_{BD} \right) \, , \\
& \langle \tilde{J}_{AB} \tilde{P}_{C} \rangle = \alpha_1 \epsilon_{ABC} \,  , \\
& \langle \tilde{P}_{A} \tilde{P}_{B} \rangle = \frac{\alpha_0}{\ell^2} \eta_{AB} \,  , \\
& \langle \tilde{T}^{ij} \tilde{T}^{kl} \rangle =  2 \alpha_0 \left( \delta^{il} \delta^{kj} - \delta^{ik} \delta^{lj} \right) \,  , \\
& \langle \tilde{T}^{IJ} \tilde{T}^{KL} \rangle =  2 \alpha_0 \left( \delta^{IL} \delta^{KJ} - \delta^{IK} \delta^{LJ} \right) \,  , \\
& \langle \tilde{T}^{ij} \tilde{S}^{kl} \rangle = - 2 \alpha_1 \left( \delta^{il} \delta^{kj} - \delta^{ik} \delta^{lj} \right) \,  , \\
& \langle \tilde{T}^{IJ} \tilde{S}^{KL} \rangle = 2 \alpha_1 \left( \delta^{IL} \delta^{KJ} - \delta^{IK} \delta^{LJ} \right) \,  , \\
& \langle \tilde{S}^{ij} \tilde{S}^{kl} \rangle =  \frac{2 \alpha_1}{\ell} \left( \delta^{il} \delta^{kj} - \delta^{ik} \delta^{lj} \right) \,  , \\
& \langle \tilde{S}^{IJ} \tilde{S}^{KL} \rangle = - \frac{2 \alpha_1}{\ell} \left( \delta^{IL} \delta^{KJ} - \delta^{IK} \delta^{LJ} \right) \,  , \\
& \langle \tilde{Q}^i_\alpha \tilde{Q}^j_\beta \rangle = 2 \left(\alpha_1 - \frac{\alpha_0}{\ell} \right) C_{\alpha \beta} \delta^{ij} \,  , \\
& \langle \tilde{Q}^I_\alpha \tilde{Q}^J_\beta \rangle = 2 \left(\alpha_1 + \frac{\alpha_0}{\ell} \right) C_{\alpha \beta} \delta^{IJ}  \,  , 
\end{split}
\end{equation}
where $\alpha_0$ and $\alpha_1$ are arbitrary constants.

In order to take the ultra-relativistic contraction of the superalgebra \eqref{osppq}, we decompose, as usual, the indices $A,B,\ldots = 0,1,2$ as in \eqref{indexdec}, which induces the decomposition \eqref{gendec}, together with \eqref{gammadec}. After that, we rescale the generators with a parameter $\sigma$ as
\begin{equation}\label{rescpq}
\tilde{H} \rightarrow \sigma H \, , \quad \tilde{K}_a \rightarrow \sigma K_a \, , \quad \tilde{S}^{ij} \rightarrow \sigma S^{ij} \, , \quad \tilde{S}^{IJ} \rightarrow \sigma S^{IJ} \, , \quad \tilde{Q}^i_\alpha \rightarrow \sqrt{\sigma} Q^i_\alpha  \, , \quad \tilde{Q}^I_\alpha \rightarrow \sqrt{\sigma} Q^I_\alpha   \, .
\end{equation}  
Then, taking the limit $\sigma \rightarrow \infty$ (and removing the tilde symbol also on the generators that we have not rescaled), we end up with the $\mathcal{N}=(p,q)$ AdS Carroll superalgebra whose non-trivial (anti)commutation relations read
\begin{equation}\label{adscarrollsuperpq}
\begin{split}
& \left[ K_a , J_{bc} \right] = \delta_{ab} K_c - \delta_{ac} K_b \, , \quad \quad \left[ J_{ab}, P_{c}\right] =\delta_{bc}P_{a}-\delta _{ac}P_{b} \, , \\
& \left[ K_{a}, P_{b}\right] = - \delta_{ab} H \, , \quad \quad \left[ P_a , P_b \right] = \frac{1}{\ell^2} J_{ab} \, , \quad \quad \left[ P_a , H \right] = \frac{1}{\ell^2} K_{a} \, , \\
& \left[ {T}^{ij} , {T}^{kl} \right] = \delta^{jk} {T}^{il} - \delta^{ik} {T}^{jl}- \delta^{jl} {T}^{ik} + \delta^{il} {T}^{jk} \, , \\
& \left[ {T}^{IJ} , {T}^{KL} \right] = \delta^{JK} {T}^{IL} - \delta^{IK} {T}^{JL}- \delta^{JL} {T}^{IK} + \delta^{IL} {T}^{JK} \, , \\
& \left[ {T}^{ij} , {S}^{kl} \right] = \delta^{jk} {S}^{il} - \delta^{ik} {S}^{jl}- \delta^{jl} {S}^{ik} + \delta^{il} {S}^{jk} \, , \\
& \left[ {T}^{IJ} , {S}^{KL} \right] = \delta^{JK} {S}^{IL} - \delta^{IK} {S}^{JL}- \delta^{JL} {S}^{IK} + \delta^{IL} {S}^{JK} \, , \\
& \left[ J_{ab},Q^i_{\alpha }\right] =-\frac{1}{2}\left( \Gamma _{ab} Q^i \right)_{\alpha } \, , \quad \quad \left[ J_{ab},Q^I_{\alpha }\right] =-\frac{1}{2}\left( \Gamma _{ab} Q^I \right)_{\alpha } \, , \\
& \left[ P_{a}, Q^i_{\alpha }\right] =- \frac{1}{2 \ell} \left( \Gamma _{a}Q^i \right) _{\alpha } \, , \quad \quad \left[ P_{a}, Q^I_{\alpha }\right] = \frac{1}{2 \ell} \left( \Gamma _{a}Q^I \right) _{\alpha } \, , 
\end{split}
\end{equation}
\begin{equation}\nonumber
\begin{split}
& \left[ {T}^{ij}, {Q}^k_{\alpha }\right] = \delta^{jk} {Q}^i_\alpha - \delta^{ik} {Q}^j_\alpha \,  , \quad \quad \left[ {T}^{IJ}, {Q}^K_{\alpha }\right] = \delta^{JK} {Q}^I_\alpha - \delta^{IK} {Q}^J_\alpha \,  , \\
& \left\{ Q^i_{\alpha }, Q^j_{\beta }\right\} = \delta^{ij} \left[- \frac{1}{\ell} \left(\Gamma^{a0} C \right)_{\alpha \beta} K_{a} + \left( \Gamma ^{0}C\right) _{\alpha \beta } H \right] + C_{\alpha \beta} S^{ij} \, , \\
& \left\{ Q^I_{\alpha }, Q^J_{\beta }\right\} = \delta^{ij} \left[ \frac{1}{\ell} \left(\Gamma^{a0} C \right)_{\alpha \beta} K_{a} + \left( \Gamma ^{0}C\right) _{\alpha \beta } H \right] - C_{\alpha \beta} S^{IJ}  \, .
\end{split}
\end{equation}
Notice that if we restrict ourselves to the special case $\mathcal{N}=(1,1)$, that is $p=q=1$, we exactly reproduce the $\mathcal{N}=(1,1)$ AdS Carroll superalgebra obtained in Section \ref{11case}, namely \eqref{adscarrollsuper11}.

In the following, we will construct a three-dimensional CS action invariant under \eqref{adscarrollsuperpq}.

\subsection{$(p,q)$ AdS Carroll supergravity}

We will now construct a three-dimensional CS supergravity action invariant under the superalgebra \eqref{adscarrollsuperpq} just introduced. We call this action $(p,q)$ AdS Carroll CS supergravity action{.}

To this aim, let us first introduce the connection $1$-form $A$ associated with \eqref{adscarrollsuperpq}, namely
\begin{equation}\label{connadscarrollsuperpq}
A = \frac{1}{2} \omega^{ab} J_{ab} + k^a K_a + V^a P_a + h H + \frac{1}{2} t^{ij} T_{ij}  + \frac{1}{2} t^{IJ} T_{IJ} + \frac{1}{2} s^{ij} S_{ij} + \frac{1}{2} s^{IJ} S_{IJ} + \psi_i Q^i + \psi_I Q^I \,  , 
\end{equation}
being $\omega^{ab}$, $k^a$, $V^a$, $h$, $t^{ij}$, $t^{IJ}$, $s^{ij}$, $s^{IJ}$, $\psi_i$, and $\psi_I$ the $1$-form fields respectively dual to the generators $J_{ab}$, $K_a$, $P_a$, $H$, $T_{ij}$, $T_{IJ}$, $S_{ij}$, $S_{IJ}$, $Q^i$, and $Q^I$ (obeying the (anti)commutation relations given in \eqref{adscarrollsuperpq}), and the related curvature $2$-form $F$, that is
\begin{equation}\label{curv2fpq}
F = \frac{1}{2} \mathcal{R}^{ab} J_{ab} + \mathcal{K}^a K_a + R^a P_a + \mathcal{H} H  + \frac{1}{2} \mathcal{T}^{ij} T_{ij} + \frac{1}{2} \mathcal{T}^{IJ} T_{IJ} + \frac{1}{2} \mathcal{S}^{ij} S_{ij} + \frac{1}{2} \mathcal{S}^{IJ} S_{IJ} + \nabla \psi_i Q^i + \nabla \psi_I Q^I \, ,
\end{equation}
with
\begin{equation}\label{curvadscarrollsuperpq}
\begin{split}
\mathcal{R}^{ab} & = d \omega^{ab} + \frac{1}{\ell^2} V^a V^b = R^{ab} + \frac{1}{\ell^2} V^a V^b  \, , \\
\mathcal{K}^a  & = d k^a + \omega^a_{\phantom{a} b} k^b + \frac{1}{\ell^2} V^a h + \frac{1}{2\ell} \bar{\psi}^i \Gamma^{a0} \psi^i - \frac{1}{2\ell} \bar{\psi}^I \Gamma^{a0} \psi^I  = \mathfrak{K}^a  + \frac{1}{2\ell} \bar{\psi}^i \Gamma^{a0} \psi^i - \frac{1}{2\ell} \bar{\psi}^I \Gamma^{a0} \psi^I \, , \\
R^a & = d V^a + \omega^a_{\phantom{a}b} V^b \, , \\
\mathcal{H} & = d h + V^a k_a - \frac{1}{2} \bar{\psi}^i \Gamma^0 \psi^i - \frac{1}{2} \bar{\psi}^I \Gamma^0 \psi^I   = \mathfrak{H} - \frac{1}{2} \bar{\psi}^i \Gamma^0 \psi^i - \frac{1}{2} \bar{\psi}^I \Gamma^0 \psi^I  \,   , \\
\mathcal{T}^{ij} & = d t^{ij} + {t^i}_k t^{kj} \, , \\
\mathcal{T}^{IJ} & = d t^{IJ} + {t^I}_K t^{KJ} \, ,\\
\mathcal{S}^{ij} & = d s^{ij} + 2 {t^i}_k s^{kj} - \bar{\psi}^i \psi^j \, , \\
\mathcal{S}^{IJ} & = d s^{IJ} + 2 {t^I}_K s^{KJ} + \bar{\psi}^I \psi^J \, ,\\
\nabla \psi^i & = d \psi^i + \frac{1}{4} \omega^{ab} \Gamma_{ab} \psi^i + \frac{1}{2 \ell} V^a \Gamma_a \psi^i + t^{ij} \psi_j \,  , \\
\nabla \psi^I & = d \psi^I + \frac{1}{4} \omega^{ab} \Gamma_{ab} \psi^I - \frac{1}{2 \ell} V^a \Gamma_a \psi^I + t^{IJ} \psi_J \,  .
\end{split}
\end{equation}

We can now move to the explicit construction of a CS action invariant under \eqref{adscarrollsuperpq}. To this aim, consider the non-vanishing components of the invariant tensor given in \eqref{invtosppq}, decompose the indices as in \eqref{indexdec}, and rescale not only the generators in compliance with \eqref{resc11} but also the coefficients appearing in \eqref{invtosppq} as in \eqref{alsc}. Consequently, the Carroll limit $\sigma \rightarrow \infty$ leads to the following non-vanishing components of an invariant tensor for the superalgebra \eqref{adscarrollsuperpq}:
\begin{equation}\label{invadscarrollsuperpq}
\begin{split}
& \langle J_{ab} J_{cd} \rangle = \alpha_0 \left( \delta_{ad} \delta_{bc} - \delta_{ac} \delta_{bd} \right) \, , \\
& \langle J_{ab} H \rangle = \alpha_1 \epsilon_{ab} \, , \\
& \langle K_a P_b \rangle = - \alpha_1 \epsilon_{ab} \, , \\
& \langle P_{a} P_{b} \rangle = \frac{\alpha_0}{\ell^2} \delta_{ab} \, , \\
& \langle {T}^{ij} {T}^{kl} \rangle =  2 \alpha_0 \left( \delta^{il} \delta^{kj} - \delta^{ik} \delta^{lj} \right) \,  , \\
& \langle {T}^{IJ} {T}^{KL} \rangle =  2 \alpha_0 \left( \delta^{IL} \delta^{KJ} - \delta^{IK} \delta^{LJ} \right) \,  , \\
& \langle {T}^{ij} {S}^{kl} \rangle = - 2 \alpha_1 \left( \delta^{il} \delta^{kj} - \delta^{ik} \delta^{lj} \right) \,  , \\
& \langle {T}^{IJ} {S}^{KL} \rangle =  2 \alpha_1 \left( \delta^{IL} \delta^{KJ} - \delta^{IK} \delta^{LJ} \right) \,  , \\
& \langle Q^i_\alpha Q^j_\beta \rangle = 2 \alpha_1  C_{\alpha \beta} \delta^{ij} \, , \\
& \langle Q^I_\alpha Q^J_\beta \rangle = 2 \alpha_1  C_{\alpha \beta} \delta^{IJ} \, .
\end{split}
\end{equation}
The invariant tensor whose components are given in \eqref{invadscarrollsuperpq} is non-degenerate when $\alpha_1 \neq 0$.

Then, substituting the connection $1$-form in \eqref{connadscarrollsuperpq} and the non-zero components of the invariant tensor \eqref{invadscarrollsuperpq} into the general expression \eqref{genCS}, we end up with the $(p,q)$ AdS Carroll CS supergravity action in three dimensions, that is
\begin{equation}\label{CSACpq}
\begin{split}
I^{(p,q)}_{CS} & = \frac{k}{4 \pi} \int_\mathcal{M} \Bigg \lbrace \frac{\alpha_0}{2} \left( \omega^a_{\phantom{a} b} R^b_{\phantom{b} a}  + \frac{2}{\ell^2} V^a R_a + 2 {t^i}_j {d t^j}_i + \frac{4}{3} {t^i}_j {t^j}_k {t^k}_i  + 2 {t^I}_J {d t^J}_I + \frac{4}{3} {t^I}_J {t^J}_K {t^K}_I   \right) \\
& + \alpha_1 \bigg[ \epsilon_{ab} R^{ab} h - 2 \epsilon_{ab} \mathfrak{K}^a V^b + \frac{1}{\ell^2} \epsilon_{ab} V^a V^b h  - 2 {t^i}_j \left( {ds^j}_i +  {t^j}_k {s^k}_i \right) + 2 {t^I}_J \left( {ds^J}_I + {t^J}_K {s^K}_I \right) \\
& + 2 \bar{\psi}^i \nabla \psi^i + 2 \bar{\psi}^I \nabla \psi^I \bigg] - d \left( \frac{\alpha_1}{2} \epsilon_{ab} \omega^{ab} h - \alpha_1 \epsilon_{ab} k^a V^b + \alpha_1 {t^i}_j {s^j}_i - \alpha_1 {t^I}_J {s^J}_I \right) \Bigg \rbrace \,  ,
\end{split}
\end{equation}
which is written in terms of the curvatures appearing in \eqref{curvadscarrollsuperpq} and it involves two coupling constants, that are $\alpha_0$ and $\alpha_1$. Up to boundary terms, the action \eqref{CSACpq} can be reworked as follows:
\begin{equation}
\begin{split}
I^{(p,q)}_{CS} & = \frac{k}{4 \pi} \int_\mathcal{M} \Bigg \lbrace \frac{\alpha_0}{2} \left( \omega^a_{\phantom{a} b} R^b_{\phantom{b} a}  + \frac{2}{\ell^2} V^a R_a + 2 {t^i}_j {d t^j}_i + \frac{4}{3} {t^i}_j {t^j}_k {t^k}_i  + 2 {t^I}_J {d t^J}_I + \frac{4}{3} {t^I}_J {t^J}_K {t^K}_I   \right) \\
& + \alpha_1 \bigg[ \epsilon_{ab} R^{ab} h - 2 \epsilon_{ab} \mathfrak{K}^a V^b + \frac{1}{\ell^2} \epsilon_{ab} V^a V^b h  - 2 {t^i}_j \left( {ds^j}_i +  {t^j}_k {s^k}_i \right) + 2 {t^I}_J \left( {ds^J}_I +  {t^J}_K {s^K}_I \right) \\
& + 2 \bar{\psi}^i \nabla \psi^i + 2 \bar{\psi}^I \nabla \psi^I \bigg] \Bigg \rbrace \, .
\end{split}
\end{equation}
As usual, the contribution proportional to $\alpha_0$ corresponds to the exotic Lagrangian, and we can see that it involves, besides the Lorentz and torsional terms, also pieces including the $1$-form fields $t^{ij}$ and $t^{IJ}$. However, it does not contain terms involving $\psi^i$ and $\psi^I$. On the other hand, the contribution proportional to $\alpha_1$ also includes pieces involving the $1$-form fields $s^{ij}$, $s^{IJ}$, $\psi^i$, and $\psi^I$.

The action \eqref{CSACpq} is invariant by construction under \eqref{adscarrollsuperpq}, and the local gauge transformations $\delta_\lambda A = d \lambda + \left[A, \lambda \right]$ with gauge parameter
\begin{equation}\label{gparpq}
\lambda = \frac{1}{2} \lambda^{ab} J_{ab} + \kappa^a K_a + \lambda^a P_a + \tau H + \frac{1}{2} \varphi^{ij} t_{ij} + \frac{1}{2} \varphi^{IJ} t_{IJ} + \frac{1}{2} \varsigma^{ij} s_{ij} + \frac{1}{2} \varsigma^{IJ} s_{IJ} + \varepsilon^i Q_i + \varepsilon^I Q_I
\end{equation}
are given by
\begin{equation}\label{gaugetrpq}
\begin{split}
\delta \omega^{ab} & = d \lambda^{ab} + \frac{2}{\ell^2} V^{[a} \lambda^{b]} \, , \\
\delta k^a  & = d \kappa^a - \lambda^a_{\phantom{a} b} k^b + \omega^a_{\phantom{a} b} \kappa^b - \frac{1}{\ell^2} \lambda^a h + \frac{1}{\ell^2} V^a \tau - \frac{1}{\ell} \bar{\varepsilon}^i \Gamma^{a0} \psi^i + \frac{1}{\ell} \bar{\varepsilon}^I \Gamma^{a0} \psi^I \, , \\
\delta V^a & = d \lambda^a - \lambda^a_{\phantom{a}b} V^b + \omega^a_{\phantom{a}b} \lambda^b \, , \\
\delta h & = d \tau - \lambda^a k_a + V^a \kappa_a + \bar{\varepsilon}^i \Gamma^0 \psi^i + \bar{\varepsilon}^I \Gamma^0 \psi^I \,   , \\
\delta t^{ij} & = d \varphi^{ij} - 2 {\varphi^{[i}}_k t^{k \vert j]} \, , \\
\delta t^{IJ} & = d \varphi^{IJ} - 2 {\varphi^{[I}}_K t^{K \vert J]} \, ,\\
\delta s^{ij} & = d \varsigma^{ij} - 2 {\varphi^{[i}}_k s^{k \vert j]} + 2 {t^{[i}}_k \varsigma^{k \vert j]} + 2 \bar{\varepsilon}^{[i} \psi^{j]} \, , \\
\delta s^{IJ} & = d \varsigma^{IJ} + 2 {\varphi^{[I}}_K s^{K \vert J]} + 2 {t^{[I}}_K \varsigma^{K \vert J]} - 2 \bar{\varepsilon}^{[I} \psi^{J]} \, , \\
\delta \psi^i & = d \varepsilon^+ - \frac{1}{4} \lambda^{ab} \Gamma_{ab} \psi^i + \frac{1}{4} \omega^{ab} \Gamma_{ab} \varepsilon^i - \frac{1}{2 \ell} \lambda^a \Gamma_a \psi^i + \frac{1}{2 \ell} V^a \Gamma_a \varepsilon^i - \varphi^{ij} \psi_j + t^{ij} \varepsilon_j \,  , \\
\delta \psi^I & = d \varepsilon^- - \frac{1}{4} \lambda^{ab} \Gamma_{ab} \psi^I + \frac{1}{4} \omega^{ab} \Gamma_{ab} \varepsilon^I + \frac{1}{2 \ell} \lambda^a \Gamma_a \psi^I - \frac{1}{2 \ell} V^a \Gamma_a \varepsilon^I  - \varphi^{IJ} \psi_J + t^{IJ} \varepsilon_J \, .
\end{split}
\end{equation}
Thus, the restriction to supersymmetry transformations gives us
\begin{equation}\label{susytrpq}
\begin{split}
\delta \omega^{ab} & =0 \, , \\
\delta k^a  & = - \frac{1}{\ell} \bar{\varepsilon}^i \Gamma^{a0} \psi^i + \frac{1}{\ell} \bar{\varepsilon}^I \Gamma^{a0} \psi^I \, , \\
\delta V^a & = 0 \, , \\
\delta h & = \bar{\varepsilon}^i \Gamma^0 \psi^i + \bar{\varepsilon}^I \Gamma^0 \psi^I \,   , \\
\delta t^{ij} & = 0 \, , \\
\delta t^{IJ} & = 0 \, ,\\
\delta s^{ij} & = 2 \bar{\varepsilon}^{[i} \psi^{j]} \, , \\
\delta s^{IJ} & = - 2 \bar{\varepsilon}^{[I} \psi^{J]} \, ,\\
\delta \psi^i & = d \varepsilon^+ + \frac{1}{4} \omega^{ab} \Gamma_{ab} \varepsilon^i + \frac{1}{2 \ell} V^a \Gamma_a \varepsilon^i + t^{ij} \varepsilon_j \,  , \\
\delta \psi^I & = d \varepsilon^- + \frac{1}{4} \omega^{ab} \Gamma_{ab} \varepsilon^I - \frac{1}{2 \ell} V^a \Gamma_a \varepsilon^I  + t^{IJ} \varepsilon_J \, .
\end{split}
\end{equation}

Finally, the equations of motion obtained from the variation of \eqref{CSACpq} with respect to the $1$-form fields $\omega^{ab}$, $k^a$, $V^a$, $h$, $t^{ij}$, $t^{IJ}$, $s^{ij}$, $s^{IJ}$, $\psi^i$, and $\psi^I$ are, respectively (up to boundary contributions),
\begin{equation}\label{eompq}
\begin{split}
\delta \omega^{ab} & : \quad \alpha_0 \mathcal{R}^{ab} + \alpha_1 \epsilon^{ab} \mathcal{H}= 0 \, , \\
\delta k^a & : \quad  \alpha_1 R^a = 0 \, , \\
\delta V^a & : \quad \frac{\alpha_0}{\ell^2} R^a + 2 \alpha_1 \epsilon_{ab} \mathcal{K}^b = 0 \,  , \\
\delta h & : \quad \alpha_1 \mathcal{R}^{ab} = 0  \, , \\
\delta t^{ij} & : \quad - \alpha_0 \mathcal{T}^{ij}  + \alpha_1 \mathcal{S}^{ij} = 0 \, , \\
\delta t^{IJ} & : \quad - \alpha_0 \mathcal{T}^{IJ}  - \alpha_1 \mathcal{S}^{IJ} = 0 \, , \\
\delta s^{ij} & : \quad \alpha_1 \mathcal{T}^{ij} = 0 \, , \\
\delta s^{IJ} & : \quad \alpha_1 \mathcal{T}^{IJ} = 0 \, , \\
\delta \psi^i & : \quad \alpha_1 \nabla \psi^i = 0 \,  , \\
\delta \psi^I & : \quad \alpha_1 \nabla \psi^I = 0 \,  ,
\end{split}
\end{equation}
and, for $\alpha_1 \neq 0$, they reduce precisely to the vanishing of the $(p,q)$ super-AdS Carroll curvature $2$-forms given in \eqref{curv2fpq}, that is to say
\begin{equation}\label{eomvacpq}
\begin{split}
& \mathcal{R}^{ab} = 0 \,  , \quad \mathcal{K}^a = 0 \, , \quad R^a = 0 \, , \quad \mathcal{H}=0 \, , \quad \mathcal{T}^{ij} = 0 \, , \quad \mathcal{T}^{IJ} = 0 \, , \quad \mathcal{S}^{ij} = 0 \, , \quad \mathcal{S}^{IJ} = 0 \, , \\
& \nabla \psi^i =0 \, , \quad \nabla \psi^I =0 \, .
\end{split}
\end{equation}
We observe that, as usual, $\alpha_1 \neq 0$ is a sufficient condition to recover \eqref{eomvacpq}, meaning that $\alpha_0$ can be consistently set to zero, making the exotic term in the CS action \eqref{CSACpq} disappear.

Let us finally mention that, if we restrict ourselves to the case $p=q=1$, we exactly reproduce the results of Section \ref{11case}, that is to say, as properly expected, the $(1,1)$ AdS Carroll supergravity theory.

\section{Study of the flat limit $\ell \rightarrow \infty$}\label{flatlimit}

In the sequel, we study the flat limit $\ell \rightarrow \infty$, which can be directly applied to the $\mathcal{N}$-extended AdS Carroll superalgebras \eqref{adscarrollsuper22}, \eqref{adscarrollsuper11}, \eqref{adscarrollsupern}, and \eqref{adscarrollsuperpq}, to the corresponding curvature $2$-forms, respectively given by \eqref{curvadscarrollsuper22}, \eqref{curvadscarrollsuper11}, \eqref{curvadscarrollsupern}, and \eqref{curvadscarrollsuperpq}, to the related CS actions \eqref{CSAC22}, \eqref{CSAC11}, \eqref{CSACn}, and \eqref{CSACpq}, to the transformation laws \eqref{gaugetr22}, \eqref{gaugetr11}, \eqref{gaugetrn}, \eqref{gaugetrpq} (and, in particular to the supersymmetry transformation laws \eqref{susytr22}, \eqref{susytr11}, \eqref{susytrn}, \eqref{susytrpq}), and to the field equations of the respective theories, namely \eqref{eom22}, \eqref{eom11}, \eqref{eomn}, and \eqref{eompq}.

\subsection{$(2,0)$ Carroll supergravity from the $\ell \rightarrow \infty$ limit}

In the limit $\ell \rightarrow \infty$, the (anti)commutation relations of the $\mathcal{N}=(2,0)$ AdS Carroll superalgebra \eqref{adscarrollsuper22} reduce to the following non-vanishing ones:
\begin{equation}\label{adscarrollsuperF22}
\begin{split}
& \left[ K_a , J_{bc} \right] = \delta_{ab} K_c - \delta_{ac} K_b \, , \\
& \left[ J_{ab}, P_{c}\right] =\delta_{bc}P_{a}-\delta _{ac}P_{b} \, , \\
& \left[ K_{a}, P_{b}\right] = - \delta_{ab} H \, , \\
& \left[ J_{ab},Q^\pm_{\alpha }\right] =-\frac{1}{2}\left( \Gamma _{ab} Q^\pm \right)_{\alpha } \, , \\
& \left[ T, Q^+_{\alpha }\right] = \left( \Gamma _{0} \right)_{\alpha \beta} Q^+_\beta \, , \\
& \left[ T, Q^-_{\alpha }\right] = - \left( \Gamma _{0} \right)_{\alpha \beta} Q^-_\beta \, , \\
& \left\{ Q^+_{\alpha }, Q^+_{\beta }\right\} = \left( \Gamma ^{0}C\right) _{\alpha \beta } \left( H + S \right) \, , \\
& \left\{ Q^-_{\alpha }, Q^-_{\beta }\right\} = \left( \Gamma ^{0}C\right) _{\alpha \beta } \left( H - S \right) \, .
\end{split}
\end{equation}
These are the (anti)commutation relations of a new $\mathcal{N}=(2,0)$, $D=3$ super-Carroll algebra, involving an extra generator $S$, which could also have been derived by considering the ultra-relativistic contraction of the $\mathcal{N}=(2,0)$, $D=3$ Poincar\'{e} superalgebra supplemented with an $so(2)$ extension consisting in the extra generator $\tilde{S}$ introduced at the relativistic level.

As $\ell \rightarrow \infty$, the $2$-form curvatures \eqref{curvadscarrollsuper22} become
\begin{equation}\label{curvadscarrollsuperF22}
\begin{split}
\mathcal{R}^{ab} & = d \omega^{ab} = R^{ab} \, , \\
\mathcal{K}^a  & = d k^a + \omega^a_{\phantom{a} b} k^b  = \mathfrak{K}^a  \, , \\
R^a & = d V^a + \omega^a_{\phantom{a}b} V^b \, , \\
\mathcal{H} & = d h + V^a k_a - \frac{1}{2} \bar{\psi}^+ \Gamma^0 \psi^+ - \frac{1}{2} \bar{\psi}^- \Gamma^0 \psi^-   = \mathfrak{H} - \frac{1}{2} \bar{\psi}^+ \Gamma^0 \psi^+ - \frac{1}{2} \bar{\psi}^- \Gamma^0 \psi^-  \,   , \\
\mathcal{T} & = d t \,   , \\
\mathcal{S} & = d s - \frac{1}{2} \bar{\psi}^+ \Gamma^0 \psi^+ + \frac{1}{2} \bar{\psi}^- \Gamma^0 \psi^- \, , \\
\nabla \psi^+ & = d \psi^+ + \frac{1}{4} \omega^{ab} \Gamma_{ab} \psi^+ - t \Gamma_0 \psi^+ \,  , \\
\nabla \psi^- & = d \psi^- + \frac{1}{4} \omega^{ab} \Gamma_{ab} \psi^- + t \Gamma_0 \psi^- \,  .
\end{split}
\end{equation}
On the other hand, by applying the $\ell \rightarrow \infty$ limit to the three-dimensional CS action \eqref{CSAC22}, we end up with 
\begin{equation}\label{CSC22}
\begin{split}
I^{(2,0)}_{CS} \vert_{\ell \rightarrow \infty} & = \frac{k}{4 \pi} \int_\mathcal{M} \Bigg \lbrace \frac{\alpha_0}{2} \left( \omega^a_{\phantom{a} b} R^b_{\phantom{b} a} - 4 t dt \right) + \alpha_1 \bigg( \epsilon_{ab} R^{ab} h - 2 \epsilon_{ab} \mathfrak{K}^a V^b + 4 t ds \\
& + 2 \bar{\psi}^+ \nabla \psi^+ + 2 \bar{\psi}^- \nabla \psi^- \bigg) - d \left( \frac{\alpha_1}{2} \epsilon_{ab} \omega^{ab} h - \alpha_1 \epsilon_{ab} k^a V^b - 2 \alpha_1 t s \right) \Bigg \rbrace \, ,
\end{split}
\end{equation}
which is written in terms of the super-Carroll curvatures appearing in \eqref{curvadscarrollsuperF22}. The latter must not be confused with the super-AdS Carroll ones given in \eqref{curvadscarrollsuper22}, since \eqref{curvadscarrollsuperF22} correspond to the flat limit of \eqref{curvadscarrollsuper22}. Here we signal that we have done a little abuse of notation.
The action \eqref{CSC22} can also be derived by using the following non-vanishing components of the invariant tensor: 
\begin{equation}
\begin{split}
& \langle J_{ab} J_{cd} \rangle = \alpha_0 \left( \delta_{ad} \delta_{bc} - \delta_{ac} \delta_{bd} \right) \, , \\
& \langle J_{ab} H \rangle = \alpha_1 \epsilon_{ab} \, , \\
& \langle K_a P_b \rangle = - \alpha_1 \epsilon_{ab} \, , \\
& \langle T T \rangle = - 2 \alpha_0\, , \\
& \langle T S \rangle = 2 \alpha_1 \, , \\
& \langle Q^+_\alpha Q^+_\beta \rangle = \langle Q^-_\alpha Q^-_\beta \rangle = 2 \alpha_1  C_{\alpha \beta} \, ,
\end{split}
\end{equation}
which are obtained by taking the limit $\ell \rightarrow \infty$ of \eqref{invadscarrollsuper22}, and the connection $1$-form for the $\mathcal{N}=(2,0)$ (flat) Carroll superalgebra \eqref{adscarrollsuperF22} in the general expression \eqref{genCS}. Notice that the exotic term, which is the one proportional to $\alpha_0$ in \eqref{CSC22}, now reduces purely to the so-called Lorentz Lagrangian.

The CS action \eqref{CSC22} is invariant by construction under the super-Carroll group associated with \eqref{adscarrollsuperF22}. 
In particular, concerning the flat limit of the gauge transformations \eqref{gaugetr22}, we get the local gauge transformations
\begin{equation}\label{gaugetrF22}
\begin{split}
\delta \omega^{ab} & = d \lambda^{ab} \, , \\
\delta k^a  & = d \kappa^a - \lambda^a_{\phantom{a} b} k^b + \omega^a_{\phantom{a} b} \kappa^b \, , \\
\delta V^a & = d \lambda^a - \lambda^a_{\phantom{a}b} V^b + \omega^a_{\phantom{a}b} \lambda^b \, , \\
\delta h & = d \tau - \lambda^a k_a + V^a \kappa_a + \bar{\varepsilon}^+ \Gamma^0 \psi^+ + \bar{\varepsilon}^- \Gamma^0 \psi^- \,   , \\
\delta t & = d \varphi \,   , \\
\delta s & = d \varsigma + \bar{\varepsilon}^+ \Gamma^0 \psi^+ - \bar{\varepsilon}^- \Gamma^0 \psi^- \, , \\
\delta \psi^+ & = d \varepsilon^+ - \frac{1}{4} \lambda^{ab} \Gamma_{ab} \psi^+ + \frac{1}{4} \omega^{ab} \Gamma_{ab} \varepsilon^+ + \varphi \Gamma_0 \psi^+ - t \Gamma_0 \varepsilon^+ \,  , \\
\delta \psi^- & = d \varepsilon^- - \frac{1}{4} \lambda^{ab} \Gamma_{ab} \psi^- + \frac{1}{4} \omega^{ab} \Gamma_{ab} \varepsilon^- - \varphi \Gamma_0 \psi^- + t \Gamma_0 \varepsilon^- \, .
\end{split}
\end{equation}
The restriction to supersymmetry transformations reads
\begin{equation}\label{susytrF22}
\begin{split}
\delta \omega^{ab} & = 0 \, , \\
\delta k^a  & = 0  \, , \\
\delta V^a & = 0 \, , \\
\delta h & = \bar{\varepsilon}^+ \Gamma^0 \psi^+ + \bar{\varepsilon}^- \Gamma^0 \psi^- \,   , \\
\delta t & = 0 \,   , \\
\delta s & = \bar{\varepsilon}^+ \Gamma^0 \psi^+ - \bar{\varepsilon}^- \Gamma^0 \psi^- \, , \\
\delta \psi^+ & = d \varepsilon^+ + \frac{1}{4} \omega^{ab} \Gamma_{ab} \varepsilon^+ - t \Gamma_0 \varepsilon^+ \,  , \\
\delta \psi^- & = d \varepsilon^- + \frac{1}{4} \omega^{ab} \Gamma_{ab} \varepsilon^- + t \Gamma_0 \varepsilon^- \, .
\end{split}
\end{equation}

Finally, the equations of motion for the action \eqref{CSC22} (flat limit of the equations of motion given in \eqref{eom22}) are
\begin{equation}\label{eomF22}
\begin{split}
\delta \omega^{ab} & : \quad \alpha_0 \mathcal{R}^{ab} + \alpha_1 \epsilon^{ab} \mathcal{H}= 0 \, , \\
\delta k^a & : \quad  \alpha_1 R^a = 0 \, , \\
\delta V^a & : \quad \alpha_1 \epsilon_{ab} \mathcal{K}^b = 0 \,  , \\
\delta h & : \quad \alpha_1 \mathcal{R}^{ab} = 0  \, , \\
\delta t & : \quad - \alpha_0 \mathcal{T} + \alpha_1 \mathcal{S} = 0  \, , \\
\delta s & : \quad \alpha_1 \mathcal{T} = 0  \, , \\
\delta \psi^+ & : \quad \alpha_1 \nabla \psi^+ = 0 \,  , \\
\delta \psi^- & : \quad \alpha_1 \nabla \psi^- = 0 \,  ,
\end{split}
\end{equation}
and we can see that, when $\alpha_1 \neq 0$, they exactly reduce to the vanishing of the curvature $2$-forms given in \eqref{curvadscarrollsuperF22}. We can also observe that, in analogy with the AdS case of Section \ref{20case}, also in the flat limit $\alpha_1 \neq 0$ results to be a sufficient condition to recover the vanishing of the curvature $2$-forms \eqref{curvadscarrollsuperF22} obtained in the flat limit, which means that one could consistently set $\alpha_0=0$ and thus neglect the exotic term (i.e., the Lorentz Lagrangian) in the CS action \eqref{CSC22}.\footnote{Let us also observe that if we now restrict ourselves to the purely bosonic part of the action \eqref{CSC22}, we get a three-dimensional CS Carroll gravity action that is different from the one obtained in \cite{Ravera:2019ize} (see also \cite{Bergshoeff:2017btm}) by considering the purely bosonic contributions, due to the presence of the bosonic $1$-form fields $t$ and $s$ dual to the generators $T$ and $S$, respectively. Nevertheless, if we consider the purely bosonic level and set $t=s=0$ through an IW contraction, we have that the aforementioned actions coincide.}

\subsection{$(1,1)$ Carroll supergravity from the $\ell \rightarrow \infty$ limit}

The limit $\ell \rightarrow \infty$ performed on the (anti)commutation relations of the $\mathcal{N}=(1,0)$ AdS Carroll superalgebra \eqref{adscarrollsuper11} leads to the following non-vanishing ones:
\begin{equation}\label{adscarrollsuperF11}
\begin{split}
& \left[ K_a , J_{bc} \right] = \delta_{ab} K_c - \delta_{ac} K_b \, , \\
& \left[ J_{ab}, P_{c}\right] =\delta_{bc}P_{a}-\delta _{ac}P_{b} \, , \\
& \left[ K_{a}, P_{b}\right] = - \delta_{ab} H \, , \\
& \left[ J_{ab},Q^\pm_{\alpha }\right] =-\frac{1}{2}\left( \Gamma _{ab} Q^\pm \right)_{\alpha } \, , \\
& \left\{ Q^\pm_{\alpha }, Q^\pm_{\beta }\right\} = \left( \Gamma ^{0}C\right) _{\alpha \beta } H \, .
\end{split}
\end{equation}
These are the (anti)commutation relations of the $\mathcal{N}=(1,1)$, $D=3$ super-Carroll algebra (see \cite{Bergshoeff:2015wma}, where \eqref{adscarrollsuperF11} corresponds to the superalgebra obtained in the $R \rightarrow \infty$ limit of the $\mathcal{N}=(1,1)$ AdS-Carroll superalgebra of Section C.4 of the same paper). It could be also obtained by considering the ultra-relativistic contraction of the $\mathcal{N}=(1,1)$, $D=3$ Poincar\'{e} superalgebra.

Taking $\ell \rightarrow \infty$, the $2$-form curvatures \eqref{curvadscarrollsuper11} reduce to
\begin{equation}\label{curvadscarrollsuperF11}
\begin{split}
\mathcal{R}^{ab} & = d \omega^{ab} = R^{ab}  \, , \\
\mathcal{K}^a  & = d k^a + \omega^a_{\phantom{a} b} k^b = \mathfrak{K}^a \, , \\
R^a & = d V^a + \omega^a_{\phantom{a}b} V^b \, , \\
\mathcal{H} & = d h + V^a k_a - \frac{1}{2} \bar{\psi}^+ \Gamma^0 \psi^+ - \frac{1}{2} \bar{\psi}^- \Gamma^0 \psi^-   = \mathfrak{H} - \frac{1}{2} \bar{\psi}^+ \Gamma^0 \psi^+ - \frac{1}{2} \bar{\psi}^- \Gamma^0 \psi^-  \,   , \\
\nabla \psi^+ & = d \psi^+ + \frac{1}{4} \omega^{ab} \Gamma_{ab} \psi^+ \,  , \\
\nabla \psi^- & = d \psi^- + \frac{1}{4} \omega^{ab} \Gamma_{ab} \psi^- \,  ,
\end{split}
\end{equation}
and the $\ell \rightarrow \infty$ limit of the CS action \eqref{CSAC11} leads us to the following three-dimensional one:
\begin{equation}\label{CSC11}
\begin{split}
I^{(1,1)}_{CS} \vert_{\ell \rightarrow \infty} & = \frac{k}{4 \pi} \int_\mathcal{M} \Bigg \lbrace \frac{\alpha_0}{2} \left( \omega^a_{\phantom{a} b} R^b_{\phantom{b} a} \right) + \alpha_1 \left( \epsilon_{ab} R^{ab} h - 2 \epsilon_{ab} \mathfrak{K}^a V^b + 2 \bar{\psi}^+ \nabla \psi^+ + 2 \bar{\psi}^- \nabla \psi^- \right) \\
& - d \left( \frac{\alpha_1}{2} \epsilon_{ab} \omega^{ab} h - \alpha_1 \epsilon_{ab} k^a V^b \right) \Bigg \rbrace \,  ,
\end{split}
\end{equation}
written in terms of the curvatures appearing in \eqref{curvadscarrollsuperF11} (again, we are doing a little abuse of notation). The action \eqref{CSC11} can also be derived by using the connection $1$-form for the $\mathcal{N}=(1,1)$ (flat) Carroll superalgebra \eqref{adscarrollsuperF11} together with the non-vanishing components of the invariant tensor 
\begin{equation}
\begin{split}
& \langle J_{ab} J_{cd} \rangle = \alpha_0 \left( \delta_{ad} \delta_{bc} - \delta_{ac} \delta_{bd} \right) \, , \\
& \langle J_{ab} H \rangle = \alpha_1 \epsilon_{ab} \, , \\
& \langle K_a P_b \rangle = - \alpha_1 \epsilon_{ab} \, , \\
& \langle Q^+_\alpha Q^+_\beta \rangle = \langle Q^-_\alpha Q^-_\beta \rangle = 2 \alpha_1  C_{\alpha \beta} \, 
\end{split}
\end{equation}
in the general expression for a three-dimensional CS action \eqref{genCS}. Analogously to what happened in the $(2,0)$ flat theory, also in the current case the exotic term, proportional to $\alpha_0$, now reduces purely to the Lorentz Lagrangian.

By construction, the CS action \eqref{CSC11} is invariant under the $(1,1)$ super-Carroll group, that is associated with the superalgebra given in \eqref{adscarrollsuperF11}. 
In particular, taking the flat limit of the gauge transformations \eqref{gaugetr11}, we get the following local gauge transformations:
\begin{equation}\label{gaugetrF11}
\begin{split}
\delta \omega^{ab} & = d \lambda^{ab} \, , \\
\delta k^a  & = d \kappa^a - \lambda^a_{\phantom{a} b} k^b + \omega^a_{\phantom{a} b} \kappa^b \, , \\
\delta V^a & = d \lambda^a - \lambda^a_{\phantom{a}b} V^b + \omega^a_{\phantom{a}b} \lambda^b \, , \\
\delta h & = d \tau - \lambda^a k_a + V^a \kappa_a + \bar{\varepsilon}^+ \Gamma^0 \psi^+ + \bar{\varepsilon}^- \Gamma^0 \psi^- \, , \\
\delta \psi^+ & = d \varepsilon^+ - \frac{1}{4} \lambda^{ab} \Gamma_{ab} \psi^+ + \frac{1}{4} \omega^{ab} \Gamma_{ab} \varepsilon^+ \,  , \\
\delta \psi^- & = d \varepsilon^- - \frac{1}{4} \lambda^{ab} \Gamma_{ab} \psi^- + \frac{1}{4} \omega^{ab} \Gamma_{ab} \varepsilon^- \, ,
\end{split}
\end{equation}
and restricting ourselves to supersymmetry transformations, we are left with
\begin{equation}\label{susytrF11}
\begin{split}
\delta \omega^{ab} & = 0 \, , \\
\delta k^a  & = 0 \, , \\
\delta V^a & = 0 \, , \\
\delta h & = \bar{\varepsilon}^+ \Gamma^0 \psi^+ + \bar{\varepsilon}^- \Gamma^0 \psi^- \,   , \\
\delta \psi^+ & = d \varepsilon^+ + \frac{1}{4} \omega^{ab} \Gamma_{ab} \varepsilon^+ \,  , \\
\delta \psi^- & = d \varepsilon^- + \frac{1}{4} \omega^{ab} \Gamma_{ab} \varepsilon^- \, .
\end{split}
\end{equation}

Concluding, the equations of motion for the action \eqref{CSC11} (flat limit of the equations of motion given in \eqref{eom11}) read as follows:
\begin{equation}\label{eomF11}
\begin{split}
\delta \omega^{ab} & : \quad \alpha_0 \mathcal{R}^{ab} + \alpha_1 \epsilon^{ab} \mathcal{H}= 0 \, , \\
\delta k^a & : \quad  \alpha_1 R^a = 0 \, , \\
\delta V^a & : \quad \alpha_1 \epsilon_{ab} \mathcal{K}^b = 0 \,  , \\
\delta h & : \quad \alpha_1 \mathcal{R}^{ab} = 0  \, , \\
\delta \psi^+ & : \quad \alpha_1 \nabla \psi^+ = 0 \,  , \\
\delta \psi^- & : \quad \alpha_1 \nabla \psi^- = 0 \,  .
\end{split}
\end{equation}
When $\alpha_1 \neq 0$, the eqs. \eqref{eomF11} exactly reduce to the vanishing of the curvature $2$-forms in \eqref{curvadscarrollsuperF11} ($\alpha_1 \neq 0$ is a sufficient condition to recover the vanishing of the curvature $2$-forms \eqref{curvadscarrollsuperF11}, meaning that one could consistently set $\alpha_0=0$, omitting the exotic term, that is the Lorentz Lagrangian, in the CS action \eqref{CSC11}).

Notice that the restriction to the purely bosonic part of the action \eqref{CSC11} yields exactly the three-dimensional CS gravity action invariant under the $D=3$ Carroll algebra \cite{LL, Bacry:1968zf}. The aforesaid CS action involving purely bosonic terms is equivalent, as argued in \cite{Bergshoeff:2017btm}, to the action found in \cite{Bergshoeff:2017btm} if we take the $D=3$ case in the same paper.

\subsection{$(\mathcal{N},0)$ Carroll supergravity theories from the $\ell \rightarrow \infty$ limit}

Taking the flat limit $\ell \rightarrow \infty$ of the (anti)commutation relations \eqref{adscarrollsupern}, we get the following non-trivial ones:
\begin{equation}\label{adscarrollsuperFn}
\begin{split}
& \left[ K_a , J_{bc} \right] = \delta_{ab} K_c - \delta_{ac} K_b \, , \\
& \left[ J_{ab}, P_{c}\right] =\delta_{bc}P_{a}-\delta _{ac}P_{b} \, , \\
& \left[ K_{a}, P_{b}\right] = - \delta_{ab} H \, , \\
& \left[ {T}^{\lambda \mu} , {T}^{\nu \rho} \right] = \delta^{\mu \nu} {T}^{\lambda \rho} - \delta^{\lambda \nu} {T}^{\mu \rho}- \delta^{\mu \rho} {T}^{\lambda \nu} + \delta^{\lambda \rho} {T}^{\mu \nu} \, , \\
& \left[ {T}^{\lambda \mu} , {U}^{\nu \rho} \right] = \delta^{\mu \rho} {U}^{\nu \lambda} - \delta^{\lambda \rho} {U}^{\nu \mu} \, , \\
& \left[ {T}'^{\lambda \mu} , {T}'^{\nu \rho} \right] = \delta^{\mu \nu} {T}'^{\lambda \rho} - \delta^{\lambda \nu} {T}'^{\mu \rho}- \delta^{\mu \rho} {T}'^{\lambda \nu} + \delta^{\lambda \rho} {T}'^{\mu \nu} \, , \\
& \left[ {T}'^{\lambda \mu} , {U}^{\nu \rho} \right] = \delta^{\mu \nu} {U}^{\lambda \rho}- \delta^{\lambda \nu} {U}^{\mu \rho} \, , \\
& \left[ {U}^{\lambda \mu} , {U}^{\nu \rho} \right] = - \delta^{\lambda \nu} {T}^{\mu \rho} - \delta^{\mu \rho} {T}'^{\lambda \nu} \, , \\
& \left[ {T}^{\lambda \mu} , {S}^{\nu \rho} \right] = \delta^{\mu \nu} {S}^{\lambda \rho} - \delta^{\lambda \nu} {S}^{\mu \rho}- \delta^{\mu \rho} {S}^{\lambda \nu} + \delta^{\lambda \rho} {S}^{\mu \nu} \, , \\
& \left[ {T}^{\lambda \mu} , {V}^{\nu \rho} \right] = \delta^{\mu \rho} {V}^{\nu \lambda} - \delta^{\lambda \rho} {V}^{\nu \mu} \, , \\
& \left[ {T}'^{\lambda \mu} , {S}'^{\nu \rho} \right] = \delta^{\mu \nu} {S}'^{\lambda \rho} - \delta^{\lambda \nu} {S}'^{\mu \rho}- \delta^{\mu \rho} {S}'^{\lambda \nu} + \delta^{\lambda \rho} {S}'^{\mu \nu} \, , \\
& \left[ {T}'^{\lambda \mu} , {V}^{\nu \rho} \right] = \delta^{\mu \nu} {V}^{\lambda \rho}- \delta^{\lambda \nu} {V}^{\mu \rho} \, , \\
& \left[ {U}^{\lambda \mu} , {S}^{\nu \rho} \right] = - \delta^{\mu \rho} {V}^{\lambda \nu} + \delta^{\mu \nu} {V}^{\lambda \rho} \, , \\
& \left[ {U}^{\lambda \mu} , {S}'^{\nu \rho} \right] = - \delta^{\lambda \rho} {V}^{\nu \mu} + \delta^{\lambda \nu} {V}^{\rho \mu} \, , \\
& \left[ {U}^{\lambda \mu} , {V}^{\nu \rho} \right] =  - \delta^{\lambda \nu} {S}^{\mu \rho} - \delta^{\mu \rho} {S}'^{\lambda \nu}  \, , \\
& \left[ J_{ab}, Q^{\pm \, \lambda} _\alpha \right] =-\frac{1}{2}\left( \Gamma _{ab} Q^{\pm \, \lambda}  \right)_{\alpha } \, , \\
& \left[ {T}^{\lambda \mu}, {Q}^{\pm \, \nu } _\alpha \right] = \frac{1}{2} \left[\delta^{\mu \nu} \left( {Q}^{+ \, \lambda } _\alpha + {Q}^{- \, \lambda } _\alpha \right) - \delta^{\lambda \nu} \left( {Q}^{+ \, \mu } _\alpha + {Q}^{- \, \mu } _\alpha \right)   \right] \,  , \\
& \left[ {T}'^{\lambda \mu}, {Q}^{\pm \, \nu } _\alpha \right] = \pm \frac{1}{2} \left[\delta^{\mu \nu} \left( {Q}^{+ \, \lambda } _\alpha - {Q}^{- \, \lambda } _\alpha \right) - \delta^{\lambda \nu} \left( {Q}^{+ \, \mu } _\alpha - {Q}^{- \, \mu } _\alpha \right)   \right]  \,  , \\
& \left[ {U}^{\lambda \mu}, {Q}^{\pm \, \nu } _\alpha \right] = \mp \frac{1}{2} \left( \Gamma_0 \right)_{\alpha \beta} \left[ \delta^{\lambda \nu} \left( {Q}^{+ \, \mu } _\beta + {Q}^{- \, \ \mu } _\beta \right) \pm \delta^{\mu \nu} \left( {Q}^{+ \, \lambda } _\beta - {Q}^{- \, \lambda } _\beta \right) \right]  \, , \\
& \lbrace {Q}^{+ \, \lambda} _\alpha , {Q}^{+ \, \mu} _\alpha \rbrace = \left( \Gamma^0 C \right)_{\alpha \beta} \left( \delta^{\lambda \mu}  {H} - {V}^{(\lambda \mu)} \right) + C_{\alpha \beta} {Y}^{[\lambda \mu]} \, , \\
& \lbrace {Q}^{+ \, \lambda} _\alpha , {Q}^{- \, \mu} _\beta \rbrace = C_{\alpha \beta} {Y}'^{[\lambda \mu]} - \left( \Gamma^0 C \right)_{\alpha \beta} {V}^{[\lambda \mu]} \, , \\
& \lbrace {Q}^{- \, \lambda} _\alpha , {Q}^{- \, \mu} _\beta \rbrace = \left( \Gamma^0 C \right)_{\alpha \beta} \left( \delta^{\lambda \mu}  {H} + {V}^{(\lambda \mu)} \right) + C_{\alpha \beta} {Y}^{[\lambda \mu]}  \, .
\end{split}
\end{equation}
The (anti)commutation relations \eqref{adscarrollsuperFn} are those of the $\mathcal{N}=(\mathcal{N},0)$, $D=3$ super-Carroll algebra (with $\mathcal{N}$ even), and one could also obtain it by taking the ultra-relativistic limit of the $\mathfrak{so}(\mathcal{N})$ extension of the $\mathcal{N}=(\mathcal{N},0)$, $D=3$ Poincar\'{e} superalgebra.

Moreover, as $\ell \rightarrow \infty$, the $2$-form curvatures \eqref{curvadscarrollsupern} become
\begin{equation}\label{curvadscarrollsuperFn}
\begin{split}
\mathcal{R}^{ab} & = d \omega^{ab} = R^{ab} \, , \\
\mathcal{K}^a  & = d k^a + \omega^a_{\phantom{a} b} k^b = \mathfrak{K}^a \, , \\
R^a & = d V^a + \omega^a_{\phantom{a}b} V^b \, , \\
\mathcal{H} & = d h + V^a k_a - \frac{1}{2} \bar{\psi}^{+ \, \lambda} \Gamma^0 \psi^{+ \, \lambda} - \frac{1}{2} \bar{\psi}^{- \, \lambda} \Gamma^0 \psi^{- \, \lambda}   = \mathfrak{H} - \frac{1}{2} \bar{\psi}^{+ \, \lambda} \Gamma^0 \psi^{+ \, \lambda} - \frac{1}{2} \bar{\psi}^{- \, \lambda} \Gamma^0 \psi^{- \, \lambda}  \,   , \\
\mathcal{T}^{\lambda \mu} & = d t^{\lambda \mu} + {t^\lambda}_\nu t^{\nu \mu} + {u'^{[\lambda}}_\nu u^{\nu \vert \mu]}  \,   , \\
\mathcal{T}'^{\lambda \mu} & = d t'^{\lambda \mu} + {t'^\lambda}_\nu t'^{\nu \mu} + {u^{[\lambda}}_{ \nu} u'^{\nu \vert \mu]} \,   , \\
\mathcal{U}^{\lambda \mu} & = d u^{\lambda \mu} + {u^\lambda}_\nu t^{\nu \mu} + {t'^\lambda}_\nu u^{\nu \mu} \,   , \\
\mathcal{S}^{\lambda \mu} & = d s^{\lambda \mu} + 2 {t^\lambda}_\nu s^{\nu \mu} + 2 {u'^{[\lambda}}_\nu v^{\nu \vert \mu]} - \frac{1}{2} \bar{\psi}^{+ \, \lambda} \psi^{+ \, \mu} - \frac{1}{2} \bar{\psi}^{- \, \lambda} \psi^{- \, \mu} - \bar{\psi}^{+ \, [ \lambda} \psi^{- \, \mu ]} \, , \\
\mathcal{S}'^{\lambda \mu} & = d s'^{\lambda \mu} + 2 {t'^\lambda}_\nu s'^{\nu \mu} + 2 {u^{[\lambda}}_\nu  v'^{\nu \vert \mu]} - \frac{1}{2} \bar{\psi}^{+ \, \lambda} \psi^{+ \, \mu} - \frac{1}{2} \bar{\psi}^{- \, \lambda} \psi^{- \, \mu} + \bar{\psi}^{+ \, [ \lambda} \psi^{- \, \mu ]} \, , \\
\mathcal{V}^{\lambda \mu} & = d v^{\lambda \mu} + {v^\lambda}_\nu t^{\nu \mu} + {t'^\lambda}_\nu v^{\nu \mu} + {u^\lambda}_\nu s^{\nu \mu} + {s'^\lambda}_\nu u^{\nu \mu} + \frac{1}{2} \bar{\psi}^{+ \,  \lambda} \Gamma^0 \psi^{+ \, \mu} - \frac{1}{2} \bar{\psi}^{- \, \lambda} \Gamma^0 \psi^{- \, \mu} \\
& + \bar{\psi}^{+ \, [ \lambda} \Gamma^0 \psi^{- \, \mu ]} \,  , \\
\nabla \psi^{+ \, \lambda} & = d \psi^{+ \, \lambda} + \frac{1}{4} \omega^{ab} \Gamma_{ab} \psi^{+ \, \lambda} + \frac{1}{2} t^{\lambda \mu} \psi^+_\mu + \frac{1}{2} t^{\lambda \mu} \psi^-_\mu + \frac{1}{2} t'^{\lambda \mu} \psi^+_\mu - \frac{1}{2} t'^{\lambda \mu} \psi^-_\mu \\
&  + u^{(\lambda \mu)} \Gamma_0 \psi^+_\mu + u^{[\lambda \mu]} \Gamma_0 \psi^-_\mu  \,  , \\
\nabla \psi^{- \, \lambda} & = d \psi^{- \, \lambda} + \frac{1}{4} \omega^{ab} \Gamma_{ab} \psi^{- \, \lambda} + \frac{1}{2} t^{\lambda \mu} \psi^+_\mu + \frac{1}{2} t^{\lambda \mu} \psi^-_\mu - \frac{1}{2} t'^{\lambda \mu} \psi^+_\mu + \frac{1}{2} t'^{\lambda \mu} \psi^-_\mu \\
& - u^{(\lambda \mu)} \Gamma_0 \psi^-_\mu - u^{[\lambda \mu]} \Gamma_0 \psi^+_\mu \, .
\end{split}
\end{equation}

Applying the $\ell \rightarrow \infty$ limit to the CS action \eqref{CSACn}, we obtain
\begin{equation}\label{CSCn}
\begin{split}
I^{(\mathcal{N},0)}_{CS} \vert_{\ell \rightarrow \infty} & = \frac{k}{4 \pi} \int_\mathcal{M} \Bigg \lbrace \frac{\alpha_0}{2} \bigg( \omega^a_{\phantom{a} b} R^b_{\phantom{b} a} + 2 {t^\lambda}_\mu {d t^\mu}_\lambda + \frac{4}{3} {t^\lambda}_\mu {t^\mu}_\nu {t^\nu}_\lambda + 2 {t'^\lambda}_\mu {d t'^\mu}_\lambda + \frac{4}{3} {t'^\lambda}_\mu {t'^\mu}_\nu {t'^\nu}_\lambda \\
& + 4 {u^\lambda}_\mu {d u'^\mu}_\lambda - 4 t_{\lambda \mu} {u'^{\lambda}}_{\nu} u^{\nu \mu} - 4 t'_{\lambda \mu} {u^{\lambda}}_{\nu} u'^{\nu \mu} \bigg) + \alpha_1 \bigg[ \epsilon_{ab} R^{ab} h - 2 \epsilon_{ab} \mathfrak{K}^a V^b \\
& - 2 {t^\lambda}_\mu \left( {ds^\mu}_\lambda +  {t^\mu}_\nu {s^\nu}_\lambda \right) - 2 {t'^\lambda}_\mu \left( {ds'^\mu}_\lambda +  {t'^\mu}_\nu {s'^\nu}_\lambda \right) - 4 {u^\lambda}_\mu {d v'^\mu}_\lambda - 2 {u'^\lambda}_\mu {u^\mu}_\nu {s^\nu}_\lambda \\
& - 2 {u^\lambda}_\mu {u'^\mu}_\nu {s'^\nu}_\lambda - 4 {u'^\lambda}_\mu {v^\mu}_\nu {t^\nu}_\lambda - 4 {u^\lambda}_\mu {v'^\mu}_\nu {t'^\nu}_\lambda + 2 \bar{\psi}^{+ \, \lambda} \nabla \psi^{+ \, \lambda} + 2 \bar{\psi}^{- \, \lambda} \nabla \psi^{- \, \lambda} \bigg] \\
& - d \left( \frac{\alpha_1}{2} \epsilon_{ab} \omega^{ab} h - \alpha_1 \epsilon_{ab} k^a V^b + \alpha_1 {t^\lambda}_\mu {s^\mu}_\lambda + \alpha_1 {t'^\lambda}_\mu {s'^\mu}_\lambda + 2 \alpha_1 {u^\lambda}_\mu {v'^\mu}_\lambda \right) \Bigg \rbrace \,  ,
\end{split}
\end{equation}
which is written in terms of the super-Carroll curvatures appearing in \eqref{curvadscarrollsuperFn} (we emphasize that the latter must not be confused with the super-AdS Carroll ones given in \eqref{curvadscarrollsupern}). Note that the exotic term in \eqref{CSCn}, that is the one proportional to $\alpha_0$, is now given by the Lorentz Lagrangian plus additional terms involving the $1$-form fields $t^{\lambda \mu}$, $t'^{\lambda \mu}$, and $u^{\lambda \mu}$. Let us further mention that the action \eqref{CSCn} can also be derived by using the non-vanishing components of the invariant tensor obtained by taking the limit $\ell \rightarrow \infty$ of \eqref{invadscarrollsupern} together with the connection $1$-form for the $\mathcal{N}=(\mathcal{N},0)$ (flat) Carroll superalgebra \eqref{adscarrollsuperFn} in the general expression \eqref{genCS}. In particular, taking the $\ell \rightarrow \infty$ limit of \eqref{invadscarrollsupern} we are left with the non-vanishing components
\begin{equation}
\begin{split}
& \langle J_{ab} J_{cd} \rangle = \alpha_0 \left( \delta_{ad} \delta_{bc} - \delta_{ac} \delta_{bd} \right) \, , \\
& \langle J_{ab} H \rangle = \alpha_1 \epsilon_{ab} \, , \\
& \langle K_a P_b \rangle = - \alpha_1 \epsilon_{ab} \, , \\
& \langle {T}^{\lambda \mu} {T}^{\nu \rho} \rangle = \langle {T}'^{\lambda \mu} {T}'^{\nu \rho} \rangle  =  2 \alpha_0 \left( \delta^{\lambda \rho} \delta^{\nu \mu} - \delta^{\lambda \nu} \delta^{\rho \mu} \right) \,  , \\
& \langle {U}^{\lambda \mu} {U}^{\nu \rho} \rangle = - 2 \alpha_0 \delta^{\lambda \nu} \delta^{\rho \mu} \, , \\
& \langle {T}^{\lambda \mu} {S}^{\nu \rho} \rangle = \langle {T}'^{\lambda \mu} {S}'^{\nu \rho} \rangle  = - 2 \alpha_1 \left( \delta^{\lambda \rho} \delta^{\nu \mu} - \delta^{\lambda \nu} \delta^{\rho \mu} \right) \,  , \\
& \langle {U}^{\lambda \mu} {V}^{\nu \rho} \rangle = 2 \alpha_1 \delta^{\lambda \nu} \delta^{\rho \mu} \, , \\
& \langle Q^{+ \, \lambda}_\alpha Q^{+ \, \mu}_\beta \rangle = \langle Q^{- \, \lambda}_\alpha Q^{- \, \mu}_\beta \rangle = 2 \alpha_1  C_{\alpha \beta} \delta^{\lambda \mu} \, .
\end{split}
\end{equation}

The action \eqref{CSCn} is invariant by construction under the super-Carroll group associated with \eqref{adscarrollsuperFn}; in particular, concerning the $\ell \rightarrow \infty$ limit of the local gauge transformations \eqref{gaugetrn}, we obtain
\begin{equation}\label{gaugetrFn}
\begin{split}
\delta \omega^{ab} & = d \lambda^{ab} \, , \\
\delta k^a  & = d \kappa^a - \lambda^a_{\phantom{a} b} k^b + \omega^a_{\phantom{a} b} \kappa^b \, , \\
\delta V^a & = d \lambda^a - \lambda^a_{\phantom{a}b} V^b + \omega^a_{\phantom{a}b} \lambda^b \, , \\
\delta h & = d \tau - \lambda^a k_a + V^a \kappa_a + \bar{\varepsilon}^{+ \, \lambda} \Gamma^0 \psi^{+ \, \lambda} + \bar{\varepsilon}^{- \, \lambda} \Gamma^0 \psi^{- \, \lambda} \,   , \\
\delta t^{\lambda \mu} & = d \varrho^{\lambda \mu} - 2 {\varrho^{[\lambda}}_\nu t^{\nu \vert \mu]} - 2 {\varphi'^{[\lambda}}_\nu u^{\nu \vert \mu]} \, , \\
\delta t'^{\lambda \mu} & = d \varrho'^{\lambda \mu} - 2 {\varrho'^{[\lambda}}_\nu t'^{\nu \vert \mu]} - 2 {\varphi^{[\lambda}}_{ \nu} u'^{\nu \vert \mu]} \, , \\
\delta u^{\lambda \mu} & = d \varphi^{\lambda \mu} - {\varphi^\lambda}_\nu t^{\nu \mu} + {u^\lambda}_\nu \varrho^{\nu \mu} - {\varrho'^\lambda}_\nu u^{\nu \mu} + {t'^\lambda}_\nu \varphi^{\nu \mu}  \, , \\
\delta s^{\lambda \mu} & = d \vartheta^{\lambda \mu} - 2 {\varrho^{[\lambda}}_\nu s^{\nu \vert \mu]} + 2 {t^{[\lambda}}_\nu \vartheta^{\nu \vert \mu]} - 2 {\varphi'^{[\lambda}}_\nu v^{\nu \vert \mu]} + 2 {u'^{[\lambda}}_\nu \varsigma^{\nu \vert \mu]} + \bar{\varepsilon}^{+ \, [\lambda} \psi^{+ \, \mu]} + \bar{\varepsilon}^{- \, [\lambda} \psi^{- \, \mu]} \\ 
& + \bar{\varepsilon}^{+ \, [ \lambda} \psi^{- \, \mu ]} + \bar{\varepsilon}^{- \, [ \lambda} \psi^{+ \, \mu ]} \, , \\
\delta s'^{\lambda \mu} & = d \vartheta'^{\lambda \mu} - 2 {\varrho'^{[\lambda}}_\nu s'^{\nu \vert \mu]} + 2 {t'^{[\lambda}}_\nu \vartheta'^{\nu \vert \mu]} - 2 {\varphi^{[\lambda}}_\nu  v'^{\nu \vert \mu]} + 2 {u^{[\lambda}}_\nu  \varsigma'^{\nu \vert \mu]} + \bar{\varepsilon}^{+ \, [\lambda} \psi^{+ \, \mu]} + \bar{\varepsilon}^{- \, [\lambda} \psi^{- \, \mu]} \\
& - \bar{\varepsilon}^{+ \, [ \lambda} \psi^{- \, \mu ]} - \bar{\varepsilon}^{- \, [ \lambda} \psi^{+ \, \mu ]} \, , \\
\delta v^{\lambda \mu} & = d \varsigma^{\lambda \mu} - {\varsigma^\lambda}_\nu t^{\nu \mu} + {v^\lambda}_\nu \varrho^{\nu \mu} - {\varrho'^\lambda}_\nu v^{\nu \mu} + {t'^\lambda}_\nu \varsigma^{\nu \mu} - {\varphi^\lambda}_\nu s^{\nu \mu} + {u^\lambda}_\nu \vartheta^{\nu \mu} - {\vartheta'^\lambda}_\nu u^{\nu \mu} + {s'^\lambda}_\nu \varphi^{\nu \mu}  \\
& - \bar{\varepsilon}^{+ \,  \lambda} \Gamma^0 \psi^{+ \, \mu} + \bar{\varepsilon}^{- \, \lambda} \Gamma^0 \psi^{- \, \mu} - \bar{\varepsilon}^{+ \, [ \lambda} \Gamma^0 \psi^{- \, \mu ]} + \bar{\varepsilon}^{- \, [ \lambda} \Gamma^0 \psi^{+ \, \mu ]} \,  , \\
\delta \psi^{+ \, \lambda} & = d \varepsilon^{+ \, \lambda} - \frac{1}{4} \lambda^{ab} \Gamma_{ab} \psi^{+ \, \lambda} + \frac{1}{4} \omega^{ab} \Gamma_{ab} \varepsilon^{+ \, \lambda} - \frac{1}{2} \varrho^{\lambda \mu} \psi^+_\mu + \frac{1}{2} t^{\lambda \mu} \varepsilon^+_\mu - \frac{1}{2} \varrho^{\lambda \mu} \psi^-_\mu + \frac{1}{2} t^{\lambda \mu} \varepsilon^-_\mu - \frac{1}{2} \varrho'^{\lambda \mu} \psi^+_\mu \\
& + \frac{1}{2} t'^{\lambda \mu} \varepsilon^+_\mu + \frac{1}{2} \varrho'^{\lambda \mu} \psi^-_\mu - \frac{1}{2} t'^{\lambda \mu} \varepsilon^-_\mu - \varphi^{(\lambda \mu)} \Gamma_0 \psi^+_\mu  + u^{(\lambda \mu)} \Gamma_0 \varepsilon^+_\mu - \varphi^{[\lambda \mu]} \Gamma_0 \psi^-_\mu  + u^{[\lambda \mu]} \Gamma_0 \varepsilon^-_\mu  \,  , \\
\delta \psi^{- \, \lambda} & = d \varepsilon^{- \, \lambda} - \frac{1}{4} \lambda^{ab} \Gamma_{ab} \psi^{- \, \lambda} + \frac{1}{4} \omega^{ab} \Gamma_{ab} \varepsilon^{- \, \lambda} - \frac{1}{2} \varrho^{\lambda \mu} \psi^+_\mu + \frac{1}{2} t^{\lambda \mu} \varepsilon^+_\mu - \frac{1}{2} \varrho^{\lambda \mu} \psi^-_\mu + \frac{1}{2} t^{\lambda \mu} \varepsilon^-_\mu + \frac{1}{2} \varrho'^{\lambda \mu} \psi^+_\mu \\
& - \frac{1}{2} t'^{\lambda \mu} \varepsilon^+_\mu - \frac{1}{2} \varrho'^{\lambda \mu} \psi^-_\mu + \frac{1}{2} t'^{\lambda \mu} \varepsilon^-_\mu + \varphi^{(\lambda \mu)} \Gamma_0 \psi^-_\mu - u^{(\lambda \mu)} \Gamma_0 \varepsilon^-_\mu + \varphi^{[\lambda \mu]} \Gamma_0 \psi^+_\mu - u^{[\lambda \mu]} \Gamma_0 \varepsilon^+_\mu \, .
\end{split}
\end{equation}
Then, restricting ourselves to the supersymmetry transformations in the limit $\ell \rightarrow \infty$, we find
\begin{equation}\label{susytrFn}
\begin{split}
\delta \omega^{ab} & = 0 \, , \\
\delta k^a  & = 0 \, , \\
\delta V^a & = 0 \, , \\
\delta h & = \bar{\varepsilon}^{+ \, \lambda} \Gamma^0 \psi^{+ \, \lambda} + \bar{\varepsilon}^{- \, \lambda} \Gamma^0 \psi^{- \, \lambda} \,   , \\
\delta t^{\lambda \mu} & = 0 \, , \\
\delta t'^{\lambda \mu} & = 0 \, , \\
\delta u^{\lambda \mu} & = 0  \, , \\
\delta s^{\lambda \mu} & = \bar{\varepsilon}^{+ \, [\lambda} \psi^{+ \, \mu]} + \bar{\varepsilon}^{- \, [\lambda} \psi^{- \, \mu]} + \bar{\varepsilon}^{+ \, [ \lambda} \psi^{- \, \mu ]} + \bar{\varepsilon}^{- \, [ \lambda} \psi^{+ \, \mu ]} \, , \\
\delta s'^{\lambda \mu} & = \bar{\varepsilon}^{+ \, [\lambda} \psi^{+ \, \mu]} + \bar{\varepsilon}^{- \, [\lambda} \psi^{- \, \mu]} - \bar{\varepsilon}^{+ \, [ \lambda} \psi^{- \, \mu ]} - \bar{\varepsilon}^{- \, [ \lambda} \psi^{+ \, \mu ]} \, , \\
\delta v^{\lambda \mu} & = - \bar{\varepsilon}^{+ \,  \lambda} \Gamma^0 \psi^{+ \, \mu} + \bar{\varepsilon}^{- \, \lambda} \Gamma^0 \psi^{- \, \mu} - \bar{\varepsilon}^{+ \, [ \lambda} \Gamma^0 \psi^{- \, \mu ]} + \bar{\varepsilon}^{- \, [ \lambda} \Gamma^0 \psi^{+ \, \mu ]} \,  , \\
\delta \psi^{+ \, \lambda} & = d \varepsilon^{+ \, \lambda} + \frac{1}{4} \omega^{ab} \Gamma_{ab} \varepsilon^{+ \, \lambda} + \frac{1}{2} t^{\lambda \mu} \varepsilon^+_\mu + \frac{1}{2} t^{\lambda \mu} \varepsilon^-_\mu + \frac{1}{2} t'^{\lambda \mu} \varepsilon^+_\mu - \frac{1}{2} t'^{\lambda \mu} \varepsilon^-_\mu + u^{(\lambda \mu)} \Gamma_0 \varepsilon^+_\mu + u^{[\lambda \mu]} \Gamma_0 \varepsilon^-_\mu  \,  , \\
\delta \psi^{- \, \lambda} & = d \varepsilon^{- \, \lambda} + \frac{1}{4} \omega^{ab} \Gamma_{ab} \varepsilon^{- \, \lambda} + \frac{1}{2} t^{\lambda \mu} \varepsilon^+_\mu + \frac{1}{2} t^{\lambda \mu} \varepsilon^-_\mu - \frac{1}{2} t'^{\lambda \mu} \varepsilon^+_\mu + \frac{1}{2} t'^{\lambda \mu} \varepsilon^-_\mu - u^{(\lambda \mu)} \Gamma_0 \varepsilon^-_\mu - u^{[\lambda \mu]} \Gamma_0 \varepsilon^+_\mu \, .
\end{split}
\end{equation}

Finally, we find that the equations of motion for the action \eqref{CSCn} (flat limit of the equations of motion given in \eqref{eomn}) read
\begin{equation}\label{eomFn}
\begin{split}
\delta \omega^{ab} & : \quad \alpha_0 \mathcal{R}^{ab} + \alpha_1 \epsilon^{ab} \mathcal{H}= 0 \, , \\
\delta k^a & : \quad  \alpha_1 R^a = 0 \, , \\
\delta V^a & : \quad  \alpha_1 \epsilon_{ab} \mathcal{K}^b = 0 \,  , \\
\delta h & : \quad \alpha_1 \mathcal{R}^{ab} = 0  \, , \\
\delta t^{\lambda \mu} & : \quad \alpha_0 \mathcal{T}^{\lambda \mu} + \alpha_1  \mathcal{S}^{\lambda \mu}  = 0 \,  , \\
\delta t'^{\lambda \mu} & : \quad \alpha_0 \mathcal{T}'^{\lambda \mu} + \alpha_1 \mathcal{S}'^{\lambda \mu}  = 0 \,  , \\
\delta u^{\lambda \mu} & : \quad  - \alpha_0 \mathcal{U}^{\lambda \mu} + \alpha_1 \mathcal{V}^{\lambda \mu} = 0    \,  , \\
\delta s^{\lambda \mu} & : \quad \alpha_1 \mathcal{T}^{\lambda \mu} = 0 \, , \\
\delta s'^{\lambda \mu} & : \quad \alpha_1 \mathcal{T}'^{\lambda \mu} = 0 \, , \\
\delta v^{\lambda \mu} & : \quad \alpha_1 \mathcal{U}^{\lambda \mu} = 0 \, , \\
\delta \psi^{+ \, \lambda} & : \quad \alpha_1 \nabla \psi^{+ \, \lambda} = 0 \,  , \\
\delta \psi^{- \, \lambda} & : \quad \alpha_1 \nabla \psi^{- \, \lambda} = 0 \,  .
\end{split}
\end{equation}
We can see that when $\alpha_1 \neq 0$, the eqs. \eqref{eomFn} reduce to the vanishing of the curvature $2$-forms given in \eqref{curvadscarrollsuperFn} ($\alpha_1 \neq 0$ is a sufficient condition to recover the vanishing of the curvature $2$-forms in \eqref{curvadscarrollsuperFn}; the coefficient $\alpha_0$ can be consistently set to zero, making the exotic term disappear from the action \eqref{CSCn}.

Let us observe that, restricting ourselves to the purely bosonic theory, we end up with the $\mathcal{N}=(\mathcal{N},0)$ Carroll gravity theories (with $\mathcal{N}$ even) in three dimensions, invariant under the $\mathcal{N}=(\mathcal{N},0)$ Carroll algebra. At the purely bosonic level, the fields $t^{\lambda \mu}$, $t'^{\lambda \mu}$, $u^{\lambda \mu}$, $s^{\lambda \mu}$, $s'^{\lambda \mu}$, and $v^{\lambda \mu}$, and the corresponding terms in the action, can also be consistently discarded by performing an IW contraction.

On the other hand, considering the special case $\mathcal{N}=(2,0)$, that is $x=1$, after some algebraic calculations, we can prove that the $(2,0)$ theory in the flat limit previously discussed in this section is exactly reproduced. 

\subsection{$(p,q)$ Carroll supergravity theories from the $\ell \rightarrow \infty$ limit}

Applying the flat limit $\ell \rightarrow \infty$ to the (anti)commutation relations given in \eqref{adscarrollsuperpq}, we get the following non-vanishing ones:
\begin{equation}\label{adscarrollsuperFpq}
\begin{split}
& \left[ K_a , J_{bc} \right] = \delta_{ab} K_c - \delta_{ac} K_b \, , \\
& \left[ J_{ab}, P_{c}\right] =\delta_{bc}P_{a}-\delta _{ac}P_{b} \, , \\
& \left[ K_{a}, P_{b}\right] = - \delta_{ab} H \, , \\
& \left[ {T}^{ij} , {T}^{kl} \right] = \delta^{jk} {T}^{il} - \delta^{ik} {T}^{jl}- \delta^{jl} {T}^{ik} + \delta^{il} {T}^{jk} \, , \\
& \left[ {T}^{IJ} , {T}^{KL} \right] = \delta^{JK} {T}^{IL} - \delta^{IK} {T}^{JL}- \delta^{JL} {T}^{IK} + \delta^{IL} {T}^{JK} \, , \\
& \left[ {T}^{ij} , {S}^{kl} \right] = \delta^{jk} {S}^{il} - \delta^{ik} {S}^{jl}- \delta^{jl} {S}^{ik} + \delta^{il} {S}^{jk} \, , \\
& \left[ {T}^{IJ} , {S}^{KL} \right] = \delta^{JK} {S}^{IL} - \delta^{IK} {S}^{JL}- \delta^{JL} {S}^{IK} + \delta^{IL} {S}^{JK} \, , \\
& \left[ J_{ab},Q^i_{\alpha }\right] =-\frac{1}{2}\left( \Gamma _{ab} Q^i \right)_{\alpha } \, , \\
& \left[ J_{ab},Q^I_{\alpha }\right] =-\frac{1}{2}\left( \Gamma _{ab} Q^I \right)_{\alpha } \, , \\
& \left[ {T}^{ij}, {Q}^k_{\alpha }\right] = \delta^{jk} {Q}^i_\alpha - \delta^{ik} {Q}^j_\alpha \,  , \\
& \left[ {T}^{IJ}, {Q}^K_{\alpha }\right] = \delta^{JK} {Q}^I_\alpha - \delta^{IK} {Q}^J_\alpha \,  , \\
& \left\{ Q^i_{\alpha }, Q^j_{\beta }\right\} = \delta^{ij}  \left( \Gamma ^{0}C\right) _{\alpha \beta } H  + C_{\alpha \beta} S^{ij} \, , \\
& \left\{ Q^I_{\alpha }, Q^J_{\beta }\right\} = \delta^{ij} \left( \Gamma ^{0}C\right) _{\alpha \beta } H  - C_{\alpha \beta} S^{IJ}  \, .
\end{split}
\end{equation}
These are the (anti)commutation relations of the $\mathcal{N}=(p,q)$, $D=3$ super-Carroll algebra{,} and we could also have obtained {the latter} by applying the Carroll contraction to the semi-direct extension of the $\mathfrak{so}(p) \oplus \mathfrak{so}(q)$ automorphism algebra by the $\mathcal{N}=(p,q)$, $D=3$ Poincar\'{e} superalgebra (see Ref. \cite{Howe:1995zm}).

Then, as $\ell \rightarrow \infty$, the $2$-form curvatures \eqref{curvadscarrollsuperpq} become
\begin{equation}\label{curvadscarrollsuperFpq}
\begin{split}
\mathcal{R}^{ab} & = d \omega^{ab} = R^{ab} \, , \\
\mathcal{K}^a  & = d k^a + \omega^a_{\phantom{a} b} k^b  = \mathfrak{K}^a  \, , \\
R^a & = d V^a + \omega^a_{\phantom{a}b} V^b \, , \\
\mathcal{H} & = d h + V^a k_a - \frac{1}{2} \bar{\psi}^i \Gamma^0 \psi^i - \frac{1}{2} \bar{\psi}^I \Gamma^0 \psi^I   = \mathfrak{H} - \frac{1}{2} \bar{\psi}^i \Gamma^0 \psi^i - \frac{1}{2} \bar{\psi}^I \Gamma^0 \psi^I  \,   , \\
\mathcal{T}^{ij} & = d t^{ij} + {t^i}_k t^{kj} \, , \\
\mathcal{T}^{IJ} & = d t^{IJ} + {t^I}_K t^{KJ} \, ,\\
\mathcal{S}^{ij} & = d s^{ij} + 2 {t^i}_k s^{kj} - \bar{\psi}^i \psi^j \, , \\
\mathcal{S}^{IJ} & = d s^{IJ} + 2 {t^I}_K s^{KJ} + \bar{\psi}^I \psi^J \, ,\\
\nabla \psi^i & = d \psi^i + \frac{1}{4} \omega^{ab} \Gamma_{ab} \psi^i + t^{ij} \psi_j \,  , \\
\nabla \psi^I & = d \psi^I + \frac{1}{4} \omega^{ab} \Gamma_{ab} \psi^I + t^{IJ} \psi_J \,  ,
\end{split}
\end{equation}
and by applying the $\ell \rightarrow \infty$ limit to the CS action \eqref{CSACpq}, we get
\begin{equation}\label{CSCpq}
\begin{split}
I^{(p,q)}_{CS} \vert_{\ell \rightarrow \infty} & = \frac{k}{4 \pi} \int_\mathcal{M} \Bigg \lbrace \frac{\alpha_0}{2} \left( \omega^a_{\phantom{a} b} R^b_{\phantom{b} a} + 2 {t^i}_j {d t^j}_i + \frac{4}{3} {t^i}_j {t^j}_k {t^k}_i  + 2 {t^I}_J {d t^J}_I + \frac{4}{3} {t^I}_J {t^J}_K {t^K}_I   \right) \\
& + \alpha_1 \bigg[ \epsilon_{ab} R^{ab} h - 2 \epsilon_{ab} \mathfrak{K}^a V^b  - 2 {t^i}_j \left( {ds^j}_i +  {t^j}_k {s^k}_i \right) + 2 {t^I}_J \left( {ds^J}_I +  {t^J}_K {s^K}_I \right) \\
& + 2 \bar{\psi}^i \nabla \psi^i + 2 \bar{\psi}^I \nabla \psi^I \bigg] - d \left( \frac{\alpha_1}{2} \epsilon_{ab} \omega^{ab} h - \alpha_1 \epsilon_{ab} k^a V^b + \alpha_1 {t^i}_j {s^j}_i - \alpha_1 {t^I}_J {s^J}_I \right) \Bigg \rbrace \,  ,
\end{split}
\end{equation}
which is written in terms of the super-Carroll curvatures appearing in \eqref{curvadscarrollsuperFpq} (let us stress that the latter must not be confused with the super-AdS Carroll ones given in \eqref{curvadscarrollsuperpq}). We can see that the exotic term appearing in \eqref{CSCpq}, namely the one proportional to $\alpha_0$, now is given by the Lorentz Lagrangian plus additional terms involving the $1$-form fields $t^{ij}$ and $t^{IJ}$. 

Notice that the action \eqref{CSCpq} can also be derived by using the non-vanishing components of the invariant tensor obtained by taking the limit $\ell \rightarrow \infty$ of \eqref{invadscarrollsuperpq} together with the connection $1$-form for the $\mathcal{N}=(p,q)$ (flat) Carroll superalgebra \eqref{adscarrollsuperFpq} in the general expression \eqref{genCS}. Specifically, the limit $\ell \rightarrow \infty$ of \eqref{invadscarrollsuperpq} gives us the following non-vanishing components:
\begin{equation}
\begin{split}
& \langle J_{ab} J_{cd} \rangle = \alpha_0 \left( \delta_{ad} \delta_{bc} - \delta_{ac} \delta_{bd} \right) \, , \\
& \langle J_{ab} H \rangle = \alpha_1 \epsilon_{ab} \, , \\
& \langle K_a P_b \rangle = - \alpha_1 \epsilon_{ab} \, , \\
& \langle {T}^{ij} {T}^{kl} \rangle =  2 \alpha_0 \left( \delta^{il} \delta^{kj} - \delta^{ik} \delta^{lj} \right) \,  , \\
& \langle {T}^{IJ} {T}^{KL} \rangle =  2 \alpha_0 \left( \delta^{IL} \delta^{KJ} - \delta^{IK} \delta^{LJ} \right) \,  , \\
& \langle {T}^{ij} {S}^{kl} \rangle = - 2 \alpha_1 \left( \delta^{il} \delta^{kj} - \delta^{ik} \delta^{lj} \right) \,  , \\
& \langle {T}^{IJ} {S}^{KL} \rangle =  2 \alpha_1 \left( \delta^{IL} \delta^{KJ} - \delta^{IK} \delta^{LJ} \right) \,  , \\
& \langle Q^i_\alpha Q^j_\beta \rangle = 2 \alpha_1  C_{\alpha \beta} \delta^{ij} \, , \\
& \langle Q^I_\alpha Q^J_\beta \rangle = 2 \alpha_1  C_{\alpha \beta} \delta^{IJ} \, .
\end{split}
\end{equation}

The CS action \eqref{CSCpq} is invariant by construction under the super-Carroll group associated with \eqref{adscarrollsuperFpq}, and, in particular, concerning the $\ell \rightarrow \infty$ limit of the local gauge transformations \eqref{gaugetrpq}, we get
\begin{equation}\label{gaugetrFpq}
\begin{split}
\delta \omega^{ab} & = d \lambda^{ab} \, , \\
\delta k^a  & = d \kappa^a - \lambda^a_{\phantom{a} b} k^b + \omega^a_{\phantom{a} b} \kappa^b \, , \\
\delta V^a & = d \lambda^a - \lambda^a_{\phantom{a}b} V^b + \omega^a_{\phantom{a}b} \lambda^b \, , \\
\delta h & = d \tau - \lambda^a k_a + V^a \kappa_a + \bar{\varepsilon}^i \Gamma^0 \psi^i + \bar{\varepsilon}^I \Gamma^0 \psi^I \,   , \\
\delta t^{ij} & = d \varphi^{ij} - 2 {\varphi^{[i}}_k t^{k \vert j]} \, , \\
\delta t^{IJ} & = d \varphi^{IJ} - 2 {\varphi^{[I}}_K t^{K \vert J]} \, ,\\
\delta s^{ij} & = d \varsigma^{ij} - 2 {\varphi^{[i}}_k s^{k \vert j]} + 2 {t^{[i}}_k \varsigma^{k \vert j]} + 2 \bar{\varepsilon}^{[i} \psi^{j]} \, , \\
\delta s^{IJ} & = d \varsigma^{IJ} + 2 {\varphi^{[I}}_K s^{K \vert J]} + 2 {t^{[I}}_K \varsigma^{K \vert J]} - 2 \bar{\varepsilon}^{[I} \psi^{J]} \, , \\
\delta \psi^i & = d \varepsilon^+ - \frac{1}{4} \lambda^{ab} \Gamma_{ab} \psi^i + \frac{1}{4} \omega^{ab} \Gamma_{ab} \varepsilon^i - \varphi^{ij} \psi_j + t^{ij} \varepsilon_j \,  , \\
\delta \psi^I & = d \varepsilon^- - \frac{1}{4} \lambda^{ab} \Gamma_{ab} \psi^I + \frac{1}{4} \omega^{ab} \Gamma_{ab} \varepsilon^I - \varphi^{IJ} \psi_J + t^{IJ} \varepsilon_J \, .
\end{split}
\end{equation}
Thus, restricting ourselves to the supersymmetry transformations in the limit $\ell \rightarrow \infty$, we are left with 
\begin{equation}\label{susytrFpq}
\begin{split}
\delta \omega^{ab} & =0 \, , \\
\delta k^a  & = 0 \, , \\
\delta V^a & = 0 \, , \\
\delta h & = \bar{\varepsilon}^i \Gamma^0 \psi^i + \bar{\varepsilon}^I \Gamma^0 \psi^I \,   , \\
\delta t^{ij} & = 0 \, , \\
\delta t^{IJ} & = 0 \, ,\\
\delta s^{ij} & = 2 \bar{\varepsilon}^{[i} \psi^{j]} \, , \\
\delta s^{IJ} & = - 2 \bar{\varepsilon}^{[I} \psi^{J]} \, ,\\
\delta \psi^i & = d \varepsilon^+ + \frac{1}{4} \omega^{ab} \Gamma_{ab} \varepsilon^i + t^{ij} \varepsilon_j \,  , \\
\delta \psi^I & = d \varepsilon^- + \frac{1}{4} \omega^{ab} \Gamma_{ab} \varepsilon^I + t^{IJ} \varepsilon_J \, .
\end{split}
\end{equation}

Concluding, the equations of motion for the action \eqref{CSCpq} (flat limit of the equations of motion given in \eqref{eompq}) read as follows:
\begin{equation}\label{eomFpq}
\begin{split}
\delta \omega^{ab} & : \quad \alpha_0 \mathcal{R}^{ab} + \alpha_1 \epsilon^{ab} \mathcal{H}= 0 \, , \\
\delta k^a & : \quad  \alpha_1 R^a = 0 \, , \\
\delta V^a & : \quad 2 \alpha_1 \epsilon_{ab} \mathcal{K}^b = 0 \,  , \\
\delta h & : \quad \alpha_1 \mathcal{R}^{ab} = 0  \, , \\
\delta t^{ij} & : \quad - \alpha_0 \mathcal{T}^{ij}  + \alpha_1 \mathcal{S}^{ij} = 0 \, , \\
\delta t^{IJ} & : \quad - \alpha_0 \mathcal{T}^{IJ}  - \alpha_1 \mathcal{S}^{IJ} = 0 \, , \\
\delta s^{ij} & : \quad \alpha_1 \mathcal{T}^{ij} = 0 \, , \\
\delta s^{IJ} & : \quad \alpha_1 \mathcal{T}^{IJ} = 0 \, , \\
\delta \psi^i & : \quad \alpha_1 \nabla \psi^i = 0 \,  , \\
\delta \psi^I & : \quad \alpha_1 \nabla \psi^I = 0 \, .
\end{split}
\end{equation}
When $\alpha_1 \neq 0$, the eqs. \eqref{eomFpq} exactly reduce to the vanishing of the curvature $2$-forms given in \eqref{curvadscarrollsuperFpq} ($\alpha_1 \neq 0$ is a sufficient condition to recover the vanishing of the curvature $2$-forms \eqref{curvadscarrollsuperFpq}, and the coefficient $\alpha_0$ can also be consistently set to zero, making the exotic term disappear from the action \eqref{CSCpq}).

Restricting ourselves to the purely bosonic theory, we end up with the $\mathcal{N}=(p,q)$ Carroll gravity theories {in} three dimensions, invariant under the $\mathcal{N}=(p,q)$ Carroll algebra. At the purely bosonic level, the fields $t^{ij}$, $t^{IJ}$, $s^{ij}$, and $s^{IJ}$, and the corresponding terms in the action, can also be consistently discarded by performing an IW contraction.

On the other hand, let us finally mention that, if we now consider the particular case $p=q=1$, we exactly reproduce the results previously obtained in this section for the $(1,1)$ theory in the flat limit.

All the studies of the flat limit presented in this section represent a new development and generalization of the previous works concerning Carroll superalgebras in three dimensions, in particular in the context of three-dimensional CS supergravity theories.

\section{Conclusions}\label{conclusions}

Motivated by the recent development of applications of Carroll symmetries (in particular, by their prominent role in the context of holography), and by the fact that, nevertheless, the study of their supersymmetric extensions in the context of supergravity models still remains poorly explored, in this paper we have presented, in a systematic fashion, the ultra-relativistic $\mathcal{N}$-extended AdS CS supergravity theories in three ($2+1$) spacetime dimensions, which are invariant under $\mathcal{N}$-extended AdS Carroll superalgebras, extending the results recently presented in \cite{Ravera:2019ize} (where the construction of the three-dimensional $\mathcal{N}=1$ CS supergravity theory invariant under the so-called AdS Carroll superalgebra, ultra-relativistic contraction of the $\mathcal{N}=1$ AdS superalgebra \cite{Bergshoeff:2015wma}, together with the study of its flat limit, has been presented for the first time). In particular, we have applied the method introduced in \cite{Concha:2016zdb} with the improvements of \cite{Ravera:2019ize} to construct the aforesaid ultra-relativistic $\mathcal{N}$-extended AdS CS supergravity theories.

We have first considered the $(2,0)$ and $(1,1)$ cases, and subsequently generalized our analysis to $\mathcal{N}=(\mathcal{N},0)$, with $\mathcal{N}$ even integer, and to $\mathcal{N}=(p,q)$, that is $\mathcal{N}=p+q$, with {$p,q>0$}.
The $\mathcal{N}$-extended AdS Carroll superalgebras have been obtained through the Carrollian (i.e., ultra-relativistic) contraction applied to an $so(2)$ extension of $\mathfrak{osp}(2|2)\otimes \mathfrak{sp}(2)$, to $\mathfrak{osp}(2|1)\otimes \mathfrak{osp}(2,1)$, to an $\mathfrak{so}(\mathcal{N})$ extension of $\mathfrak{osp}(2|\mathcal{N})\otimes \mathfrak{sp}(2)$, and to the direct sum of an $\mathfrak{so}(p) \oplus \mathfrak{so}(q)$ algebra and $\mathfrak{osp}(2|p)\otimes \mathfrak{osp}(2,q)$, respectively. 

{An} $\mathcal{N}=(2,0)$ AdS Carroll superalgebra in three dimensions was previously introduced in \cite{Bergshoeff:2015wma}. Nevertheless, the latter does not allow for a non-degenerate invariant tensor, meaning that one cannot construct a well-defined CS action based on this superalgebra. To overcome this point, we have considered an $so(2)$ extension of $\mathfrak{osp}(2|2)\otimes \mathfrak{sp}(2)$ and performed the ultra-relativistic contraction on it, ending up with a new $\mathcal{N}=(2,0)$ AdS Carroll superalgebra endowed with a non-degenerate invariant tensor. This has allowed us to develop the three-dimensional CS supergravity action invariant under this $\mathcal{N}=(2,0)$ AdS Carroll superalgebra. We have called this action the $(2,0)$ AdS Carroll CS supergravity action. We have done an analogous analysis in the $(1,1)$ case, and subsequently generalized our study to $\mathcal{N}=(\mathcal{N},0)$ (with $\mathcal{N}$ even) and to $\mathcal{N}=(p,q)$ (with {$p,q>0$}).
In particular, after having introduced the ultra-relativistic superalgebras, we have constructed the respective CS supergravity theories in three-dimensions by exploiting the non-vanishing components of the corresponding invariant tensor.
The aforementioned actions are all based on a non-degenerate, invariant bilinear form (i.e., an invariant metric), and each of them is characterized by two coupling constants and involve an exotic contribution.
The results presented in this paper were also open problems suggested in Ref. \cite{Bergshoeff:2015wma}, and they represent the $\mathcal{N}$-extended generalization of \cite{Ravera:2019ize}. Interestingly, one can observe that the CS formulation in the $\mathcal{N}$-extended cases $\mathcal{N}=(\mathcal{N},0)$ and $\mathcal{N}=(p,q)$ requires the presence of $\mathfrak{so}(\mathcal{N})$ and $\mathfrak{so}(p) \oplus \mathfrak{sp}(q)$ generators, respectively, also at the ultra-relativistic level, that is in the Carroll limit; thus, what happens at the relativistic level for three-dimensional $\mathcal{N}$-extended CS Poincar\'{e} and AdS supergravity theories (see \cite{Howe:1995zm}), that is the need to introduce the aforementioned extra generators (together with their dual $1$-form fields) in the theory in order to obtain a non-degenerate invariant tensor, has repercussions also on (and still holds at) the ultra-relativistic level.

We have also analyzed the flat limit $\ell \rightarrow \infty$ of the aforementioned models, in which we have recovered the ultra-relativistic $\mathcal{N}$-extended (flat) CS supergravity theories invariant under $\mathcal{N}$-extended super-Carroll algebras. The flat limit has been applied at the level of the superalgebras, CS actions, supersymmetry transformation laws, and field equations. Also all the studies of the flat limit presented in Section \ref{flatlimit} represent a new development and generalization of the previous works presented in the literature concerning Carroll (super)algebras in three dimensions, in particular in the context of three-dimensional CS (super)gravity theories.

The recently discovered relations among the Carrollian world and flat holography suggest that this work might represents a starting point to go further in the analysis of supersymmetry invariance of flat supergravity in the presence of a non-trivial boundary, along the lines of \cite{Concha:2018ywv}. Besides, now, having well-defined three-dimensional CS (super)gravity theories respectively invariant under the $\mathcal{N}$-extended AdS-Carroll and Carroll (super)algebras, it would be intriguing to go beyond and study the asymptotic symmetry of these models, following, for instance, the prescription given in Ref. \cite{Concha:2018zeb}. It would also be interesting to further extend our analysis to more general amount of supersymmetry, involving also odd $\mathcal{N}$ cases, and to higher-dimensional models (recently, a study exploring the Carroll limit corresponding to M2- as well as M3-branes propagating
over $D=11$ supergravity backgrounds in M-theory has been presented \cite{Roychowdhury:2019aoi}), where Carrollian (super)gravity theories still remain poorly explored. Finally, all these ultra-relativistic theories constructed \`{a} la CS could have some applications in the context of Carrollian fluids (and their relations with flat holography, see Refs. \cite{Ciambelli:2018xat, Ciambelli:2018wre, Ciambelli:2018ojf, Campoleoni:2018ltl}).

\section{Acknowledgments}

L.R. is grateful to L. Andrianopoli and B.L. Cerchiai for useful discussions and support.



\begin{thebibliography}{}

\bibitem{LL}
  J.-M. Levy-Leblond, 
  ``Une nouvelle limite non-relativiste du groupe de Poincar\'{e},'' 
  Annales de l'Institut Henri Poincar\'{e} (A) Physique th\'{e}orique \textbf{3} (1965) no. 1, 1-12. \\
  http://eudml.org/doc/75509

\bibitem{Bacry:1968zf}
  H.~Bacry and J.~Levy-Leblond,
  ``Possible kinematics,''
  J.\ Math.\ Phys.\  {\bf 9} (1968) 1605.

\bibitem{Gibbons:2002tv}
  G.~Gibbons, K.~Hashimoto and P.~Yi,
  ``Tachyon condensates, Carrollian contraction of Lorentz group, and fundamental strings,''
  JHEP {\bf 0209} (2002) 061
  [hep-th/0209034].

\bibitem{Hofman:2014loa}
  D.~M.~Hofman and B.~Rollier,
  ``Warped Conformal Field Theory as Lower Spin Gravity,''
  Nucl.\ Phys.\ B {\bf 897} (2015) 1
  [arXiv:1411.0672 [hep-th]].

\bibitem{Bagchi:2013bga}
  A.~Bagchi,
  ``Tensionless Strings and Galilean Conformal Algebra,''
  JHEP {\bf 1305} (2013) 141
  [arXiv:1303.0291 [hep-th]].
  
\bibitem{Bagchi:2015nca}
  A.~Bagchi, S.~Chakrabortty and P.~Parekh,
  ``Tensionless Strings from Worldsheet Symmetries,''
  JHEP {\bf 1601} (2016) 158
  [arXiv:1507.04361 [hep-th]].
  
\bibitem{Bagchi:2016yyf}
  A.~Bagchi, S.~Chakrabortty and P.~Parekh,
  ``Tensionless Superstrings: View from the Worldsheet,''
  JHEP {\bf 1610} (2016) 113
  [arXiv:1606.09628 [hep-th]].
  
\bibitem{Bagchi:2017cte}
  A.~Bagchi, A.~Banerjee, S.~Chakrabortty and P.~Parekh,
  ``Inhomogeneous Tensionless Superstrings,''
  JHEP {\bf 1802} (2018) 065
  [arXiv:1710.03482 [hep-th]].
  
\bibitem{Bagchi:2018wsn}
  A.~Bagchi, A.~Banerjee, S.~Chakrabortty and P.~Parekh,
  ``Exotic Origins of Tensionless Superstrings,''
  arXiv:1811.10877 [hep-th].        

\bibitem{Roychowdhury:2019aoi}
  D.~Roychowdhury,
  ``Carroll membranes,''
  JHEP {\bf 1910} (2019) 258
  [arXiv:1908.07280 [hep-th]]. 

\bibitem{Hartong:2015xda}
  J.~Hartong,
  ``Gauging the Carroll Algebra and Ultra-Relativistic Gravity,''
  JHEP {\bf 1508} (2015) 069
  [arXiv:1505.05011 [hep-th]].

\bibitem{Bergshoeff:2016soe}
  E.~Bergshoeff, D.~Grumiller, S.~Prohazka and J.~Rosseel,
  ``Three-dimensional Spin-3 Theories Based on General Kinematical Algebras,'' 
  JHEP {\bf 1701} (2017) 114 
  [arXiv:1612.02277 [hep-th]].   
  
\bibitem{Bergshoeff:2017btm}
  E.~Bergshoeff, J.~Gomis, B.~Rollier, J.~Rosseel and T.~ter Veldhuis,
  ``Carroll versus Galilei Gravity,''
  JHEP {\bf 1703} (2017) 165
  [arXiv:1701.06156 [hep-th]].     

\bibitem{Bergshoeff:2015wma}
  E.~Bergshoeff, J.~Gomis and L.~Parra,
  ``The Symmetries of the Carroll Superparticle,''
  J.\ Phys.\ A {\bf 49} (2016) no.18,  185402
  [arXiv:1503.06083 [hep-th]].

\bibitem{Matulich:2019cdo} 
  J.~Matulich, S.~Prohazka and J.~Salzer,
  ``Limits of three-dimensional gravity and metric kinematical Lie algebras in any dimension,''
  arXiv:1903.09165 [hep-th].

\bibitem{Figueroa-OFarrill:2018ilb}
  J.~Figueroa-O'Farrill and S.~Prohazka,
  ``Spatially isotropic homogeneous spacetimes,''
  JHEP {\bf 1901} (2019) 229
  [arXiv:1809.01224 [hep-th]]. 

\bibitem{Figueroa-OFarrill:2019sex}
  J.~Figueroa-O'Farrill, R.~Grassie and S.~Prohazka,
  ``Geometry and BMS Lie algebras of spatially isotropic homogeneous spacetimes,''
  arXiv:1905.00034 [hep-th].  

\bibitem{Bagchi:2009my}
  A.~Bagchi and R.~Gopakumar,
  ``Galilean Conformal Algebras and AdS/CFT,''
  JHEP {\bf 0907} (2009) 037
  [arXiv:0902.1385 [hep-th]].  
  
\bibitem{Christensen:2013lma}
  M.~H.~Christensen, J.~Hartong, N.~A.~Obers and B.~Rollier,
  ``Torsional Newton-Cartan Geometry and Lifshitz Holography,''
  Phys.\ Rev.\ D {\bf 89} (2014) 061901
  [arXiv:1311.4794 [hep-th]].
  
\bibitem{Christensen:2013rfa}
  M.~H.~Christensen, J.~Hartong, N.~A.~Obers and B.~Rollier,
  ``Boundary Stress-Energy Tensor and Newton-Cartan Geometry in Lifshitz Holography,'' 
  JHEP {\bf 1401} (2014) 057
  [arXiv:1311.6471 [hep-th]].    
  
\bibitem{Hartong:2014oma}
  J.~Hartong, E.~Kiritsis and N.~A.~Obers,
  ``Lifshitz space-times for Schr\"{o}dinger holography,''
  Phys.\ Lett.\ B {\bf 746} (2015) 318
  [arXiv:1409.1519 [hep-th]].
  
\bibitem{Bergshoeff:2014uea}
  E.~A.~Bergshoeff, J.~Hartong and J.~Rosseel,
  ``Torsional Newton-Cartan geometry and the Schr\"{o}dinger algebra,''
  Class.\ Quant.\ Grav.\  {\bf 32} (2015) no.13,  135017
  [arXiv:1409.5555 [hep-th]].
  
\bibitem{Hartong:2015wxa}
  J.~Hartong, E.~Kiritsis and N.~A.~Obers,
  ``Field Theory on Newton-Cartan Backgrounds and Symmetries of the Lifshitz Vacuum,''
  JHEP {\bf 1508} (2015) 006
  [arXiv:1502.00228 [hep-th]].
  
\bibitem{Bagchi:2010eg}
  A.~Bagchi,
  ``Correspondence between Asymptotically Flat Spacetimes and Nonrelativistic Conformal Field Theories,'' 
  Phys.\ Rev.\ Lett.\  {\bf 105} (2010) 171601 
  [arXiv:1006.3354 [hep-th]].  

\bibitem{Bagchi:2012cy}
  A.~Bagchi and R.~Fareghbal,
  ``BMS/GCA Redux: Towards Flatspace Holography from Non-Relativistic Symmetries,''
  JHEP {\bf 1210} (2012) 092
  [arXiv:1203.5795 [hep-th]].  
  
\bibitem{Bagchi:2016bcd}
  A.~Bagchi, R.~Basu, A.~Kakkar and A.~Mehra,
  ``Flat Holography: Aspects of the dual field theory,''
  JHEP {\bf 1612} (2016) 147
  [arXiv:1609.06203 [hep-th]].  

\bibitem{Lodato:2016alv}
  I.~Lodato and W.~Merbis,
  ``Super-BMS$_{3}$ algebras from $ \mathcal{N}=2 $ flat supergravities,''
  JHEP {\bf 1611} (2016) 150
  [arXiv:1610.07506 [hep-th]].

\bibitem{Bagchi:2019xfx}
  A.~Bagchi, A.~Mehra and P.~Nandi,
  ``Field Theories with Conformal Carrollian Symmetry,''
  arXiv:1901.10147 [hep-th].

\bibitem{Duval:2014uva}
  C.~Duval, G.~W.~Gibbons and P.~A.~Horvathy,
  ``Conformal Carroll groups and BMS symmetry,''
  Class.\ Quant.\ Grav.\  {\bf 31} (2014) 092001
  [arXiv:1402.5894 [gr-qc]].

\bibitem{Duval:2014lpa}
  C.~Duval, G.~W.~Gibbons and P.~A.~Horvathy,
  ``Conformal Carroll groups,''
  J.\ Phys.\ A {\bf 47} (2014) no.33,  335204
  [arXiv:1403.4213 [hep-th]].

\bibitem{Ciambelli:2018xat}
  L.~Ciambelli, C.~Marteau, A.~C.~Petkou, P.~M.~Petropoulos and K.~Siampos,
  ``Covariant Galilean versus Carrollian hydrodynamics from relativistic fluids,''
  Class.\ Quant.\ Grav.\  {\bf 35} (2018) no.16,  165001
  [arXiv:1802.05286 [hep-th]].  
  
\bibitem{Ciambelli:2018wre}
  L.~Ciambelli, C.~Marteau, A.~C.~Petkou, P.~M.~Petropoulos and K.~Siampos,
  ``Flat holography and Carrollian fluids,''
  JHEP {\bf 1807} (2018) 165
  [arXiv:1802.06809 [hep-th]].
  
\bibitem{Ciambelli:2018ojf}
  L.~Ciambelli and C.~Marteau,
  ``Carrollian conservation laws and Ricci-flat gravity,''
  Class.\ Quant.\ Grav.\  {\bf 36} (2019) no.8,  085004
  [arXiv:1810.11037 [hep-th]].  
    
\bibitem{Campoleoni:2018ltl}
  A.~Campoleoni, L.~Ciambelli, C.~Marteau, P.~M.~Petropoulos and K.~Siampos,
  ``Two-dimensional fluids and their holographic duals,''
  arXiv:1812.04019 [hep-th]. 

%
%
%
  
\bibitem{Ravera:2019ize}
  L.~Ravera,
  ``AdS Carroll Chern-Simons supergravity in 2 + 1 dimensions and its flat limit,''
  Phys.\ Lett.\ B {\bf 795} (2019) 331
  [arXiv:1905.00766 [hep-th]].  
  
\bibitem{Concha:2016zdb}
  P.~K.~Concha, O.~Fierro and E.~K.~Rodríguez,
  ``Inönü-Wigner contraction and $D=2+1$ supergravity,''
  Eur.\ Phys.\ J.\ C {\bf 77} (2017) no.1,  48
  [arXiv:1611.05018 [hep-th]]. 

\bibitem{IW}
  E. Inönü, E.P. Wigner, 
  ``On the Contraction of Groups and Their Representations,''
  Proc. Nat. Acad. Sci USA \textbf{39} (1953) 510.

\bibitem{WW}
  E. Weimar-Woods, 
  ``Contractions, Generalized Inönü-Wigner contractions and deformations of 
  finite-dimensional Lie algebras,'' Rev. Mod. Phys. \textbf{12} (2000) 1505.

\bibitem{Concha:2018jxx}
  P.~Concha, D.~M.~Peñafiel and E.~Rodríguez,
  ``On the Maxwell supergravity and flat limit in 2 + 1 dimensions,''
  Phys.\ Lett.\ B {\bf 785} (2018) 247
  [arXiv:1807.00194 [hep-th]].

\bibitem{Howe:1995zm}
  P.~S.~Howe, J.~M.~Izquierdo, G.~Papadopoulos and P.~K.~Townsend,
  ``New supergravities with central charges and Killing spinors in (2+1)-dimensions,''
  Nucl.\ Phys.\ B {\bf 467} (1996) 183
  [hep-th/9505032].

\bibitem{Giacomini:2006dr}
  A.~Giacomini, R.~Troncoso and S.~Willison,
  ``Three-dimensional supergravity reloaded,''
  Class.\ Quant.\ Grav.\  {\bf 24} (2007) 2845
  [hep-th/0610077].  

\bibitem{deAzcarraga:2011pa}
  J.~A.~de Azcarraga and J.~M.~Izquierdo,
  ``(p,q) D=3 Poincare supergravities from Lie algebra expansions,''
  Nucl.\ Phys.\ B {\bf 854} (2012) 276
  [arXiv:1107.2569 [hep-th]].

\bibitem{DK}
  S. Deser, J.H. Kay, 
  ``Topologically massive supergravity,''
  Phys. Lett. \textbf{B} 120 (1983) 97.

\bibitem{Deser}
  S. Deser, 
  ``Cosmological Topological Supergravity, Quantum Theory of Gravity: Essays in honor
  of the 60th Birthday of Bryce S,''
 (DeWitt. Published by Adam Hilger Ltd., Bristol, 1984).

\bibitem{PvN}
  P. van Nieuwenhuizen,
  ``Three-dimensional conformal supergravity and Chern-Simons terms,''
  Phys. Rev. D \textbf{32} (1985) 872.

\bibitem{AT1}
  A. Achucarro, P.K. Townsend, 
  ``A Chern-Simons action for three-dimensional anti-De Sitter supergravity theories,''
  Phys. Lett. B \textbf{180} (1986) 89.

\bibitem{RPvN}
  M. Rocek, P. van Nieuwenhuizen,
  ``$N \geq 2$ supersymmetric Chern-Simons terms as $d = 3$ extended conformal supergravity,''
  Class. Quant. Grav. \textbf{3} (1986) 43.

\bibitem{Witten}
  E. Witten, 
  ``(2+1)-Dimensional gravity as an exactly soluble system,''
  Nucl. Phys. B \textbf{311} (1988) 46.
  
\bibitem{AT2}
  A. Achucarro, P.K. Townsend, 
  ``Extended supergravities in $d = (2 + 1)$ as Chern-Simons theories,''
  Phys. Lett. B \textbf{229} (1989) 383.  

\bibitem{NG}
  H. Nishino, S.J. Gates Jr., 
  ``Chern-Simons theories with supersymmetries in three dimensions,''
  Mod. Phys. A \textbf{8} (1993) 3371.



\bibitem{Banados:1996hi}
  M.~Banados, R.~Troncoso and J.~Zanelli,
  ``Higher dimensional Chern-Simons supergravity,''
  Phys.\ Rev.\ D {\bf 54} (1996) 2605
  [gr-qc/9601003].


  
\bibitem{Concha:2019icz}
  P.~Concha,
  ``$\mathcal{N}$-extended Maxwell supergravities as Chern-Simons theories in three spacetime dimensions,''
  Phys.\ Lett.\ B {\bf 792} (2019) 290
  [arXiv:1903.03081 [hep-th]].  
  
\bibitem{Lukierski:2006tr}
  J.~Lukierski, I.~Prochnicka, P.~C.~Stichel and W.~J.~Zakrzewski,
  ``Galilean exotic planar supersymmetries and nonrelativistic supersymmetric wave equations,''
  Phys.\ Lett.\ B {\bf 639} (2006) 389
  [hep-th/0602198].  

\bibitem{Concha:2019mxx}
  {P.~Concha, L.~Ravera and E.~Rodríguez,
  ``Three-dimensional Maxwellian extended Bargmann supergravity,''
  arXiv:1912.09477 [hep-th].}

  
\bibitem{Concha:2018ywv}
  P.~Concha, L.~Ravera and E.~Rodríguez,
  ``On the supersymmetry invariance of flat supergravity with boundary,''
  JHEP {\bf 1901} (2019) 192
  [arXiv:1809.07871 [hep-th]].  

\bibitem{Concha:2018zeb}
  P.~Concha, N.~Merino, O.~Miskovic, E.~Rodríguez, P.~Salgado-Rebolledo and O.~Valdivia,
  ``Asymptotic symmetries of three-dimensional Chern-Simons gravity for the Maxwell algebra,''
  JHEP {\bf 1810} (2018) 079
  [arXiv:1805.08834 [hep-th]]. 
  
\end{thebibliography}
\end{document}